\documentclass[
    reprint, 
    aps,
    preprintnumbers,
    twocolumn,
    prb,
    superscriptaddress
]{revtex4-2}

\usepackage[utf8]{inputenc}
\usepackage{booktabs}
\usepackage[multiple]{footmisc}
\usepackage{lipsum}
\usepackage{rotating}
\usepackage{lipsum}

\usepackage{amssymb}
\usepackage{amsbsy}
\usepackage{amsmath}
\usepackage{bm}
\usepackage{bbm}
\usepackage{xfrac}
\usepackage{mathtools}

\usepackage{tikz}
\usepackage[T1]{fontenc}
\usepackage{etoolbox}
\usepackage{graphics}
\usepackage{multirow}
\usepackage{url}
\usepackage{hyperref}
\usepackage{placeins}
\usepackage[normalem]{ulem}

\usepackage{color}

\usepackage{adjustbox}

\graphicspath{{mainplots/}} 

\newcommand*\vc[1]{\mathbf{#1}}
\newcommand*\tx[1]{\mathrm{#1}}
\newcommand*\wn{cm$^{-1}$}
\newcommand\tsT{\rule{0pt}{2.6ex}} 
\newcommand\tsB{\rule[-1.2ex]{0pt}{0pt}} 

\newcommand{\1}[1]{#1}

\begin{document}

\title[]{Fast Sampling of Protein Conformational Dynamics}

\author{Michael A. Sauer}
\author{Souvik Mondal}
\author{Brandon Neff}
\author{Sthitadhi Maiti}
\author{Matthias Heyden}
\email{mheyden1@asu.edu}
\affiliation{School of Molecular Sciences, Arizona State University, Tempe, AZ 85287, U.S.A.}

\date{\today}

\begin{abstract}
Protein function does not depend solely on structure but often relies on dynamical transitions between distinct conformations. 
Despite this fact, our ability to characterize or predict protein dynamics is substantially less developed compared to state-of-the-art protein structure prediction.
Molecular simulations provide unique opportunities to study protein dynamics, but the timescales associated with conformational changes generate substantial challenges. 
Enhanced sampling algorithms with collective variables can greatly reduce the computational cost of sampling slow processes. 
However, defining collective variables suitable to enhance sampling of protein conformational transitions is nontrivial. 
Low-frequency vibrations have long been considered as promising candidates for collective variables, but their identification so far relied on assumptions inherently invalid at low frequencies.
We recently introduced an analysis of molecular vibrations that does not rely on such approximations and remains accurate at low frequencies. 
Here, we modified this approach to efficiently isolate low-frequency vibrations in proteins and applied it to a set of five proteins of varying complexity.
We demonstrate that our approach is not only highly reproducible, but results in collective variables that consistently enhance the sampling of protein conformational transitions and associated free energy surfaces on timescales compatible with high-throughput applications. 
This enables the efficient generation of protein conformational ensembles, which will be key for future prediction algorithms aiming beyond static protein structures.
\end{abstract}

\maketitle

\section*{Introduction}
Recent advances in protein structure prediction driven by machine learning ({\em e.g.}, Alphafold~\cite{senior2020improved,jumper2021highly,gao2022af2complex,abramson2024accurate}, ESMfold~\cite{lin2023evolutionary}, RoseTTAFold~\cite{baek2021accurate,krishna2024generalized}) have dramatically improved capabilities for structure-based designs of enzymes and other proteins ({\em e.g.}, RFdiffusion~\cite{watson2023novo}).
However, protein function is often not the result of a single stable conformation but involves reversible conformational transitions between two or more distinct states, {\em i.e.}, protein dynamics~\cite{petrovic2018conformational,campitelli2020role,ayaz2023}.

The characterization or prediction of protein dynamics requires more intricate information on the associated free energy surfaces than current machine-learning models can provide.
Stability differences between distinct conformational states that reversibly inter-convert are small (by definition) and thus challenging to predict reliably.
Nevertheless, machine learning (ML) approaches are likely going to play a critical role for the inclusion of protein dynamics in computational enzyme design strategies.
A remaining challenge is the lack of reliable and diverse training data for protein dynamics. 
Compared to hundreds of thousands of experimentally determined protein structures in the PDB~\cite{PDB}, the dynamics of proteins are characterized in detail for only a limited set of systems.

With few exceptions ({\em e.g.}, time-resolved x-ray crystallography~\cite{branden2021advances,hekstra2023emerging}), many experiments characterize protein conformational dynamics only in terms of distance fluctuations between individually labeled sites~\cite{nettels2024single}. 
Even sophisticated multi-color single-molecule FRET experiments~\cite{hohng2014maximizing,gotz2016multicolor,chung2018protein,yoo2020fast} that track multiple distances simultaneously, do not provide fine-grained structural information that is easily used to train ML models of protein dynamics.

Classical molecular dynamics (MD) simulations play a critical role in expanding the available data on protein conformational transitions at an atomic-level. 
Enhanced sampling simulations along suitable CVs can drastically accelerate sampling of slow conformational dynamics~\cite{barducci2008well,zwier2015westpa}.
However, without prior knowledge of suitable collective variables (CVs) to drive conformational transitions in biased simulations, MD simulations of proteins are typically under-sampled (with notable exceptions~\cite{lindorff2011fast,lindorff2016picosecond,shaw2021,zimmerman2021sars,ayaz2023}) and remain too costly to run on timescales associated with relevant biomolecular functions (often milliseconds to seconds).
 
Determining a suitable set of low-dimensional CVs to explore the dynamics of a given protein has remained a pervasive challenge.
Selecting CVs typically requires prior knowledge of the conformational changes to be observed and, ideally, CVs align with the most likely transition pathways.
Identifying suitable CVs without prior knowledge is an active field of research~\cite{brotzakis2018enhanced,bonati2023unified,shmilovich2023girsanov,rydzewski2022reweighted,mehdi2024enhanced,shen2024adaptive}.  

We recently developed a theoretical framework that allows for the analysis of anharmonic low-frequency vibrations in molecules including proteins: frequency-selective anharmonic (FRESEAN) mode analysis~\cite{sauer2023frequency}. 
In contrast to low-frequency harmonic or quasi-harmonic normal modes, which have historically been envisioned as potential CVs to explore protein conformational space~\cite{brooks1983harmonic,yang2007well,mahajan2017jumping,costa2023}, FRESEAN mode analysis does not rely on approximations that are invalid at the lowest frequencies.
Instead, we demonstrated that FRESEAN mode analysis can isolate collective degrees of freedom that describe structural distortions associated with minimal restraining forces~\cite{sauer2023frequency}.
We speculated that such motions may be associated with conformational dynamics~\cite{sauer2023frequency} and performed enhanced sampling MD simulations for a test system, alanine dipeptide, that utilized vibrational modes obtained from FRESEAN mode analysis as CVs to speed up sampling of the peptide conformational space~\cite{mondal2024exploring}.

Here, we expand this work to a set of proteins  with known conformational dynamics that provide a suitable test set.
We combine our approach with coarse-grained representations of all-atom protein simulations designed to retain information on the lowest-frequency vibrations. 
In the following, we show that FRESEAN mode analysis of short MD simulations yields highly reproducible vibrational modes at low frequencies. 
Further, enhanced sampling simulations using these modes as CVs reliably sample known conformational transitions in each test protein.
Our approach does not require prior knowledge of expected conformational transitions and can be performed in a single day using standard high performance computing hardware.
Such throughput enables, for example, generating large data sets for protein conformational ensembles to train the next generation of ML models, which are capable to predict combined relationships between protein sequence, structure, and dynamics.

\section*{Anharmonic Low-Frequency Modes in Proteins}

To test our hypothesis that anharmonic low-frequency vibrations describe natural collective variables (CVs) for enhanced sampling simulations of protein dynamics, we focused on five well-studied protein systems: hen egg-white lysozyme (HEWL)~\cite{costa2023}, HIV-1 protease (HIV-1 Pr)~\cite{huang2018replica}, myeloid cell leukemia 1 (MCL-1)~\cite{benabderrahmane2020insights}, ribose binding protein (RBP)~\cite{ren2021unraveling}, and the Kirsten rat sarcoma virus (KRAS) protein~\cite{chen2021mutation} (bound to GDP). 
This diverse set includes a homo-dimeric complex (HIV-1 Pr) and a multi-domain protein (RBP) in addition to single-domain proteins (HEWL, MCL-1, KRAS).
In this group, HEWL stands out due four internal disulfide bridges that stabilize its structure.

For each protein, we applied FREquency-SElective ANharmonic (FRESEAN) mode analysis~\cite{sauer2023frequency} to extract anharmonic low-frequency modes from all-atom molecular dynamics (MD) trajectories (see Methods and \1{Supporting Information, SI}).
We derived FRESEAN mode analysis {\em via} a time-correlation formalism of atomic velocity fluctuations that identifies, at any given frequency, collective vibrational modes based on contributions to the vibrational spectrum~\cite{sauer2023frequency} ({\em i.e.}, the vibrational density of states, VDoS, defined in \1{Eq. S2 in the SI}). 
A full set of orthogonal normal modes can be generated for any sampled frequency, but we focus our analysis here exclusively on modes generated at zero frequency (implied in the remainder of the paper).
Zero frequency modes describe diffusion and relaxation processes as well as the lowest frequency vibrations of the system.
Notably, harmonic approximations are intrinsically invalid in this regime~\cite{sauer2023frequency}, which motivated us to develop FRESEAN mode analysis.
As a time-correlation formalism that does not assume a single set of orthogonal modes~\cite{sauer2023frequency}, FRESEAN mode analysis does not rely on harmonic approximations.

To minimize computational cost, we introduce here a coarse-grained representation (two beads per residue, see \1{SI} for details) of all-atom protein trajectories as input for FRESEAN mode analysis.
As shown in \1{Figure~S2 of the SI}, this representation preserves collective low-frequency vibrations.

\section*{Reproducible Detection of Low-Frequency Modes}
To employ anharmonic low-frequency vibrations as reliable CVs in enhanced sampling simulations, they must be easily reproduced from independent sets of simulations.
To test this, we applied FRESEAN mode analysis to five replicas of 20~ns equilibrium trajectories for each protein (each replica equilibrated with randomized velocities, see \1{SI} for details). 

Modes 1-6 (at zero frequency) describe translational and rotational diffusion of the entire protein (see \1{Figure~S3 in the SI}) and are excluded from collective variable selection.
Modes with indices 7 and larger describe vibrations whose contributions to the zero frequency VDoS decrease with increasing mode index.
We analyzed contributions of individual modes to the vibrational spectrum at all frequencies by projecting fluctuations in our trajectories onto each mode~\cite{sauer2023frequency}.
For each system, we plotted these contributions for modes 1-10 (including translations and rotations) in \1{Figure~S4 in the SI}.
Notably, the spectra in \1{Figure~S4} are free from other spectral signatures, {\em e.g.}, vibrations at higher frequencies. 
This highlights that the selected modes truly isolate low-frequency vibrations in the simulated proteins (in contrast to harmonic or quasi-harmonic modes, see Ref.~\citenum{sauer2023frequency}).
Zero-frequency contributions of low-frequency vibrations (modes 7-10) result from the low-frequency tail of a peak at 5-13~\wn, depending on the system.
This is characteristic for anharmonic low-frequency vibrations and informs our CV selection for enhanced sampling simulations.
In the following, we focus our analysis on modes 7-9, {\em i.e.}, the three vibrations with the largest zero-frequency contribution to the VDoS. 

\begin{figure}[t!]
\centering
\includegraphics[width=0.35\textwidth]{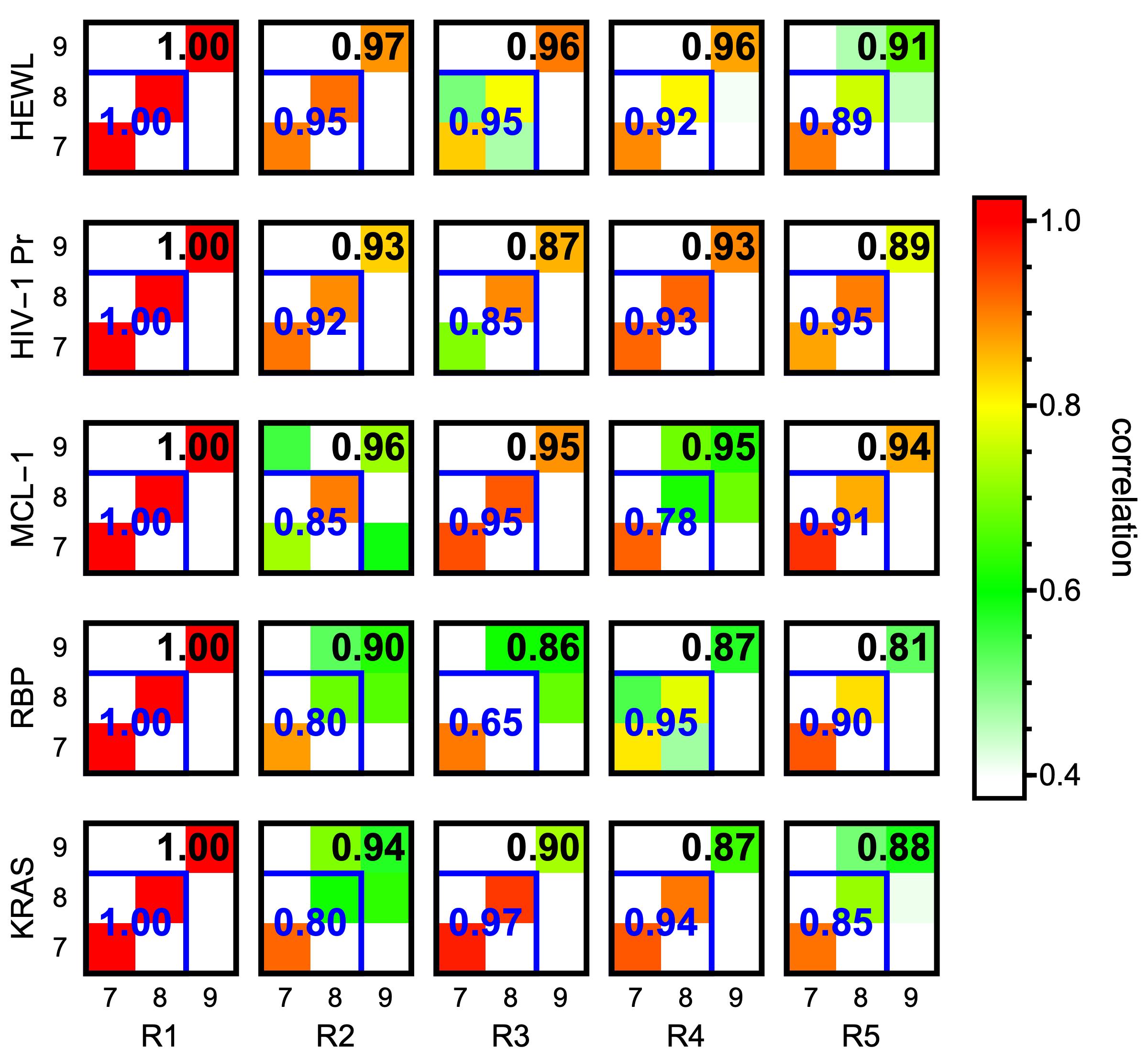}
\caption{
Matrices describing correlations between FRESEAN modes 7-9 in replicas R1 and R1 to R5 for each protein system (self-correlations with replica 1 are 1.0 by definition and only shown for clarity, see \1{Figure S5 in the SI} for all correlations).
Colors of individual squares indicate the direct correlation between a pair of modes.
Correlation coefficients between 2D (3D) sub-spaces described by modes 7-8 (7-9) in a pair of simulations are shown as blue (black) numerals.
}
\label{f:matrix}
\end{figure}

A key question is whether our analysis provides reproducible assignments of anharmonic low-frequency modes.
We investigate this in Figure~\ref{f:matrix} using correlation coefficients between pairs of modes 7-9 from the five 20~ns replica simulations generated for each system (colored squares).
For clarity, only correlations with replica R1 are shown (see \1{Figure~S5 in the SI} for all correlations).

Correlations close to 1.0 (red) along the diagonal indicate that modes are reproduced in identical order. 
In some cases, correlation coefficients along the diagonal are smaller but accompanied by non-zero off-diagonal correlations. 
In these cases, it is more insightful to test whether the 2D and 3D sub-spaces described by modes 7-8 and 7-9 are reproduced (see \1{Eqs.~S12 and S13 in the SI}). 
These correlations are indicated in Figure~\ref{f:matrix} as blue and black numerical values, respectively.
Overall, the correlation analysis demonstrates that low-frequency vibrational modes are reproduced with high confidence between distinct replicas, despite the short length of the simulations (20~ns).
In particular the 2D sub-spaces spanned by modes 7-8, which we used as CVs for enhanced sampling simulations below, frequently exhibit correlations $>$0.9. 

Notably, this is not true for alternative approaches such as quasi-harmonic or principal component modes that aim to describe molecular vibrations based on a single co-variance matrix of atomic displacements.
Correlations between pairs of low-frequency quasi-harmonic modes obtained from the same trajectories, as well as 2D and 3D sub-spaces spanned by two or three modes, rarely exceed 0.5 (\1{Figures~S6 \& S7 in the SI}).
Interestingly, extending the length of simulations to the microsecond timescale improves the reproducibility of co-variance matrix eigenvectors only moderately.
This is shown in \1{Figures~S8 \& S9 in the SI} for principal component modes of HEWL computed from five 1~microsecond replica simulations.
In contrast, the high reproducibility of low-frequency mode assignments with FRESEAN mode analysis demonstrates that, in addition to its ability to isolate anharmonic low-frequency vibrations~\cite{sauer2023frequency}, its underlying time-correlation formalism does not depend on rare events.

\section*{Fast and Consistent Sampling of Protein Dynamics}

\begin{figure}[t!]
\centering
\includegraphics[width=0.48\textwidth]{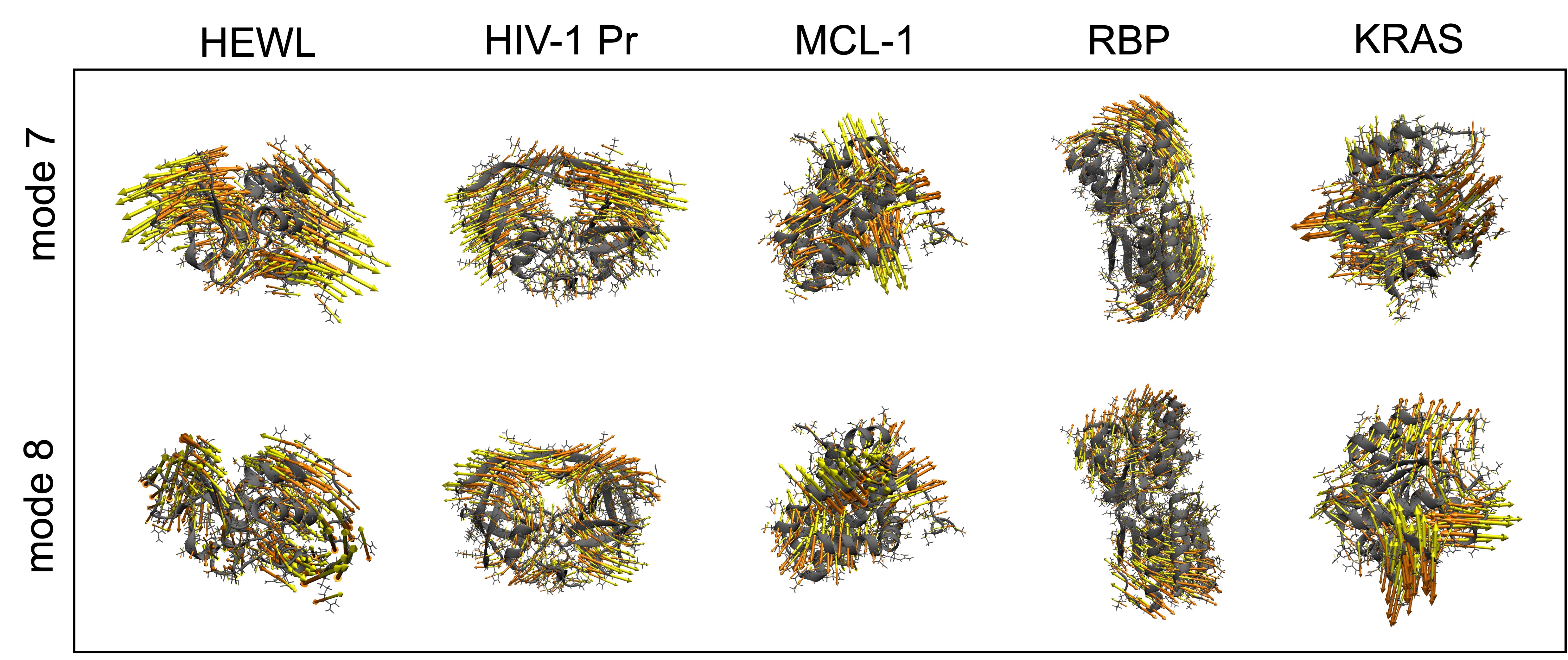}
\caption{
Visualization of anharmonic low-frequency vibrational modes used as CVs for enhanced sampling simulations. 
The displayed modes were selected for their contributions to the VDoS at zero frequency using FRESEAN mode analysis and are shown as displacement vectors (yellow and orange) relative to a reference structure.
}
\label{f:modes}
\end{figure}

Next, we tested the ability of anharmonic low-frequency modes identified with FRESEAN mode analysis to enhance sampling of conformational dynamics in molecular dynamics simulations.
Specifically, we used modes 7-8, {\em i.e.}, the two most prominent vibrations contributing to the VDoS at zero frequency, as CVs in enhanced sampling simulations with the well-tempered metadynamics protocol~\cite{barducci2008well}
(see \1{SI} for details).
For each protein, we visualized modes 7-8 (replica R1) as displacement vectors relative to a reference structure in Figure~\ref{f:modes} to illustrate the associated collective motions.

For each system, we performed metadynamics simulations using modes 7-8 as CVs that were obtained independently from the five replica simulations described above.
Each simulation was stopped after 100~ns (less than 24~hours on a single GPU) and analyzed for sampled conformational changes.
To test whether these simulations provided consistent information on protein conformational dynamics, we first analyzed the biased simulation trajectories.
Despite the presence of the biasing term, these trajectories provide insight into whether characteristic large amplitude motions are consistently sampled.
We analyzed the latter using geometric variables (see \1{Table~S2 and Figure~S10 in the SI}) that have been introduced previously for each system~\cite{costa2023,huang2018replica,chen2021mutation,ren2021unraveling,benabderrahmane2020insights}.
These geometric variables contain prior knowledge of the expected dynamics, but they are not part of our enhanced sampling protocol.
We use them to facilitate comparisons with previous work and to characterize the efficiency of our sampling scheme.

\begin{figure}[t!]
\centering
\includegraphics[width=0.49\textwidth]{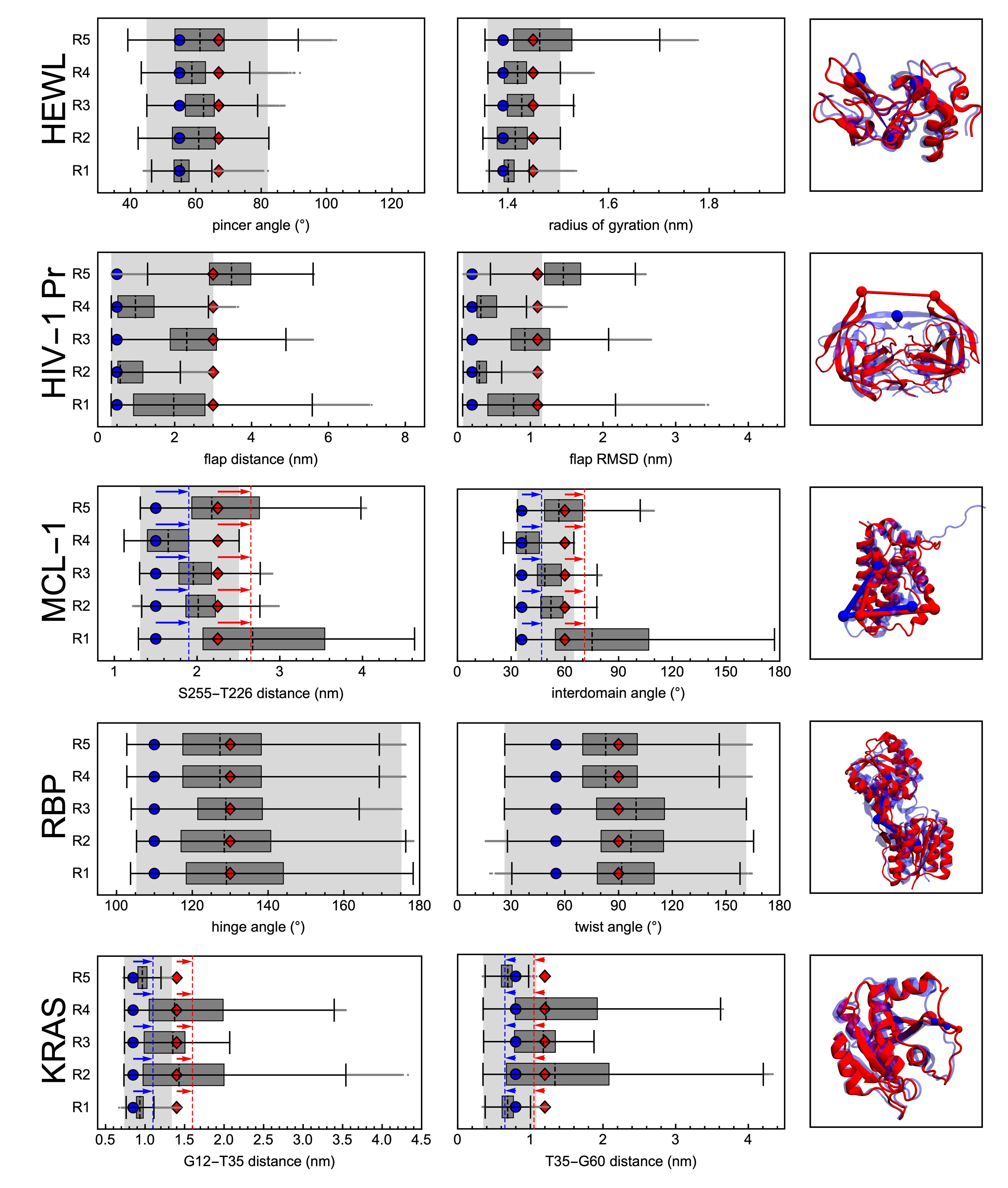}
\caption{
Box and whisker plots for each system illustrate the sampled conformational space in each metadynamics simulation replica R1 to R5.
Plotted are distributions of two geometric variables, defined for each system in \1{Table~S2 of the SI}, computed directly from the biased trajectories (left and mid panels,
see \1{SI} for definition of box and whiskers).
Blue circles and red diamonds indicate conformational states reported in the literature (for MCL-1 and KRAS, arrows and dashed lines indicate constant shifts derived from free energy surfaces in Fig.~\ref{f:fes}).
Insets on the right illustrate representative super-imposed structures from our simulations that correspond to "closed" and "open" states.
}
\label{f:whisker}
\end{figure}

The results are shown in Figure~\ref{f:whisker}, where the extent of motion along the geometric variables is shown as box and whisker plots. 
A gray background indicates the range of motion consistently described by each of the five independent replicas.
Blue and red symbols show the range of previously reported conformations for each system (for simplicity, we refer to them as "closed" and "open", respectively; see \1{SI} for details) and the insets display representative structures~\cite{desimone2013characterization,huang2018replica,chen2021mutation,ren2021unraveling,benabderrahmane2020insights}.
Both conformations and transitions between them are consistently explored in each simulation replica for HEWL, HIV-1 Protease and RBP. 
For MCL-1 and KRAS, we observed constant shifts between coordinates reported in the literature~\cite{benabderrahmane2020insights,chen2021mutation} and free energy minima in our simulations (see next section).
These shifts and free energy minima observed in our simulations are indicated as arrows and vertical dashed lines in Figure~\ref{f:whisker}.
Taking this into account, we find that for MCL-1 four out of five replicas sample both sates, while one replica (R4) narrowly misses the shifted "open" state.
For KRAS, we find that all five replicas sample the full range of T35-G60 distances associated with the "closed" to "open" transition (including the shifts).
However, two replicas (R1 and R5) do not fully sample the G12-T35 distance of the shifted "open" state.
Thus, considering all systems and five respective replicas, 22 out of 25 metadynamics trajectories (88\%) sample known conformational transitions in less than 100~ns (using CVs extracted from unbiased 20~ns simulations).
We note that extending the metadynamics trajectories to 155~ns results in full sampling of "closed" to "open" transitions for every replica of each protein.

A comparison to five replicas of unbiased 100~ns control simulations for each system in \1{Figure S12 of the SI} demonstrates the impact of our enhanced sampling scheme. 
For all systems, enhanced sampling along anharmonic low-frequency modes increased the consistently sampled range of motion at least two-fold (closer to sixfold for HIV-1 Protease and RBP).
Despite the overall much more constricted dynamics in the unbiased control simulations, individual trajectories for HEWL and KRAS sampled one full transition, indicating the unpredictability of rare events.

\section*{Sampling Free Energy Surfaces}
FESs as a function of the respective CVs obtained directly from the five independent metadynamics simulations are shown in \1{Figure S13 in the SI}.
The independently sampled CVs used in each simulation are similar as shown in Figure~\ref{f:matrix} but not identical, which hinders direct comparisons between replicas. 
However, after unbiasing the metadynamics trajectories, we projected weighted conformational ensembles on the geometric variables introduced in Figure~\ref{f:whisker} to construct FESs with a common set of variables~\cite{branduardi2012}.
In contrast to the biased distributions shown in Figure~\ref{f:whisker}, the resulting FESs in \1{Figure~S14 in the SI} allow us to distinguish high and low free energy conformations sampled during the metadynamics simulations.
Correlations between geometric variables (sampled states fall on diagonal) indicate redundant information.
Notably, we do not observe such correlations for the CVs used to enhance sampling in our simulations, {\em i.e.}, modes 7-8 obtained from FRESEAN mode analysis at zero frequency, which are orthogonal collective motions by definition.

\begin{figure*}[t!]
\centering
\includegraphics[width=0.9\textwidth]{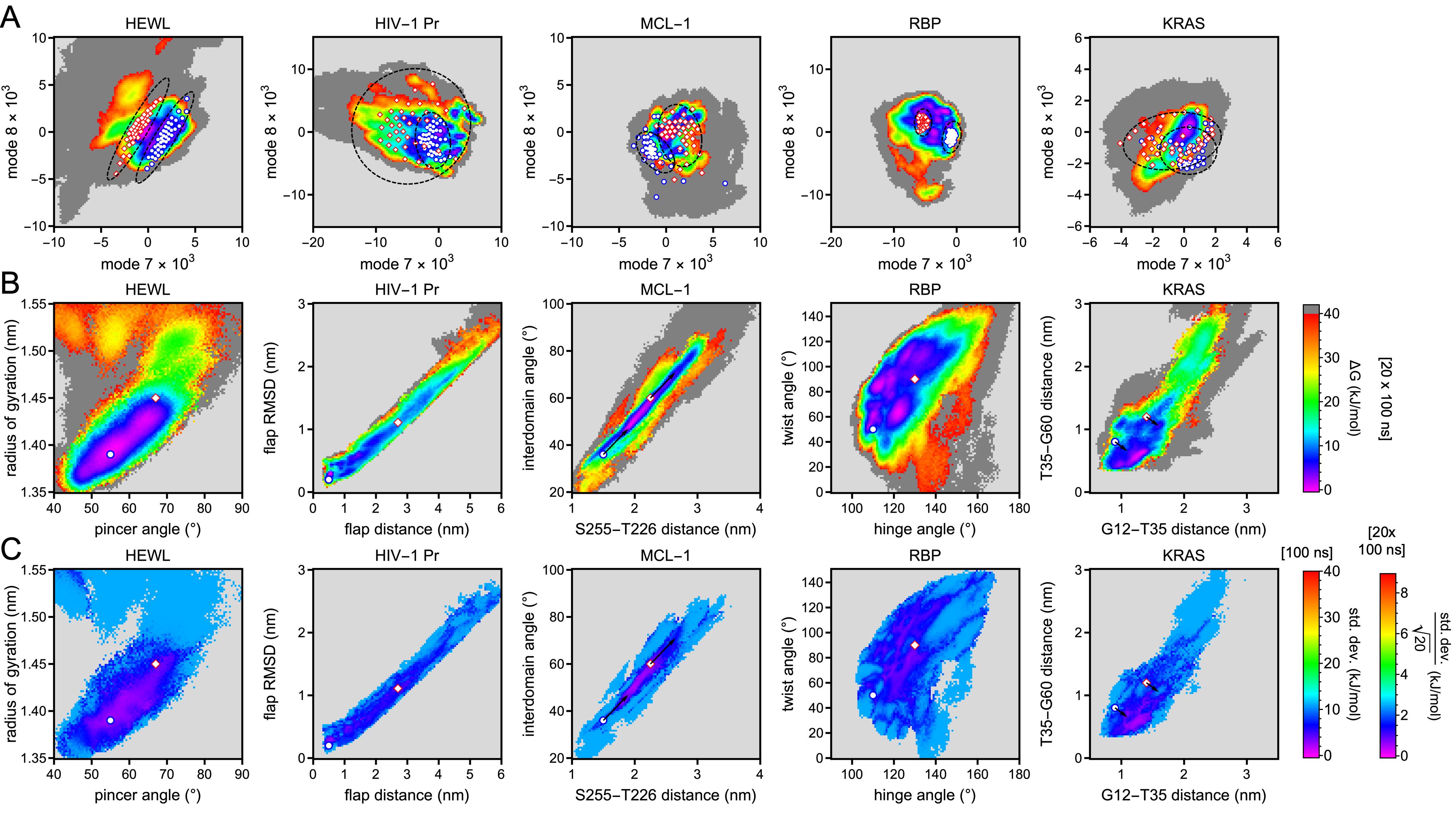}
\caption{
Free energy surfaces and statistical errors obtained from 20 replica simulations for each system. (A) FESs as a function of CVs defined by anharmonic low-frequency vibrations. Open circles (blue) and diamonds (red) indicate 40 microstates of the "closed" and "open" ensembles (including constant shifts for MCL-1 and KRAS). Ellipses highlight the separation/overlap between both ensembles. (B) FESs as a function of geometric variables with the "closed" and "open" states indicated by one open circles (blue) and diamonds (red). For MCL-1 and KRAS, arrows indicate constant shifts between prior reports of these states and free energy minima in our simulations. (C) Statistical uncertainties of the FESs in geometric variable space with "closed" and "open" states and shifts indicated as in (B).
}
\label{f:fes}
\end{figure*}

Comparisons of FESs obtained for distinct replica simulations of the same protein \1{(Figure~S14 in the SI)} show that sampling remains insufficient to fully characterize the conformational ensembles. 
To overcome this limitation, we increased the number of replica metadynamics simulations for each system to 20 (starting structures sampled every 5~ns from unbiased 100~ns trajectories), each with 100~ns of metadynamics sampling and parallel processing.
Here, we used identical vibrational modes as CVs (obtained for R1 above) in each replica to allow for direct comparisons and averaging of thermodynamic ensembles (see \1{SI} for details).
In Figure~\ref{f:fes}, we plot the resulting FESs (obtained from averaged weighted probability distributions, see \1{SI}) both in the space defined by vibrational modes (panel A) and geometric variables (panel B).
Statistical errors, {\em i.e.}, the standard deviation describing the uncertainty of a single 100~ns simulation and the standard error of the mean for all 20 replica simulations, are shown as a function of the geometric variables in panel C.
These statistical errors indicate uncertainties of less than $\pm$10~kJ/mol for a single 100~ns simulation and less than $\pm$3~kJ/mol after averaging.
This is further emphasized in \1{Figure S15 in the SI}, in which we plot a 1-dimensional minimum free energy path for the "closed" to "open" transition in HEWL including statistical error bars.
Such small uncertainties are remarkable given that error bars for protein FESs are often not even reported.~\cite{costa2023,huang2018replica,chen2021mutation,ren2021unraveling,benabderrahmane2020insights}
The generation of converged FESs for five distinct protein systems, obtained with the same general protocol, highlights the role of anharmonic low-frequency vibrations as drivers of conformational change.

We indicate previously reported conformations~\cite{desimone2013characterization,huang2018replica,chen2021mutation,ren2021unraveling,benabderrahmane2020insights} (referred to as "closed" and "open" in analogy to Figure~\ref{f:whisker}) with blue circles and red diamonds in panels B and C of Figure~\ref{f:fes}.
Our simulations accurately reproduce previously reported conformations (apart from constant shifts for MCL-1 and KRAS), while revealing additional details on the FES that have so far not yet been described.
The success of low-frequency vibrations to enhance sampling of previously observed conformational changes is readily explained upon projection of the "closed" and "open" state ensembles into our CV space.
Notably, any pair of geometric variables (a single point in Figure~\ref{f:fes}B) describes an ensemble of conformations in alternative representations. 
In panel A of Figure~\ref{f:fes}, we indicate 40 conformations representative of the "closed" and "open" states for each system (selected by k-means clustering) in CV spaces defined by low-frequency vibrations.
With the exception of KRAS, the "closed" and "open" conformations are well separated in CV space (highlighted by ellipses), confirming that the low-frequency vibrations used here as CVs are directly related to conformational dynamics reported previously for these proteins in the literature.\cite{costa2023,huang2018replica,ren2021unraveling,benabderrahmane2020insights}
The lack of separation between the "closed" and "open" states of KRAS in our CV space indicates that the low-frequency vibrations detected for KRAS describe independent collective motions that are not directly coupled to conformational dynamics reported previously for this system.
Nevertheless, sampling is enhanced for KRAS due to indirect effects, {\em e.g.}, the biased motion along vibrational modes allows the system to access alternative transition pathways with lower free energy barriers.
\section*{Conclusion and Outlook}
In this study, we introduced anharmonic low-frequency vibrations as natural CVs for enhanced sampling simulations of protein conformational transitions. 
Such vibrations can be extracted reproducibly from short unbiased simulation trajectories on the nanosecond timescale using an frequency-selective anharmonic mode analysis\cite{sauer2023frequency}.
Our approach enables enhanced sampling simulations of protein conformational dynamics without requiring prior knowledge of the type motions involved. 

We follow longstanding conceptual ideas on the relationship between low-frequency vibrations, barrier crossing events, and conformational transitions\cite{brooks1983harmonic,levitt1985protein,chennubhotla2005elastic,yang2007well,mahajan2017jumping}.
The key advance here is the possibility to isolate true low-frequency vibrations without simplifying the system of implying harmonic approximations that are invalid in the low-frequency limit.
The absence of high-frequency fluctuations (see \1{Figure~S4 in the SI}) along the identified modes ensures that only minimal biasing forces are sufficient to alter the protein conformation.
Consequently, biasing simulations along the identified modes consistently speeds up protein conformational sampling.

Averaging over multiple parallel simulations on timescales compatible with processing times of less than 24 hours on GPUs for many systems can produce converged free energy surfaces and corresponding conformational ensembles.
This high throughput enables the generation of large datasets as well as systematic studies on individual proteins, {\em e.g.}, by comparing the dynamics of distinct mutants.
Such fast and efficient strategies to sample protein conformational ensembles are critical for future revolutions in the development of predictive models that combine protein structure and dynamics.
State-of-the-art predictors for stable protein structures\cite{senior2020improved,jumper2021highly,baek2021accurate,lin2023evolutionary,watson2023novo,abramson2024accurate,krishna2024generalized} were enabled by the existence of extensive datasets ranging from known protein structures to vast amounts of genomic sequences\cite{PDB,uniprot2019uniprot}.
So far, similar predictions for protein conformational ensembles have currently only been achieved for intrinsically disordered proteins, using datasets generated with computationally efficient coarse-grained polymer simulations\cite{janson2023direct,tesei2024conformational,lotthammer2024direct}.
The enhanced sampling scheme presented here, which can be applied without requiring prior knowledge of expected conformational transitions, provides an efficient approach to enable similar advances for folded proteins.
\subsection*{Methods}

Simulation protocols for all-atom molecular dynamics simulations of all systems are described in the \1{SI}, combined with a description of their coarse-grained representations, details on the FRESEAN mode analysis, and the well-tempered metadynamics simulations.
Briefly, all simulations were performed with GROMACS 2022.5 and started from crystal structures reported in PDB entries 1HEL for HEWL, 1BVE for HIV-1 Protease, 3WIX for MCL-1, 1DRJ for RBP, and 5WCC for KRAS.
Force fields employed for each system were selected in accordance with prior work in the literature to allow for direct comparisons.
Each protein was protonated assuming neutral pH conditions and solvated in a cubic box in aqueous solution with 150~mM NaCl (see \1{Table S1 in the SI} for details).

With position restraints applied to non-hydrogen protein atoms, the energy of each system was minimized and five replicas of each system were equilibrated in the NPT ensemble with independently sampled initial velocities.
For each replica of each system, we then performed unrestrained NPT simulations for 20~ns to perform the FRESEAN mode analysis. 
For the analysis, protein trajectories (coordinates \& velocities stored every 20~fs) were: 1) rotated into a reference coordinate system based on the crystal structure; and 2) coarse-grained into beads (1 bead for glycines, two beads for each other amino acid).
A matrix of mass-weighted time velocity cross correlation functions was computed for all coarse-grained degrees of freedom with a maximum correlation time of 2~ps. 
We time-symmetrized each matrix term and performed a Fourier transform into the frequency domain with a 10~\wn Gaussian window. 

The eigenvectors (modes) of the zero-frequency matrix describe displacement vectors corresponding to translational motion (modes 1-3), rotational motion (modes 4-6), and anharmonic low-frequency vibrations (modes 7+ with non-zero eigenvalues). 
We used modes 7 and 8 to define CVs for 100~ns well-tempered metadynamics simulations with the PLUMED-2.8.2 plug-in for GROMACS.
The CVs are evaluated by: 1) aligning the protein to the equilibrated starting structure as a reference; 2) switching to coarse-grained representation; 3) computing displacement vectors relative to the reference structure; 3) projecting the displacement vectors onto the mode vector.
Metadynamics simulations were performed for each replica of each system (5 systems $\times$ 5 replicas $\times$ 100~ns). 

In addition, we performed for replica 1 of each system an additional 100~ns unbiased simulations from which we extracted 20 sets of coordinates \& velocities every 5~ns.
These states were then used to start an additional set of 20 metadynamics simulations for each system (5 systems $\times$ 20 simulations $\times$ 100~ns) to analyze the convergence of free energy surfaces.

\subsection*{Software Availability}
Source code for FRESEAN mode analysis and input files for WT-metadynamics simulations are available at \url{https://github.com/HeydenLabASU-collab/FRESEAN-metadynamics}.

\begin{acknowledgements}
This work is supported by the National Science Foundation (CHE-2154834) and the National Institute of General Medical Sciences (R01GM148622). The authors acknowledge Research Computing at Arizona State University for providing high performance computing resources~\cite{sol} that have contributed to the research results reported within this work.
\end{acknowledgements}

\pagebreak
\widetext
\newpage
\begin{center}
\textbf{\large Supporting Information: Fast Sampling of Protein Conformational Dynamics}
\end{center}
\setcounter{equation}{0}
\setcounter{figure}{0}
\setcounter{table}{0}
\setcounter{page}{1}
\makeatletter
\renewcommand{\theequation}{S\arabic{equation}}
\renewcommand{\thefigure}{S\arabic{figure}}
\renewcommand{\thetable}{S\arabic{table}}

\subsection{Simulation Protocol}
All simulations were performed using the GROMACS 2022.5 software package.\cite{abraham15} Simulations were performed on five systems, including Hen Egg White Lysozyme (HEWL; 1HEL.pdb), HIV-1 Protease (HIV-1 PR; 1HHP.pdb), Myeloid cell leukemia-1 (MCL-1; 3WIX.pdb), K-Ras (5W22.pdb) and Ribose-Binding Protein (RBP; 1DRJ.pdb, G134R mutant). 
Molecular mechanics force fields used and the number of water molecules and ions included in each system are listed in Table~\ref{t:sys} and follow previous simulation studies of these systems in the literature.\cite{desimone2013characterization,huang2018replica,chen2021mutation,ren2021unraveling,benabderrahmane2020insights}
All simulations were performed with the TIP3P water model~\cite{jorgensen83} for Amber-family force fields and the modified TIP3P model (also known as TIPS3P) for CHARMM-family force fiels.\cite{mackerell98}
In accordance to previous literature, RBP, more precisely its G134R mutant, carried no charge and was simulated in pure water.\cite{ren2021unraveling} 

\begin{table}[ht]
\centering
\begin{adjustbox}{width=0.95\textwidth}
\small
\begin{tabular}{l|ccccc}
    \hline
     Protein Name & PDB & Force Field & Water & Na$^{+}$ & Cl$^{-}$ \\ [0.5ex] 
     \hline\hline
     Hen Egg White Lysozyme (HEWL) & 1HEL.pdb & Amber99sb~\cite{hornak2006amber99sb} & 25644 & 73 & 81\\ 
     HIV-1 Protease (HIV-1 Pr) & 1HHP.pdb & Amber14sb~\cite{maier15amber14sb} & 25342 & 73 & 77\\
     Myeloid Cell Leukemia-1 (MCL-1) & 3WIX.pdb & CHARMM36m~\cite{charmm36m}& 16056 & 46 & 47\\
     K-Ras (KRAS) & 5W22.pdb & Amber14sb~\cite{maier15amber14sb} & 22933 & 68 & 61\\
     Ribose Binding Protein (RBP) & 1DRJ.pdb & CHARMM36m~\cite{charmm36m}& 21977 & 0 & 0 \\
\end{tabular}
\end{adjustbox}
\caption{Simulation Parameters for Studied Systems} 
\label{t:sys}
\end{table} 
    
First, the energy of each system was minimized using the steepest descent algorithm for 1000 steps. 
Then, the system was equilibrated in the isobaric-isothermal (NPT) ensemble at 300~K and 1~bar for 100~ps.
For the equilibration simulations, we used a 1~fs integration time step, a velocity rescaling thermostat~\cite{bussi07} with a 1.0~ps time constant, and a stochastic cell rescaling barostat~\cite{bernetti2020} with a time constant of 2.0~ps. 
    
This was followed by an unbiased 20~ns simulation in the NPT ensemble. 
In this simulation, we used a 2~fs integration timestep, a Nos{\'e}-Hoover thermostat~\cite{nose1984,hoover1985} with a 1.0~ps time constant, and a Parrinello-Rahman barostat~\cite{parrinello1981} with a 2.0~ps time constant. 
All covalent bonds involving hydrogens were constrained using the LINCS algorithm.\cite{hess1997}.
Short-ranged electrostatic and Lennard-Jones interactions were treated with a 10~\AA\ real-space cutoff with energy and pressure corrections for dispersion interactions. 
Long-ranged electrostatic interactions were treated with the Particle Mesh Ewald algorithm~\cite{darden1993} using a 1.2~\AA\ grid. 
Coordinates and velocities were stored every 20~fs for subsequent analysis.
    
This simulation protocol, starting with the equilibration of the energy minimized structure, was repeated five times for each system using resampled starting velocities from a Maxwell distribution with a unique random seed. 
We then performed FREquency-SElective ANharmonic (FRESEAN) mode analysis~\cite{sauer2023frequency} for each 20~ns trajectory to assess the reproducibility of the analysis protocol.
    
\subsection{Well-Tempered Metadynamics Simulations}
    
Well-tempered metadynamics (WT-metadyn)~\cite{barducci2008well} simulations were performed for each independent replica using the PLUMED 2.8 software package.\cite{plumed,tribello2014}
FRESEAN modes 7 and 8 at 0~THz were used as collective variables (CVs). 
Details on the coarse-grained representation of our CVs and the conversion to an all-atom representation for use in WT-metadyn simulations are described in the following sections. 
WT-metadyn simulations were run for 100~ns in the NPT ensemble with a 2~fs timestep. 
All other parameters (thermostat, barostat, bond constraints, short- and long-ranged interactions) were treated in the same way as for the unbiased 20~ns NPT simulations. 
Gaussian functions were added to the biasing potential every 1~ps with an initial height of 0.1~kJ/mol and a standard deviation of 0.001. 
The unit-less bias factor in the PLUMED implementation of metadynamics was set to 10. 
The free energy surface is recovered by inverting the sum of all Gaussian hills deposited throughout the simulation. 

\subsection{Coarse-Grained Representation of Proteins}
    
\begin{figure}[h!]
\centering
\includegraphics[width=1\textwidth]{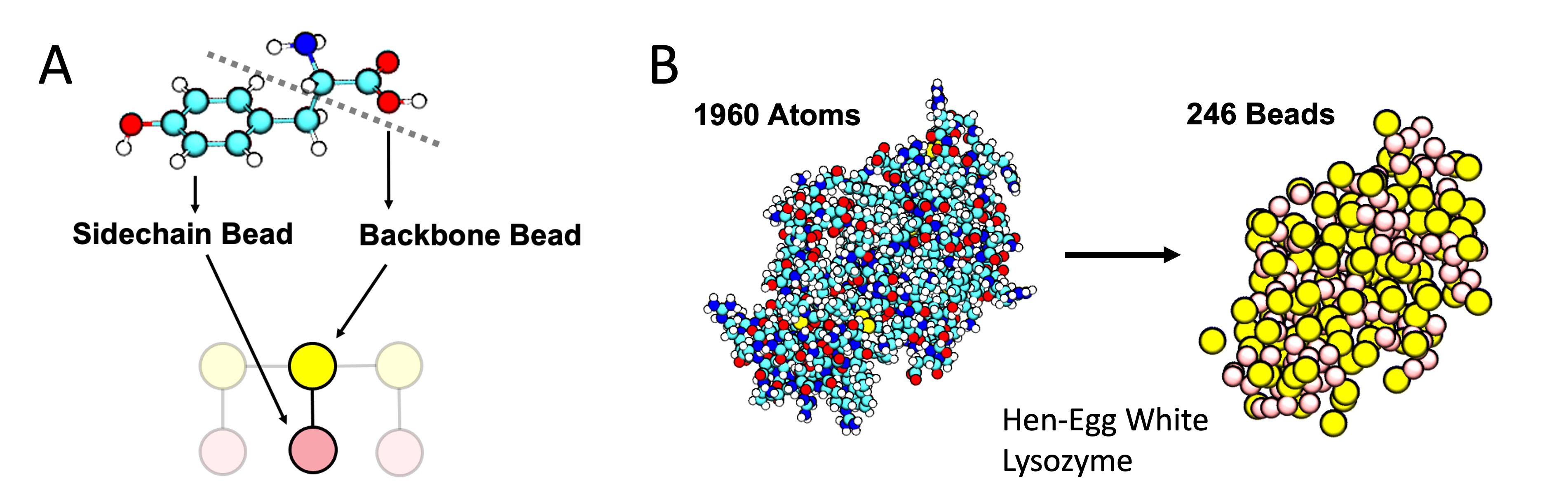}
\caption{Visualization of the two-bead coarse grained  representation of amino acids. (A) An example of the coarse-graining scheme for tyrosine, which shows which atoms are associated with either the backbone or side chain. 
In the coarse-grained representation, the corresponding backbone (yellow) and side chain (pink) beads correspond to the COM of the corresponding group of atoms.
(B) Application of the coarse-graining scheme to HEWL. 
The coarse-graining scheme reduces the number of particles (all-atom: 1960 atoms; coarse-grained: 246 beads) and the dimensions of the velocity cross correlation matrices computed during FRESEAN mode analysis (see Ref.~\citenum{sauer2023frequency} and later section for additional information).}
\label{f:cg}
\end{figure}

We defined the following coarse-grained representation of protein coordinates and velocities, which describes each amino acid (except glycine) with two beads: one bead represents the center of mass (COM) of all backbone atoms and one bead represents the COM of all side chain atoms (see Figure~\ref{f:cg}). 
Glycine is treated as a single bead.

To perform WT-metadyn simulations along coarse-grained FRESEAN modes as CVs (modes 7 and 8 at zero frequency), we need to translate between the coarse-grained representation of FRESEAN modes (used in our analysis of unbiased simulations) and the all-atom representation used in the WT-metadyn simulations. 
One option is to: A) evaluate the COM position for each group of atoms that correspond to a given bead during each step of the simulation; B) align the resulting coarse-grained structure to a coarse-grained reference structure; C) evaluate bead displacements relative to the reference; D) project a combined vector of all bead displacements on the coarse-grained FRESEAN modes used as CVs. 

Here, we used an equivalent approach that is more straightforward to implement in PLUMED by converting the coarse-grained FRESEAN modes into a corresponding all-atom representation prior to the simulation. 
In this representation, per-bead components of a coarse-grained FRESEAN mode are distributed over the atoms contributing to each bead so that the sum of atomic components equals the corresponding per-bead components.
During each step of the simulation, we then: A) align the all-atom protein structure to an all-atom reference structure; B) calculate atomic displacements relative to the reference; C) project a combined vector of all atomic displacements on the all-atom representations of FRESEAN modes used as CVs.

\subsection{All-Atom {\em vs.} Coarse-Grained Vibrational Density of States}

To quantify whether coarse-grained protein trajectories reproduce low-frequency vibrations compared to an all-atom representation, we evaluated the vibrational density of states (VDoS) using both representations for an unbiased 20 ns HEWL trajectory. 
To resolve high-frequency vibrations prominently observed in the all-atom representation, we stored coordinates and velocities every 4~fs for this analysis.

\begin{figure}[h!]
\centering
\includegraphics[width=1\textwidth]{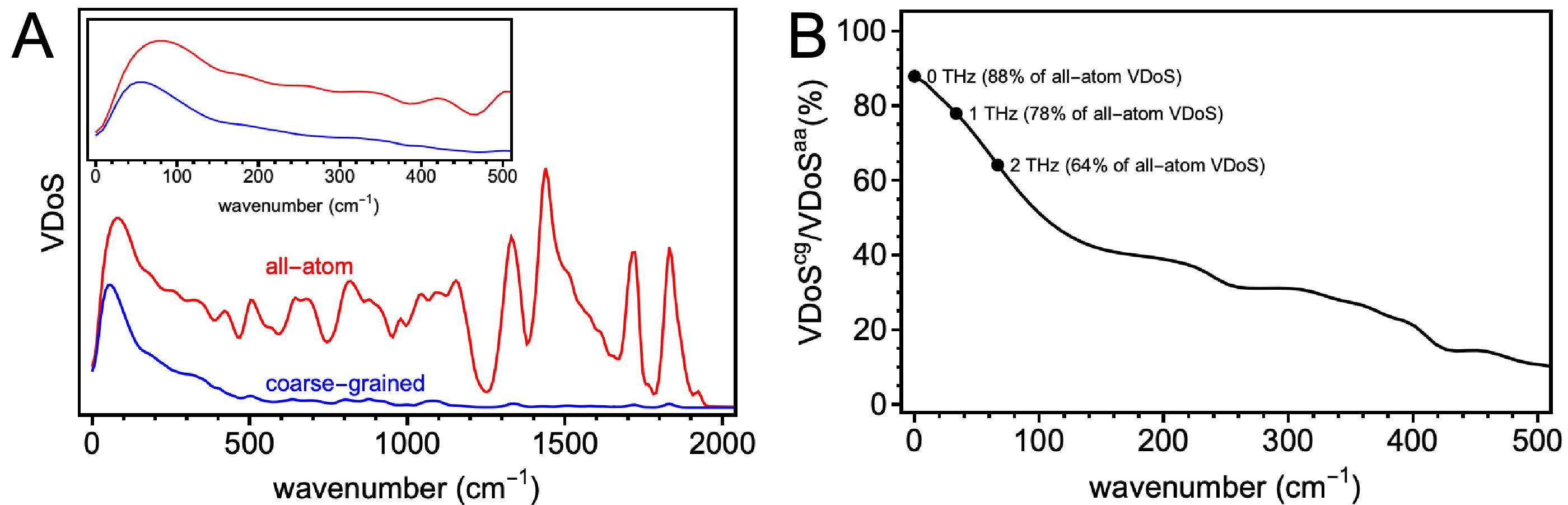}
\caption{(A) Vibrational Density of States (VDoS) calculated for the all-atom and coarse-grained representations of a 20 ns simulation of Hen Egg White Lysozyme (HEWL). The inset highlights frequencies $<$500~\wn. (B) Percentage of the all-atom VDoS information captured in the coarse-grained representation described by the ratio $\tx{VDoS}^\tx{cg}(\omega)/\tx{VDoS}^\tx{aa}(\omega)$.}
\label{f:aacg}
\end{figure}

For both representations, we define weighted velocities of either atoms or beads:
\begin{equation}
    \tilde{\vc{v}_i} = \sqrt{m_i} \cdot \vc{v}_i
    \label{e:wv}
\end{equation}
Here, $\vc{v}_i$ is the velocity vector of atom $i$ in the all-atom representation, or the velocity vector of bead $i$ in the coarse-grained representation, {\em i.e.}, the COM velocity of its contributing atoms; $m_i$ is either the mass of atom $i$ in the all-atom representation or the mass of bead $i$ in the coarse-grained representation, {\em i.e.}, the sum of its contributing atomic masses.
We can define the VDoS of the protein via the Fourier transform of the sum over $N$ (= number of atoms or beads) time auto correlation functions of weighted velocities $\tilde{\vc{v}_i}$.
\begin{equation}
    \tx{VDoS}(\omega) = \frac{1}{2\pi} \cdot \frac{2}{k_B T} \int_{-\infty}^{+\infty}\tx{exp}(\tx{i}\omega \tau) \sum_i^{N} \langle \tilde{\vc{v}_i}(t) \cdot \tilde{\vc{v}_i}(t+\tau)\rangle_t \, d\tau
    \label{e:vdos}
\end{equation}
Here, $\langle ... \rangle_t$ indicates the ensemble average over the simulation time.
With this definition, the integral of the VDoS over positive frequencies corresponds to the number of degrees of freedom described by the velocities $\vc{v}_i$.
\begin{equation}
    N_\tx{DOF} = \int_0^{+\infty} \tx{VDoS}(\omega) \, d\omega
\end{equation}
For HEWL, which has 1960 atoms and 959 constraints (bonds involving hydrogens), $N_\tx{DOF}^\tx{aa} = 3 \times 1960 - 959 = 4921$ in its all-atom representation. 
In the coarse-grained representation with 246 beads as shown in Figure~\ref{f:cg}, this integral of the VDoS reduces to $N_\tx{DOF}^\tx{cg} = 3 \times 246 - 4.4 = 733.6$ (here, 4.4 is the observed number of effective constraints after converting from the all-atom representation with 959 bond constraints to the coarse-grained representation).
The ratio $N_\tx{DOF}^\tx{cg}/N_\tx{DOF}^\tx{aa}$ describes how much information is lost overall due to coarse-graining. 
The ratio $\tx{VDoS}^\tx{cg}(\omega)/\tx{VDoS}^\tx{aa}(\omega)$ quantifies this information loss as a function of frequency.
We plotted both, the all-atom and coarse-grained VDoS, in panel A of Figure \ref{f:aacg} for frequencies, and their ratio in panel~B.

The all-atom VDoS describes all vibrations in the protein including bond vibrations etc. within the backbone and side chains that are most prominent in the vibrational fingerprint region. 
These vibrations have no impact on the COMs used for our coarse-grained representation.
Thus the coarse-grained VDoS is primarily missing vibrational degrees of freedom at high frequencies while low-frequency vibrations corresponding to relative vibrations of secondary structure units, side chain motions etc. are largely preserved.
Correspondingly, we find that 88\% of the zero frequency motions are captured in the coarse-grained representation, while the information loss sets in gradually with increasing frequency.

Since, our approach to identify CVs to enhance conformational sampling is based solely on zero frequency modes, Figure~\ref{f:aacg} indicates that the coarse-grained protein representation contains effectively all information needed for FRESEAN mode analysis at zero frequency.

\subsection{FREquency SElective ANharmonic (FRESEAN) Mode Analysis}

FRESEAN mode analysis is described in detail in our previous work in Ref.~\citenum{sauer2023frequency}. 
We formulate the velocity cross correlation matrix with scaled velocities as in Eq.~\ref{e:wv}. 
For simplicity, we use the index $i$ to refer to a single degree of freedom, {\em i.e.}, the $x$-, $y$-, or $z$-component of a coarse-grained bead velocity vector.
\begin{equation}
\tilde{\tx{v}_i} = \sqrt{m} \cdot \tx{v}_i
\label{e:wv2}
\end{equation}
While analyzing a simulation trajectory, we rotate coordinates and velocities of the protein trajectory into a common reference frame and compute the time cross-correlation matrix between mass-weighted velocities for all degrees of freedom $i$ and $j$.
\begin{equation}
C_{\tilde{\tx{v}},ij}(\tau)=\langle \tilde{\tx{v}}_i(t) \tilde{\tx{v}}_j(t+\tau)\rangle_t
\label{e:corr}
\end{equation}
The Fourier transformation of each matrix element results in a frequency-dependent cross-correlation matrix with the following elements.
\begin{equation}
    C_{\tilde{\tx{v}},ij}(\omega) = \frac{1}{2\pi}\int_{-\infty}^{+\infty}\tx{exp}(\tx{i}\omega \tau) C_{\tilde{\tx{v}},ij}(\tau) d\tau
    \label{e:ft}
\end{equation}
Notably, the trace of this matrix describes the Fourier transform of the sum of time auto correlations, which is identical to the vibrational density of states (VDoS) defined in Eq.~\ref{e:vdos}.
\begin{equation}
\tx{VDoS}(\omega)=\frac{2}{k_B T} \sum_i^{3N} C_{\tilde{\mathrm{v}},ii}(\omega) 
\label{e:vdos2}
\end{equation}
This does not change if we diagonalize the matrix $\vc{C}_{\tilde{\mathrm{v}}}(\omega)$ (with elements $C_{\tilde{\mathrm{v}},ij}$) at each sampled frequency because the corresponding coordinate transformations are unitary.
Therefore, we obtain an equivalent expression for the VDoS as the sum of eigenvalues $\lambda_i$ at each frequency.
\begin{equation}
\tx{VDoS}(\omega)=\frac{2}{k_B T} \sum_i^{3N} \lambda_i(\omega) 
\label{e:vdos3}
\end{equation}
Consequently, at any specific frequency $\omega'$, the eigenvalues $\lambda_i(\omega')$ describe contributions of mass-weighted velocity fluctuations along the corresponding eigenvectors $\vc{Q}_i(\omega')$ to $\tx{VDoS}(\omega')$.
An important observation is that the majority of eigenvalues $\lambda_i(\omega')$ are zero.
Sorting the eigenvalue/eigenvector-pairs by the magnitude of $\lambda_i(\omega')$ (large to small) thus unambiguously identifies the collective degrees of freedom, {\em i.e.}, eigenvectors describing displacements relative to a reference structure, that contribute to the VDoS at frequency $\omega'$.

\begin{figure}[h!]
\centering
\includegraphics[width=1\textwidth]{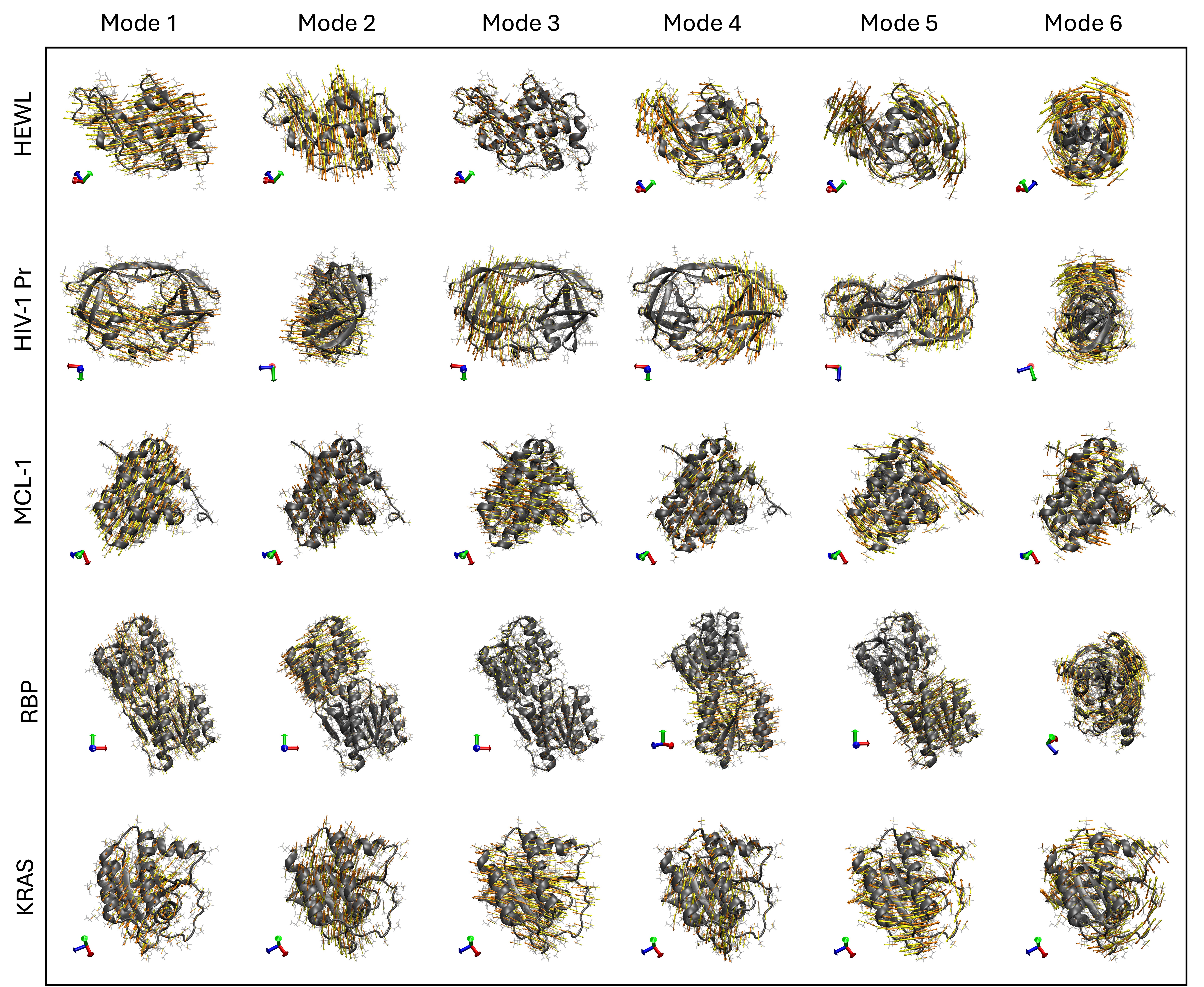}
\caption{Visualization of FRESEAN modes 1-6 (yellow/orange arrows) at zero frequency for the five studied systems. Modes 1-3 describe protein translations, while modes 4-6 describe rigid body rotations. Arrows in the bottom left of each rendering denote the protein orientation selected separately for clarity.}
\label{f:m16}
\end{figure}

At zero frequency, it is not surprising that eigenvalues $\lambda_{1-3}(\omega=0)$, {\em i.e.}, contributions to $\tx{VDoS}(\omega=0)$, are related to eigenvectors describing translations of the entire protein. 
Likewise, eigenvectors associated with eigenvalues $\lambda_{4-6}(\omega=0)$ describe essentially rigid body rotations. 
For HIV-1 Protease and RBP, we further observe some decoupling of the translational/rotational motions of the two distinct monomers and domains, respectively.
We illustrated these eigenvectors, {\em i.e.}, zero-frequency modes 1-6, as displacement vectors for each coarse-grained bead superimposed with the visualization of an all-atom reference structure in Figure~\ref{f:m16}.
The diffusive character of the underlying motions is apparent from the analysis in the next section.

\subsection{1D-VDoS for Eigenvector Projections}

To illustrate that eigenvectors generated at zero frequency with FRESEAN mode analysis isolate diffusive motion (modes 1-6) and low-frequency vibrations (modes 7+), we computed the VDoS separately for the collective one-dimensional degrees of freedom described by the corresponding eigenvectors $\vc{Q}_i \, \left[=\vc{Q}_i(\omega=0)\right]$ for $i$ from 1 to 10.

\begin{figure}[h!]
\centering
\includegraphics[width=0.75\textwidth]{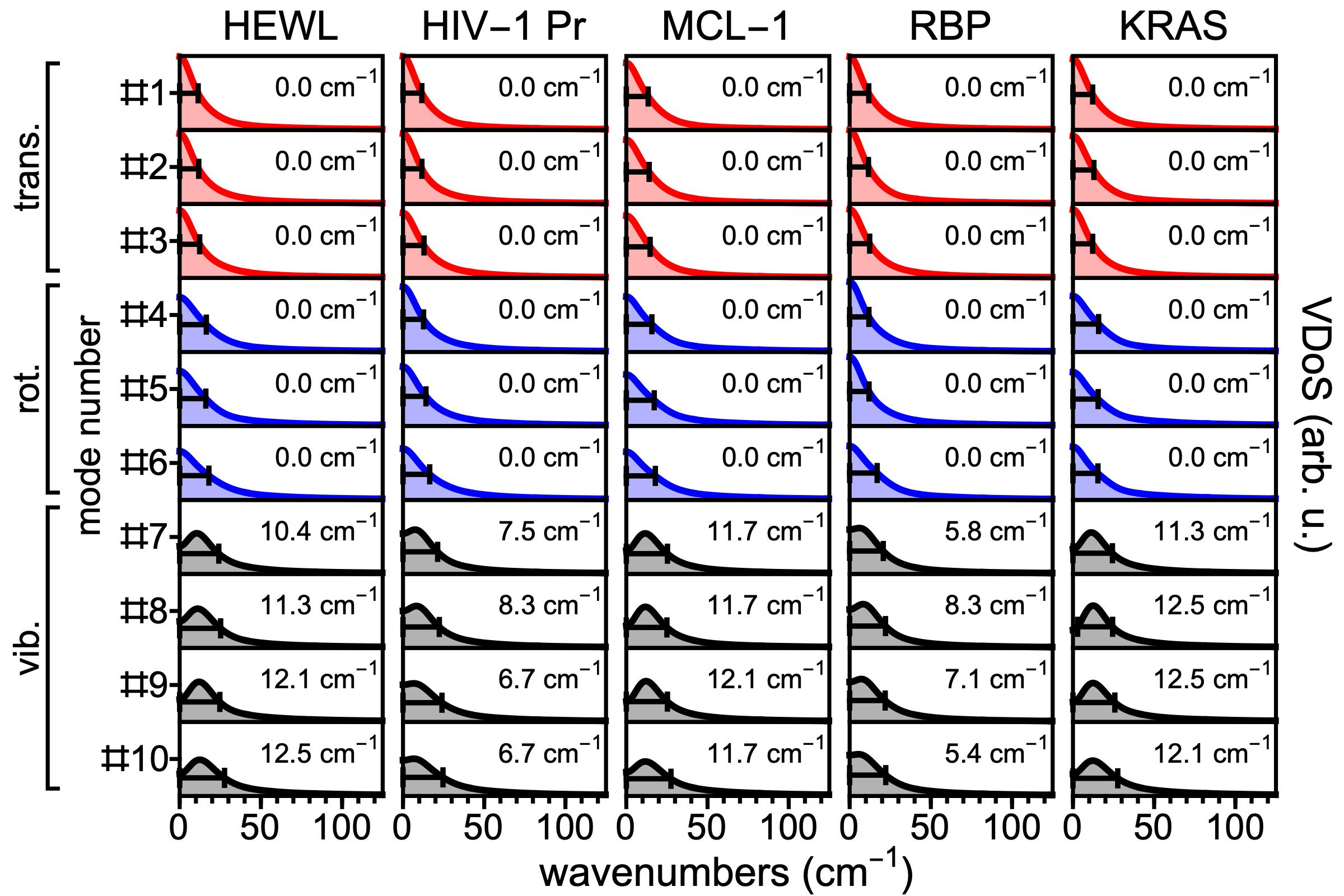}
\caption{Visualization of the 1D-VDoS for individual FRESEAN modes 1-10 obtained at zero frequencies for all five systems. 
The 1D-VDoS for translational, rotational and low-frequency vibrational modes are shown in red, blue and black, respectively.
Numbers indicate the peak position and a horizontal bar indicates the full width at half maximum.}
\label{f:1d}
\end{figure}

To do so, we first project weighted velocities ($\tilde{\mathrm{v}}(t)$) from our simulation trajectories on a given eigenvector $\vc{Q}_i$ with its $3N$ components $q_i^j$.
\begin{equation}
\dot{q}_i(t) = \sum_j q_i^j \tilde{\tx{v}}_j(t)
\end{equation}
This allows us to define a mass-weighted time correlation function for fluctuations along $\vc{Q}_i$, which we Fourier transform to obtain the VDoS for this single collective degree of freedom.
\begin{equation}
\tx{VDoS}_{Q_i}\left(\omega\right) = \frac{2}{k_B T} \left[ \frac{1}{2 \pi} \int_{-\infty}^\infty \mathrm{exp} \left(\tx{i} \omega \tau\right) \, \langle \dot{q}_i(t) \dot{q}_i(t+\tau)\rangle_t \, d\tau\right]
\end{equation}
We plotted the resulting one-dimensional VDoS for modes 1-10 of each system in Figure~\ref{f:1d}. 
The diffusive motion along modes 1-6 is evident through the peak position at 0~\wn, whose amplitude is proportional to the corresponding diffusion coefficient.
For modes 7-10, we observe low-frequency vibrations with peak intensities between 5 and 13~\wn\ that contribute to the zero frequency VDoS via a low-frequency tail.
We note that additional peaks are absent, highlighting the succesful isolation of low-frequency vibrations, in contrast to similar projections on low-frequency vibrational modes obtained with conventional methods that rely on harmonic approximations (see Ref.~\citenum{sauer2023frequency}).

\subsection{Reproducibility of FRESEAN Mode Analysis}

A key feature of FRESEAN mode analysis is not only its ability to isolate low-frequency as shown in the previous section and Ref.~\citenum{sauer2023frequency}, but a high reproducibility of low-frequency eigenvectors obtained from distinct simulations.

\begin{figure}[h!]
\centering
\includegraphics[width=1\textwidth]{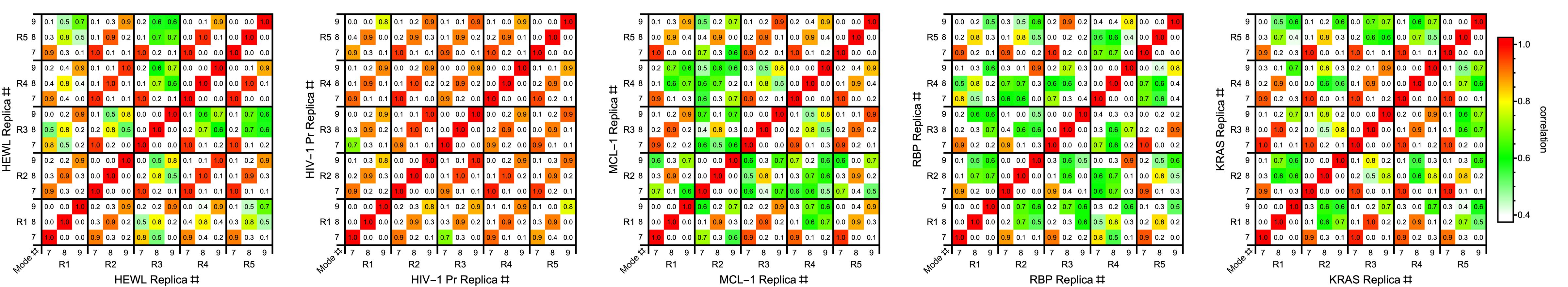}
\caption{Full correlation matrices between all pairs of zero-frequency FRESEAN modes 7-9 obtained for replicas R1 to R5 for each protein system. Colors of individual squares indicate the direct correlation between a pair of modes.
}
\label{f:c-all}
\end{figure}

To compare vibrational modes obtained from separate simulation trajectories, we compute the correlation coefficients between the corresponding normalized eigenvectors as a scalar product.
\begin{equation}
C_{(i,j)}^{(k,l)} = \left|Q_i^{(k)} \cdot Q_j^{(l)}\right|
\label{e:coeff}
\end{equation}
Here, the indices $i$ and $j$ indicate the eigenvector index and the indices $k$ and $l$ indicate the index of the simulation replica.
Correlations between modes 7-9 in replica 1 and modes 7-9 in replica R1 to R5 are shown in Figure~1 of the main text. 
In Figure~\ref{f:c-all}, for completeness, we show the corresponding correlations for all combinations of modes and simulation replicas.

In addition to comparisons between pairs of modes it is often useful to compare the sub-spaces described by a small set of modes obtained from two distinct simulations. 
This is indicated for the two-dimensional sub-spaces spanned by modes 7 and 8 and the three-dimensional sub-spaces spanned by modes 7, 8, and 9 as numerical values in Figure~1 of the main text. 
To define the correlation between sub-spaces, we project eigenvectors obtained from simulation replica $k$ into the low-dimensional space defined by eigenvectors obtained from simulation replica $l$. 
We then compute the average length of the projected vectors in these low-dimensional projections, which is 1 if the sub-spaces are equivalent, {\em i.e.}, each eigenvector obtained from simulation replica $k$ can be fully represented as a linear combination of the selected eigenvectors obtained from simulation replica $l$.
\begin{eqnarray}
    C_\tx{2D}^{\left(k,l\right)} &=& \left(\sum_{i\in{7,8}} \sqrt{\sum_{j\in{7,8}} \left(Q_i^{(k)} \cdot Q_j^{(l)} \right)^2}\right)/2 \\
    C_\tx{3D}^{\left(k,l\right)} &=& \left(\sum_{i\in{7,8,9}} \sqrt{\sum_{j\in{7,8,9}} \left(Q_i^{(k)} \cdot Q_j^{(l)} \right)^2}\right)/3
\end{eqnarray}

\subsection{Quasi-Harmonic and Principal Component Modes}

In contrast to FRESEAN mode analysis, other traditional methods to identify low-frequency vibrations not only fail to isolate low-frequency vibrations due to their reliance on harmonic approximations (see Ref.~\citenum{sauer2023frequency}) but also lack reproducibility between distinct simulations. 
We demonstrate this here for quasi-harmonic normal modes obtained from the same simulation trajectories used for FRESEAN mode analysis.

\begin{figure}[h!]
\centering
\includegraphics[width=0.5\textwidth]{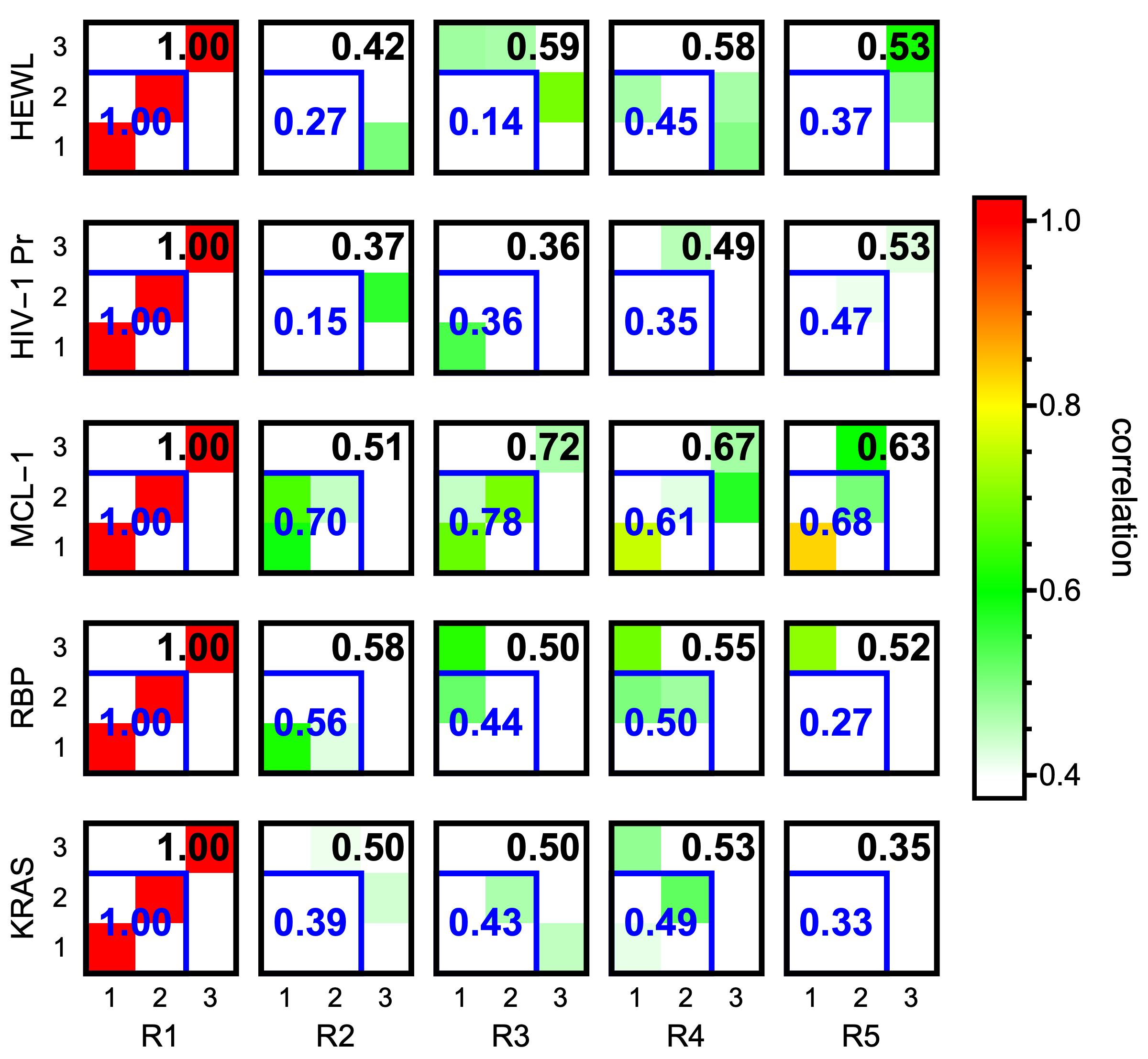}
\caption{Matrices describing correlations between quasi-harmonic normal modes 1-3 in replicas R1 and R1 to R5 for each protein system (self-correlations with replica 1 are 1.0 by definition and only shown for clarity).
Colors of individual squares indicate the direct correlation between a pair of modes.
Correlation coefficients between 2D (3D) sub-spaces described by modes 1-2 (1-3) in a pair of simulations are shown as blue (black) numerals.}
\label{f:qh1}
\end{figure}

Quasi-harmonic normal mode analysis is based on the mass-weighted co-variance matrix of atomic displacements relative to an average structure (after alignment with a reference structure).
The elements of the displacement co-variance matrix are defined in Eq.~\ref{e:covar}.
\begin{equation}
    C_{ij}^\tx{covat} = \sqrt{m_i \cdot m_j} \cdot \langle \left(x_i - \langle x_i \rangle \right) \cdot \left(x_j - \langle x_j \rangle \right) \rangle_t
    \label{e:covar}
\end{equation}
In a harmonic system, the eigenvectors of this matrix correspond to the harmonic normal modes of the system and the eigenvalues, $\lambda_i$, describe the (mass-weighted) variance of projections onto each normal mode.
The square-root of the latter is inversely proportional to the harmonic frequency of each normal mode.
\begin{equation}
    \omega_i^\tx{QH} = \sqrt{\frac{k_B T}{\lambda_i}}
\end{equation}
Since the co-variance matrix only describes intramolecular distortions, none of the eigenvectors describe translational or rotational motion. 
Thus we perform the correlation analysis described in the previous section for modes 1-3, {\em i.e.}, the lowest frequency vibrational modes obtained from quasi-harmonic normal mode analysis.
In Figure~\ref{f:qh1}, we show the correlations between quasi-harmonic normal modes obtained from replica R1 to the five simulation replicas R1 to R5 for each system, in analogy to the corresponding Figure~1 in the main text.

\begin{figure}[h!]
\centering
\includegraphics[width=1\textwidth]{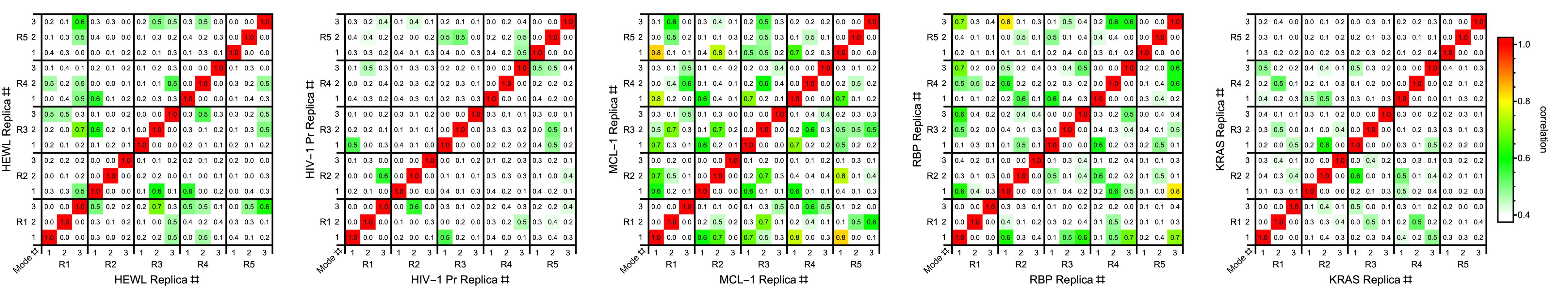}
\caption{Full correlation matrices between all pairs of quasi-harmonic normal modes 1-3 obtained for replicas R1 to R5 for each protein system. Colors of individual squares indicate the direct correlation between a pair of modes.}
\label{f:qh2}
\end{figure}

It is apparent that the pair-wise corerlations between modes as well as the correlations between 2D and 3D sub-spaces are drastically reduced compared to low-frequency modes obtained from FRESEAN mode analysis.
The correlations between pairs of quasi-harmonic modes rarely reach values >0.5 and even the correlations between 2D and 3D sub-spaces, which allow for more flexibility, {\em e.g.}, mixing modes or changing their order, are not improving this picture.
Thus, it is clear that FRESEAN mode analysis is substantially more stable and depends less on rare events.
For completeness, we also show the correlations between all combinations of simulation replicas in Figure~\ref{f:qh2} in analogy to Figure~\ref{f:c-all} for the FRESEAN modes.

To further emphasize the difference between FRESEAN mode analysis and methods based on co-variances of atomic displacements, we selected one system, HEWL, to increase the sampling time per simulation replica from 20~ns to 1000~ns, {\em i.e.}, 1~microsecond.
Applying the same formalism as for quasi-harmonic normal mode analysis, we now obtain principal component modes. 
The distinction is that for long simulations, the assumption that the system explores a local potential energy minimum is no longer applied.
This principal component analysis is useful to identify collective degrees of freedom involved in large conformational changes sampled in the simulation trajectories.
The eigenvalues of the co-variance matrix are now directly interpreted as variances of projections along each eigenvector.

\begin{figure}[h!]
\centering
\includegraphics[width=0.5\textwidth]{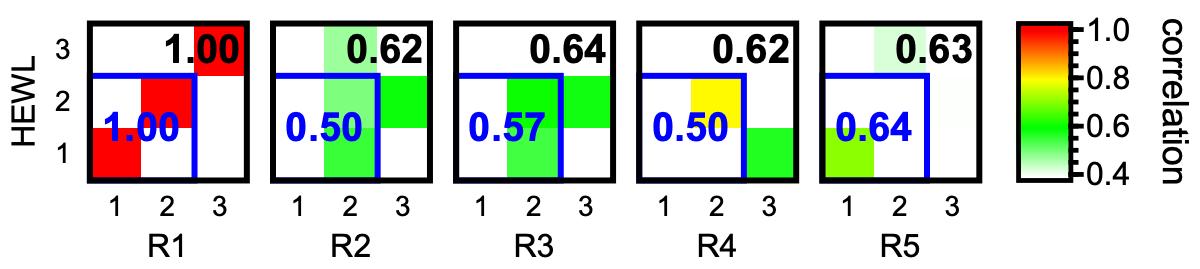}
\caption{Matrices describing correlations between principal component modes 1-3 in replicas R1 and R1 to R5 (1~microsecond simulations each) for HEWL (self-correlations with replica 1 are 1.0 by definition and only shown for clarity).
Colors of individual squares indicate the direct correlation between a pair of modes.
Correlation coefficients between 2D (3D) sub-spaces described by modes 1-2 (1-3) in a pair of simulations are shown as blue (black) numerals.}
\label{f:pc1}
\end{figure}

\begin{figure}[h!]
\centering
\includegraphics[width=0.4\textwidth]{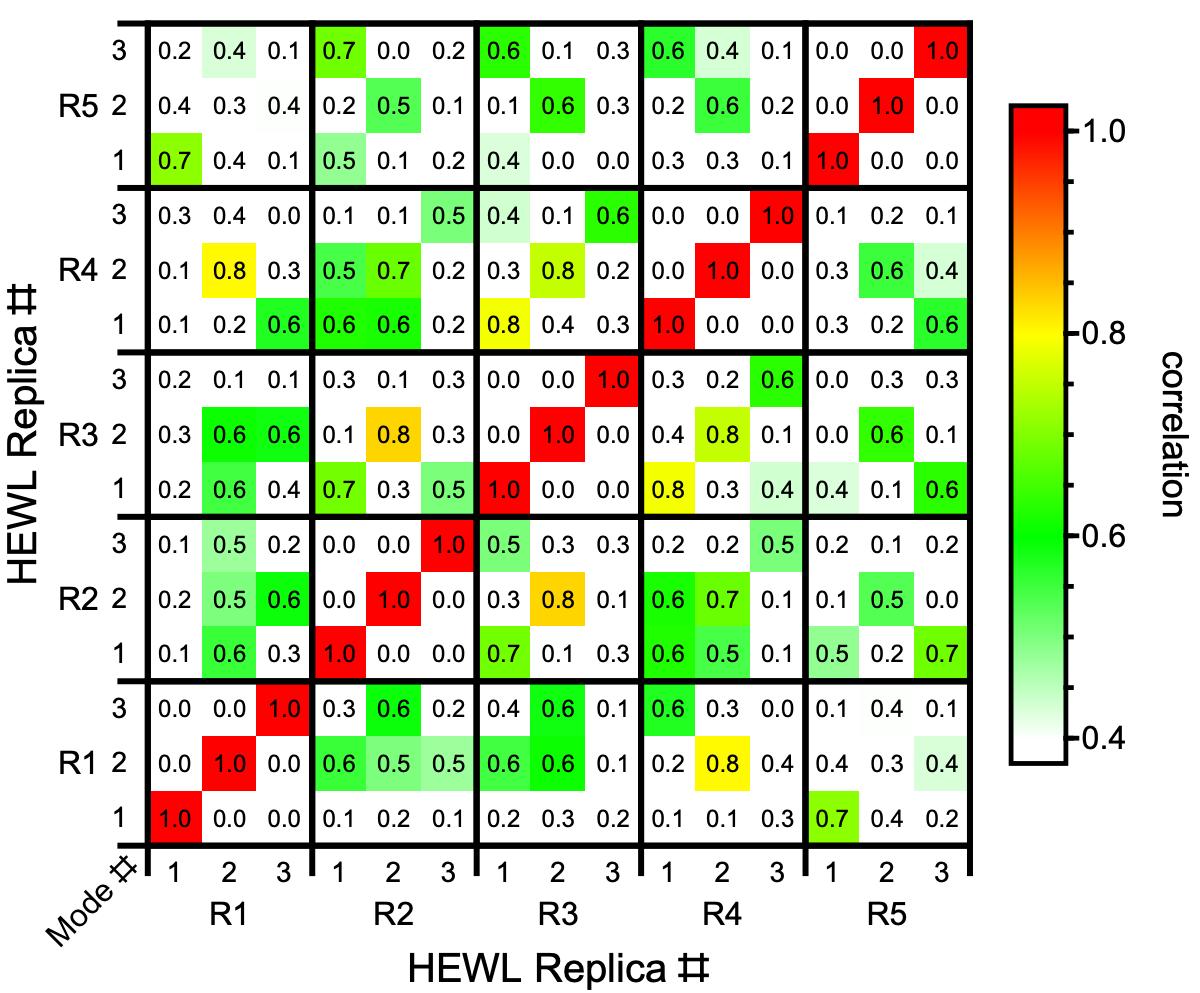}
\caption{Full correlation matrices describing the correlation between all pairs of principal component modes 1-3 between replicas R1 to R5 for HEWL. Colors of individual squares indicate the direct correlation between a pair of modes.}
\label{f:pc2}
\end{figure}

We show the pair-wise correlations of principal component modes 1-3 in simulation replica R1 with replica R1 to R5 (all simulated for 1~microsecond) as well as the correlations between 2D and 3D sub-spaces in Figure~\ref{f:pc1}. 
We observe improvements compared to the quasi-harmonic mode analysis, which indicates relatively well-defined conformational dynamics in this system. 
However, the correlations between the extracted modes are still substantially smaller than for the FRESEAN modes shown in Figure~1 in the main text despite a 50-fold increase in the simulation time.
As before, we show the complete set of correlations between all simulation replicas in Figure~\ref{f:pc2}.

\subsection{Definition and Visualization of Geometric Variables}

\begin{figure}[h!]
\centering
\includegraphics[width=1.0\textwidth]{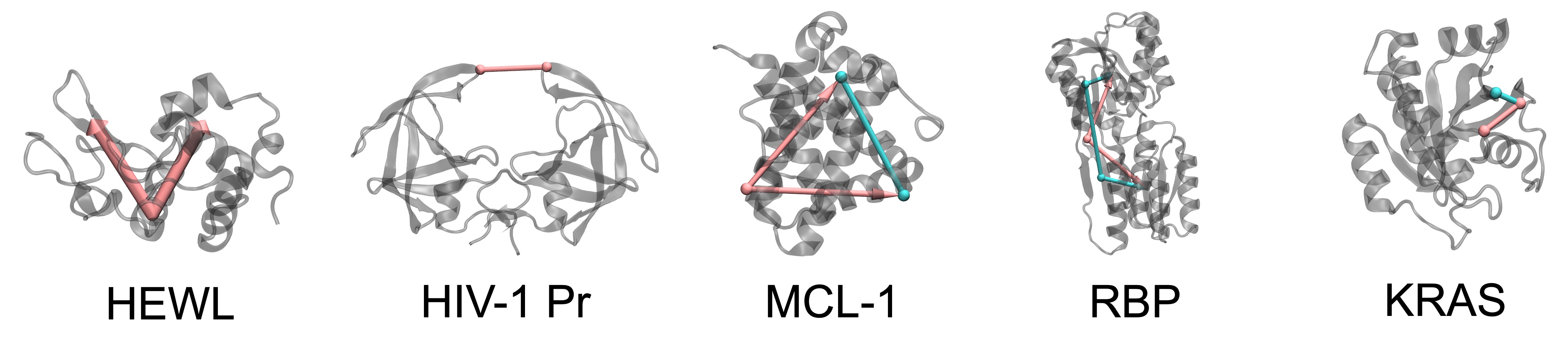}
\caption{Visualization of geometric collective variables (distances and angles shown as colored cylinders with spherical end points) used for the analysis of enhanced sampling simulations in the context of the protein structures (gray cartoons). Colors (cyan and magenta) correspond to the text color used in Table~\ref{t:geo}.}
\label{f:geo}
\end{figure}

As described in the main text, we use geometric variables introduced previously in the literature for the five simulated proteins to analyze the performance of our enhanced sampling scheme. 
The exact definitions of these collective variables are provided in Table~\ref{t:geo}.
We further provide a visualization of all distances and angles in the context of the respective protein structures in Figure~\ref{f:geo}.

\begin{table}[h!]
\centering
\begin{adjustbox}{width=0.8\textwidth}
\begin{tabular}{l|c|l}
     Protein & Geometric Variable & Definition/Notes \tsB \\
     \hline\hline    
     HEWL & \color{magenta} pincer angle \color{black} & angle between C$_\alpha$-atom centers of 3 residue groups: \tsT \\
     && \,\,\, 1. [28-31,111-114] \\
     && \,\,\, 2. [90-93] \\
     && \,\,\, 3. [44-45,51-52] \tsB \\
     & radius of gyration & radius of gyration computed with all protein atoms \tsB \\
     \hline
     HIV-1 Pr & \color{magenta}flap distance\color{black} & distance between C$_\alpha$-atoms of G51 in each monomer \tsT \tsB \\
     & flap RMSD & RMSD of residues I50-G52 in both monomers \tsB \\
     \hline
     MCL-1 & \color{cyan}S255-T226\color{black} & distance between C$_\alpha$-atoms of residues S225 and T226 \tsT \tsB \\
     & \color{magenta}interdomain angle\color{black} & angle between C$_\alpha$-atoms of residues S225, D241, T226 \tsB \\
     \hline
     RBP & \color{cyan}twist angle\color{black} & dihedral angle between COMs of 4 groups of residues: \tsT \\
     && \,\,\, 1. [1-100,236-259] \\ 
     && \,\,\, 2. [99-100,236-237,258-259] \\
     && \,\,\, 3. [108-109,230-231,269-270], \\ 
     && \,\,\, 4. [108-231,269-271] \tsB \\
     & \color{magenta}hinge angle\color{black} & angle between COMs of 3 groups of residues: \\
     && \,\,\, 1. [1-100,236-259] \\ 
     && \,\,\, 2. [101-107,232-235,260-268] \\
     && \,\,\, 3. [108-231,269-271] \tsB \\
     \hline
     KRAS & \color{magenta}G12-T35\color{black} & distance between C$_\alpha$-atoms of residues G12 and T35 \tsT \tsB \\
     & \color{cyan}T35-G60\color{black} & distance between C$_\alpha$-atoms of residues T35 and G60 
\end{tabular}
\end{adjustbox}
\caption{Geometric Variables for Analysis} 
\label{t:geo}
\end{table} 

\subsection{"Closed" and "Open" Conformations}

\begin{figure}[h!]
\centering
\includegraphics[width=1.0\textwidth]{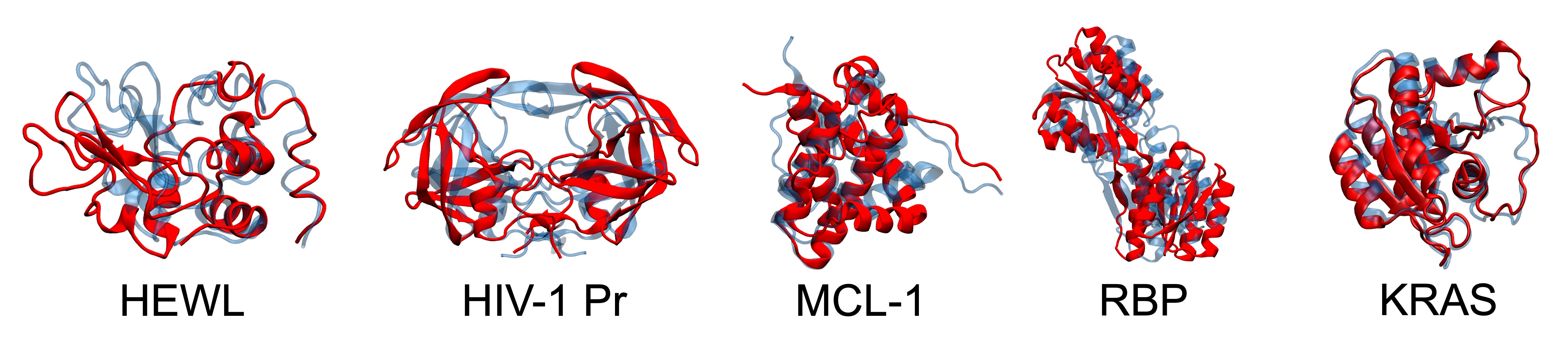}
\caption{Visualization of the "closed" (blue) and "open" (red) states as secondary structure cartoons for each protein.}
\label{f:states}
\end{figure}

In the main text, we use the terms "closed" and "open" as a common denomination for previously observed conformations of each protein.
The corresponding structures are shown in blue (closed) and red (open) in Figure~\ref{f:states}.
These states are also indicated as blue and red symbols in Figures~3 and 4 of the main text as well as in Figures~\ref{f:unbiased}, \ref{f:fes5geo}, and \ref{f:fes1d} below.

\subsection{Averaged Free Energy Surfaces}
To obtain the converged free energy surfaces shown in Figure~4 of the main text, we ran unbiased 100~ns NPT simulations for each protein and extract from it 20 configurations (every 5 ns). 
These are then used as starting points for 20 metadynamics simulations using the zero frequency modes 7 and 8 obtained from FRESEAN mode analysis for replica R1 for each system, while otherwise using the same metadynamics simulation protocol as described for replicas R1 to R5.
To obtain averaged free energy surfaces from multiple metadynamics simulations, we first compute for each simulation an unbiased probability distribution as a function of its CVs, {\em e.g.}, $CV_1$ and $CV_2$.
\begin{equation}
    p(CV_1,CV_2) = \frac{1}{\Omega} \, \tx{exp} \left[ \frac{E_\tx{bias}(CV_1,CV_2)}{k_B T} \right]
    \label{e:prob}
\end{equation}
Here, $E_\tx{bias}$ is the biasing potential generated as a sum of Gaussians during metadynamics sampling (describes a negative free energy) and $\Omega$ is the corresponding partition function, which here serves simply as a normalization constant.
\begin{equation}
    \Omega = \int_{-\infty}^{+\infty}\int_{-\infty}^{+\infty} \tx{exp} \left[ \frac{E_\tx{bias}(CV_1,CV_2)}{k_B T} \right] \, dCV_1 \, dCV_2
\end{equation}
We then compute the average probability distribution $\left\langle p(CV_1,CV_2) \right\rangle$ over the 20 simulations before converting into a free energy surface.
\begin{equation}
    \left\langle p(CV_1,CV_2) \right\rangle = \sum_{i=1}^{20} p_i(CV_1,CV_2)/20
\end{equation}
\begin{equation}
    \Delta G(CV_1,CV_2) = - k_B T \tx{ln} \left[ \left\langle p(CV_1,CV_2) \right\rangle \right]
\end{equation}

\subsection{Free Energy Surfaces in Distinct Variable Spaces}
To define free energy surfaces as a function of geometric variables, {\em e.g.},  as described in Table~\ref{t:geo}, using biased simulation trajectories generated with bias potentials applied to a distinct set of CVs, {\em e.g.}, low-frequency vibrations, we generate a weighted ensemble of structures based on the probabilities given in Eq.~\ref{e:prob}.
For each time frame $t_i$ of the trajectory, we compute the CVs used to define $E_\tx{bias}$ and evaluate $p(CV_1,CV_2)$ to obtain its relative weight.
\begin{equation}
    w(t_i) = p\left[CV_1(t_i),CV_2(t_i)\right]
    \label{e:w1}
\end{equation}
We then compute the unbiased probability distribution for any other set of variables, {\em e.g.}, $h_1$ and $h_2$ that can be expressed in terms of the system coordinates at time $t_i$.
\begin{equation}
    p(h_1,h_2) = \frac{\sum_i^{n_t} w(t_i) \cdot \delta\left[h_1'(t_i) - h_1\right] \cdot \delta\left[h_2'(t_i) - h_2\right]}{\sum_i^{n_t} w(t_i)}
    \label{e:w2}
\end{equation}
Here, $\delta$ indicates the Kronecker delta function with $\delta[0]=1$ and $\delta[x\neq0]=0$.
In practice, the delta function is replaced by a histogram bin of finite size. 
The probability function $p(h_1,h_2)$ now describes a weighted ensemble as a function of the variables $h_1$ and $h_2$ for which we can define the free energy surface.
\begin{equation}
    \Delta G(h_1,h_2) = - k_B T \tx{ln} \left[ p(h_1,h_2) \right]
    \label{e:w3}
\end{equation}

\subsection{Box \& Whisker Plot Definition}

The box-and-whisker plots shown in Figure~3 of the main text and Figure~\ref{f:unbiased} below are defined as follows. 
The dark gray box represents the inter-quartile range, which is defined from the 25th percentile to the 75th percentile. 
The dashed vertical line denotes the median value of the dataset. 
The lower whisker position is defined as the dataset minimum or 1.5x the inter-quartile range, whichever is smaller. 
The higher whisker position is determined in a similar way, except with the 75th percentile and the dataset maximum. 
Outliers are defined as points outside of the whisker range.

\subsection{Unbiased Control Simulations}

\begin{figure}[h!]
\centering
\includegraphics[width=0.4\textwidth]{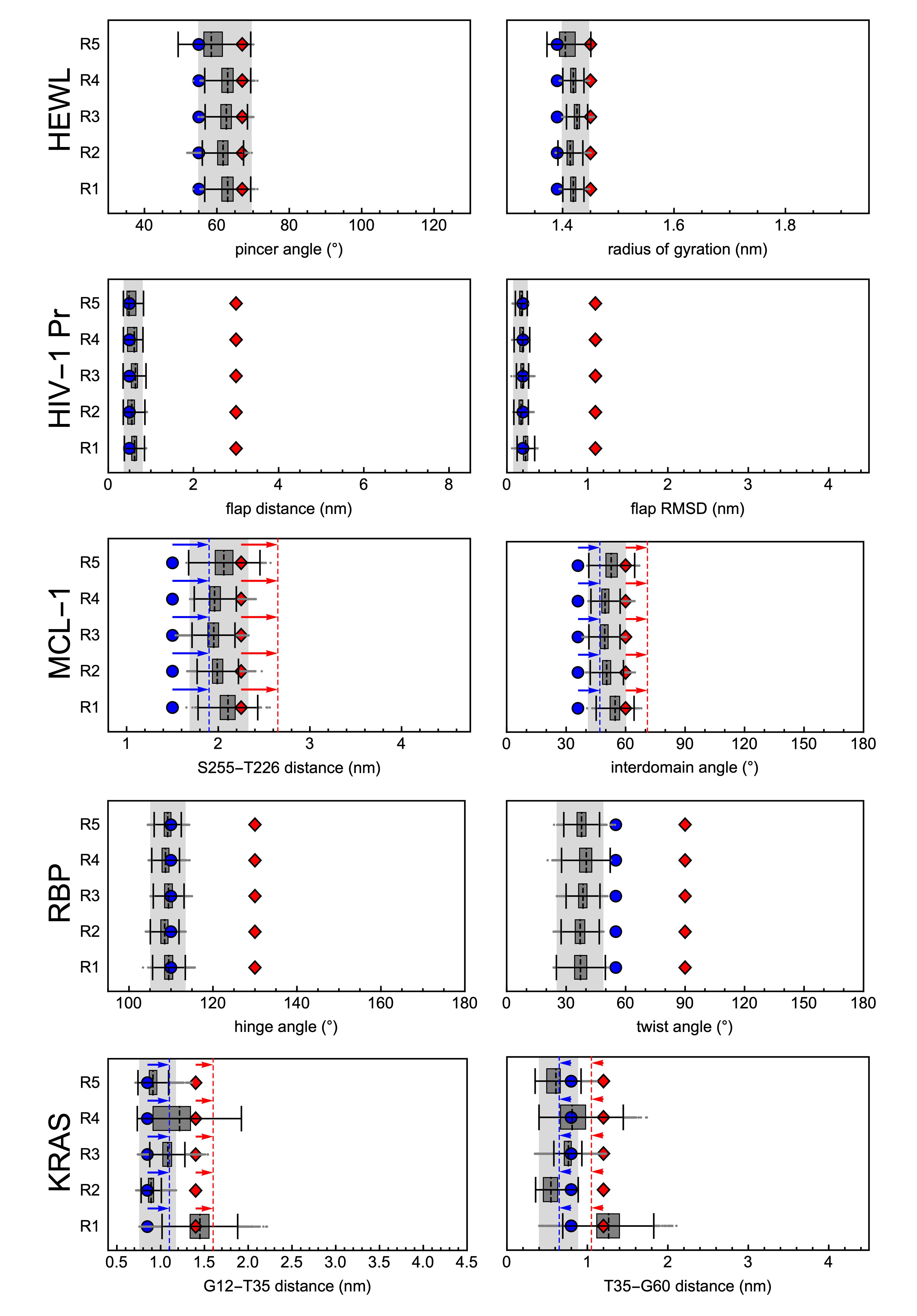}
\caption{
Box and whisker plots for each system illustrate the sampled conformational space in unbiased control simulation replicas R1 to R5.
Plotted are distributions of two geometric variables computed from the unbiased trajectories and defined in Table~\ref{t:geo} and Figure~\ref{f:geo} for each system (see above for definition of box and whisker plots).
Blue circles and red diamonds indicate conformational states reported in the literature (for MCL-1 and KRAS, arrows and dashed lines indicate constant shifts of free energy minima observed in Figure~4 of the main text).
}
\label{f:unbiased}
\end{figure}

To verify that our metadynamics simulations with low-frequency vibrational modes as CVs indeed achieve enhanced sampling, we performed unbiased control simulations with otherwise identical simulation parameters (five replicas R1 to R5 for each system with 100~ns sampling time). 
We then computed box and whisker plots as a function of the geometrical variables defined in Table~\ref{t:geo} to illustrate the sampled dynamics shown in Figure~\ref{f:unbiased}, which can be compared directly to the corresponding results for our metadynamics simulations in Figure~3 of the main text.

\newpage

\subsection{Free Energy Surfaces for Individual Replicas in CV-Space}

\begin{figure}[h!]
\centering
\includegraphics[width=0.9\textwidth]{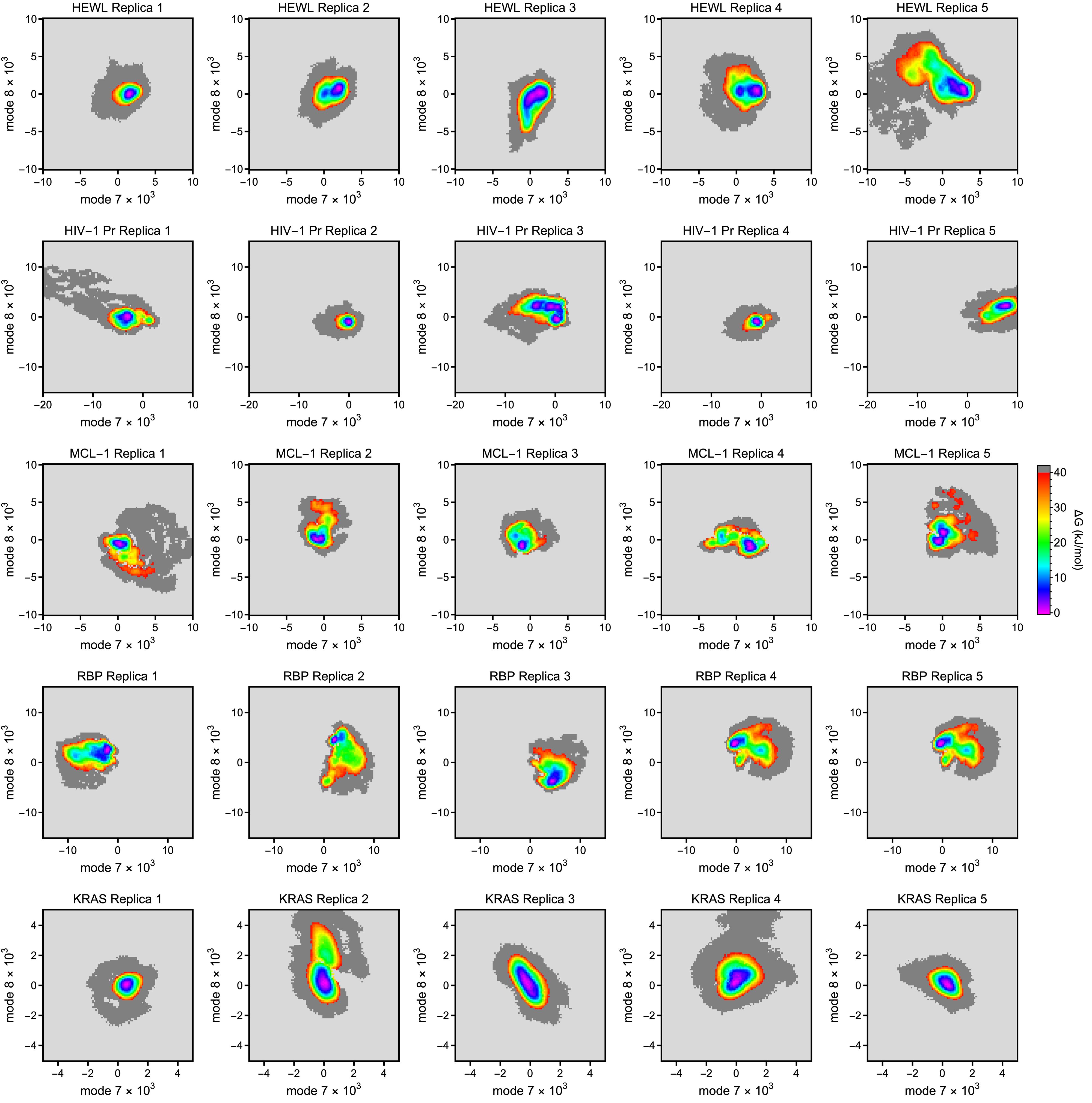}
\caption{
Free energy surfaces of replicas R1 to R5 for each protein system obtained from 100~ns metadynamics simulations as a function of vibrational modes used as CVs.
The specific CVs for each replica were obtained from independent 20~ns simulations using FRESEAN mode analysis and are thus not identical despite the high correlations shown in Figure~1 of the main text. 
Even trivial differences that have no impact on enhanced sampling can significantly alter the appearance of each replica FES, which is why a direct comparison is not possible.
For example, switches in sign of the corresponding eigenvectors have no impact on enhanced sampling or the correlation coefficient defined in Eq.~\ref{e:coeff}, but the corresponding FES would be inverted. 
Likewise, distinct linear combinations of the two selected eigenvectors have no impact on enhanced sampling or correlations of the two-dimensional sub-space defined by them, but the FES would be rotated.
}
\label{f:fes5cv}
\end{figure}

\newpage

\subsection{Free Energy Surfaces for Individual Replicas in Geometric Space}
\begin{figure}[h!]
\centering
\includegraphics[width=0.9\textwidth]{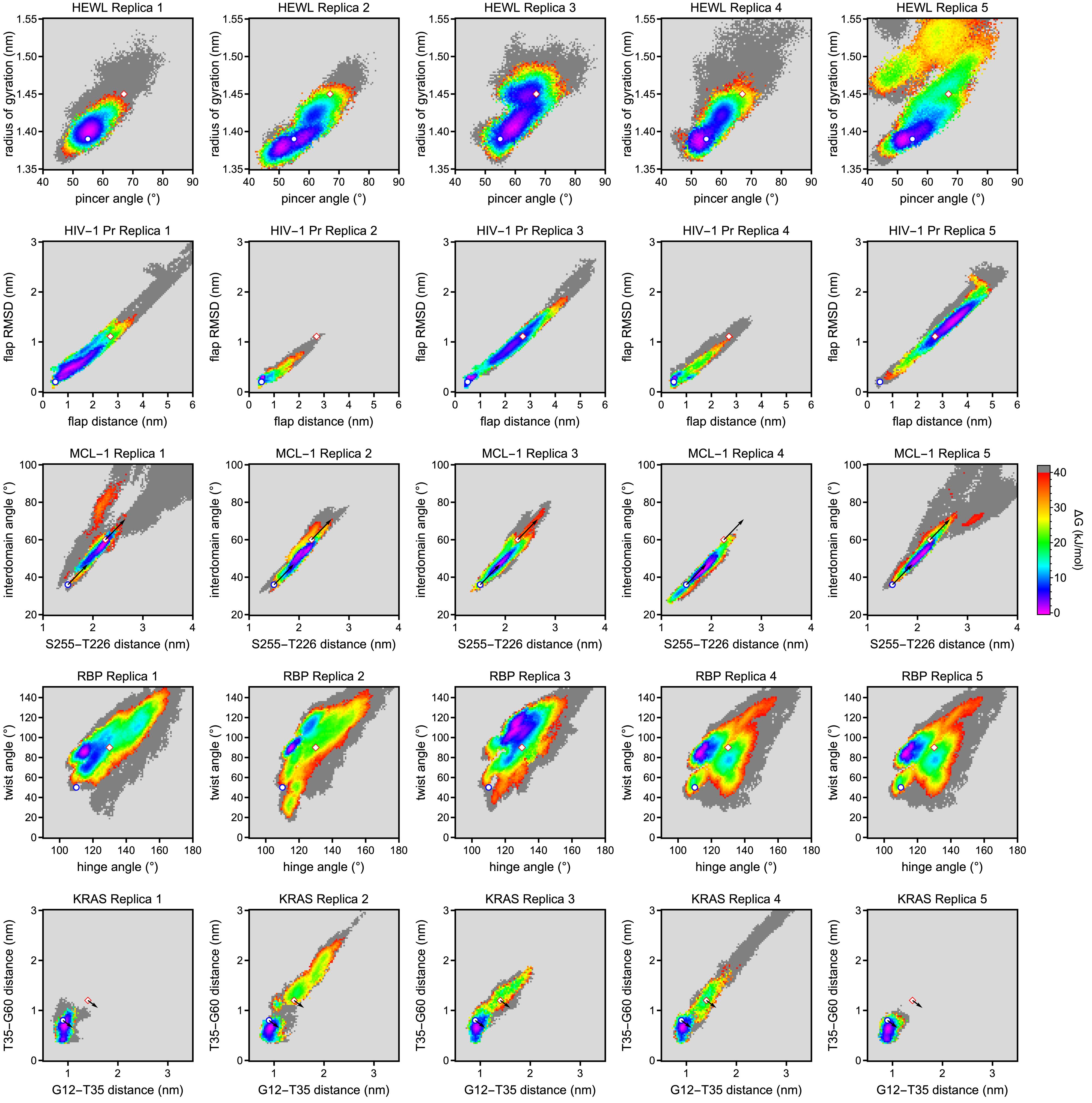}
\caption{
FESs of replicas R1 to R5 for each protein system obtained from 100~ns metadynamics simulations as a function of geometric variables defined in Table~\ref{t:geo} and Figure~\ref{f:geo}.
These FESs were obtained from weighted ensembles as described in Eqs.~\ref{e:w1} to \ref{e:w3} and can be compared between the five replicas for each system despite the use of distinct CVs during metadynamics (see caption of Figure~\ref{f:fes5cv}).
While common themes are apparent for different replicas for each system, we performed a second set of 20 replica metadynamics simulations (using identical CVs) to obtain the converged free energy surfaces shown in Figure~4 of the main text.
}
\label{f:fes5geo}
\end{figure}

\newpage

\subsection{Minimum Free Energy Path for Conformational Transition in HEWL}

\begin{figure}[h!]
\centering
\includegraphics[width=1\textwidth]{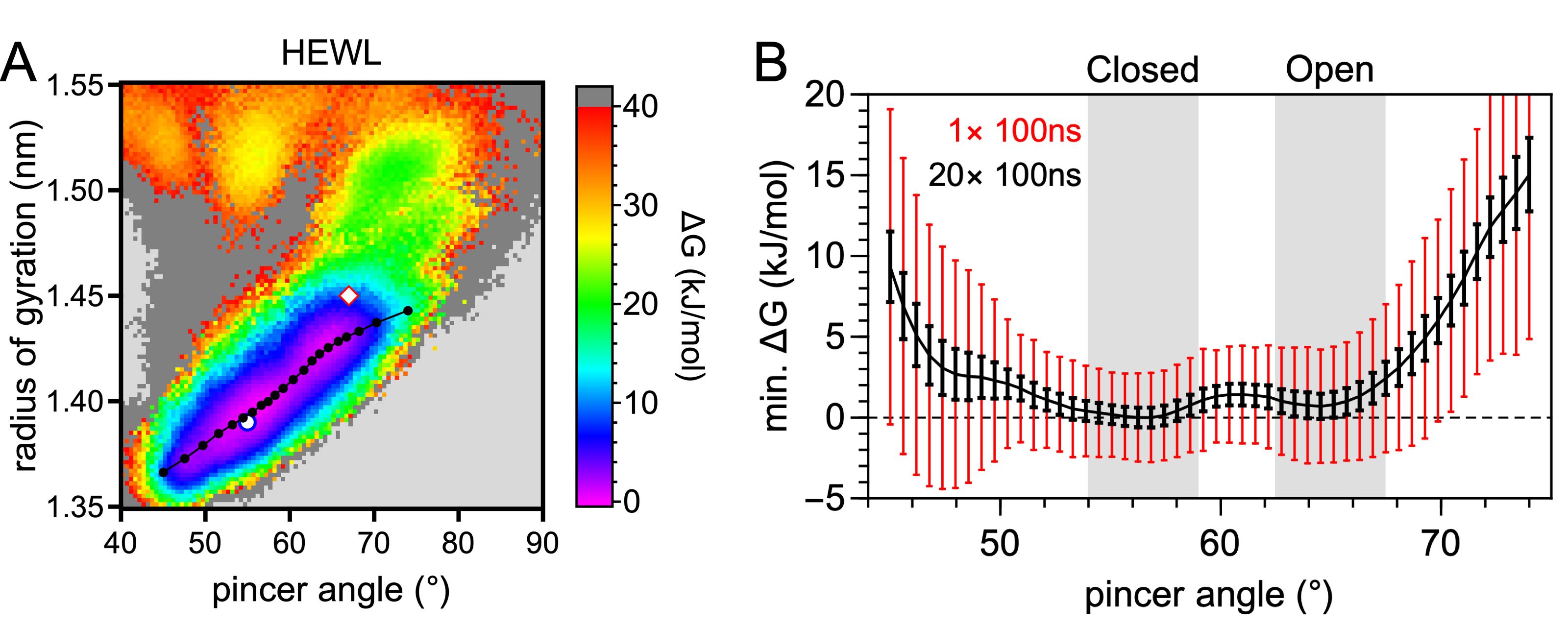}
\caption{To further visualize the statistical convergence of free energy surfaces obtained from 20 independent metadynamics simulations using low-frequency modes obtained from FRESEAN mode analysis as CVs, we re-plot in panel A the converged free energy surface for HEWL as a function of its geometric variables (same as in Figure~4 of the main text), while indicating the minimum free energy pathway connecting the "closed"  and "open" conformations (dashed line).
In panel B, we plot the average free energy profile along the minimum free energy path as a function of the pincer angle. 
Gray boxes are centered on the location of "closed" and "open" states as reported in the literature.
These do not exactly correspond to the minima of our free energy surface, but fall within larger regions in which we find the FES to be approximately flat.
In addition, we show error bars representing the standard deviation (red) and the error of the mean obtained after averaging over all 20 simulations (black).
}
\label{f:fes1d}
\end{figure}


\begin{thebibliography}{65}%
\makeatletter
\providecommand \@ifxundefined [1]{%
 \@ifx{#1\undefined}
}%
\providecommand \@ifnum [1]{%
 \ifnum #1\expandafter \@firstoftwo
 \else \expandafter \@secondoftwo
 \fi
}%
\providecommand \@ifx [1]{%
 \ifx #1\expandafter \@firstoftwo
 \else \expandafter \@secondoftwo
 \fi
}%
\providecommand \natexlab [1]{#1}%
\providecommand \enquote  [1]{``#1''}%
\providecommand \bibnamefont  [1]{#1}%
\providecommand \bibfnamefont [1]{#1}%
\providecommand \citenamefont [1]{#1}%
\providecommand \href@noop [0]{\@secondoftwo}%
\providecommand \href [0]{\begingroup \@sanitize@url \@href}%
\providecommand \@href[1]{\@@startlink{#1}\@@href}%
\providecommand \@@href[1]{\endgroup#1\@@endlink}%
\providecommand \@sanitize@url [0]{\catcode `\\12\catcode `\$12\catcode
  `\&12\catcode `\#12\catcode `\^12\catcode `\_12\catcode `\%12\relax}%
\providecommand \@@startlink[1]{}%
\providecommand \@@endlink[0]{}%
\providecommand \url  [0]{\begingroup\@sanitize@url \@url }%
\providecommand \@url [1]{\endgroup\@href {#1}{\urlprefix }}%
\providecommand \urlprefix  [0]{URL }%
\providecommand \Eprint [0]{\href }%
\providecommand \doibase [0]{https://doi.org/}%
\providecommand \selectlanguage [0]{\@gobble}%
\providecommand \bibinfo  [0]{\@secondoftwo}%
\providecommand \bibfield  [0]{\@secondoftwo}%
\providecommand \translation [1]{[#1]}%
\providecommand \BibitemOpen [0]{}%
\providecommand \bibitemStop [0]{}%
\providecommand \bibitemNoStop [0]{.\EOS\space}%
\providecommand \EOS [0]{\spacefactor3000\relax}%
\providecommand \BibitemShut  [1]{\csname bibitem#1\endcsname}%
\let\auto@bib@innerbib\@empty
\bibitem [{\citenamefont {Senior}\ \emph {et~al.}(2020)\citenamefont {Senior},
  \citenamefont {Evans}, \citenamefont {Jumper}, \citenamefont {Kirkpatrick},
  \citenamefont {Sifre}, \citenamefont {Green}, \citenamefont {Qin},
  \citenamefont {{\v{Z}}{\'\i}dek}, \citenamefont {Nelson}, \citenamefont
  {Bridgland} \emph {et~al.}}]{senior2020improved}%
  \BibitemOpen
  \bibfield  {author} {\bibinfo {author} {\bibfnamefont {A.~W.}\ \bibnamefont
  {Senior}}, \bibinfo {author} {\bibfnamefont {R.}~\bibnamefont {Evans}},
  \bibinfo {author} {\bibfnamefont {J.}~\bibnamefont {Jumper}}, \bibinfo
  {author} {\bibfnamefont {J.}~\bibnamefont {Kirkpatrick}}, \bibinfo {author}
  {\bibfnamefont {L.}~\bibnamefont {Sifre}}, \bibinfo {author} {\bibfnamefont
  {T.}~\bibnamefont {Green}}, \bibinfo {author} {\bibfnamefont
  {C.}~\bibnamefont {Qin}}, \bibinfo {author} {\bibfnamefont {A.}~\bibnamefont
  {{\v{Z}}{\'\i}dek}}, \bibinfo {author} {\bibfnamefont {A.~W.}\ \bibnamefont
  {Nelson}}, \bibinfo {author} {\bibfnamefont {A.}~\bibnamefont {Bridgland}},
  \emph {et~al.},\ }\bibfield  {title} {\bibinfo {title} {Improved protein
  structure prediction using potentials from deep learning},\ }\href@noop {}
  {\bibfield  {journal} {\bibinfo  {journal} {Nature}\ }\textbf {\bibinfo
  {volume} {577}},\ \bibinfo {pages} {706} (\bibinfo {year}
  {2020})}\BibitemShut {NoStop}%
\bibitem [{\citenamefont {Jumper}\ \emph {et~al.}(2021)\citenamefont {Jumper},
  \citenamefont {Evans}, \citenamefont {Pritzel}, \citenamefont {Green},
  \citenamefont {Figurnov}, \citenamefont {Ronneberger}, \citenamefont
  {Tunyasuvunakool}, \citenamefont {Bates}, \citenamefont {{\v{Z}}{\'\i}dek},
  \citenamefont {Potapenko} \emph {et~al.}}]{jumper2021highly}%
  \BibitemOpen
  \bibfield  {author} {\bibinfo {author} {\bibfnamefont {J.}~\bibnamefont
  {Jumper}}, \bibinfo {author} {\bibfnamefont {R.}~\bibnamefont {Evans}},
  \bibinfo {author} {\bibfnamefont {A.}~\bibnamefont {Pritzel}}, \bibinfo
  {author} {\bibfnamefont {T.}~\bibnamefont {Green}}, \bibinfo {author}
  {\bibfnamefont {M.}~\bibnamefont {Figurnov}}, \bibinfo {author}
  {\bibfnamefont {O.}~\bibnamefont {Ronneberger}}, \bibinfo {author}
  {\bibfnamefont {K.}~\bibnamefont {Tunyasuvunakool}}, \bibinfo {author}
  {\bibfnamefont {R.}~\bibnamefont {Bates}}, \bibinfo {author} {\bibfnamefont
  {A.}~\bibnamefont {{\v{Z}}{\'\i}dek}}, \bibinfo {author} {\bibfnamefont
  {A.}~\bibnamefont {Potapenko}}, \emph {et~al.},\ }\bibfield  {title}
  {\bibinfo {title} {Highly accurate protein structure prediction with
  alphafold},\ }\href@noop {} {\bibfield  {journal} {\bibinfo  {journal}
  {Nature}\ }\textbf {\bibinfo {volume} {596}},\ \bibinfo {pages} {583}
  (\bibinfo {year} {2021})}\BibitemShut {NoStop}%
\bibitem [{\citenamefont {Gao}\ \emph {et~al.}(2022)\citenamefont {Gao},
  \citenamefont {Nakajima~An}, \citenamefont {Parks},\ and\ \citenamefont
  {Skolnick}}]{gao2022af2complex}%
  \BibitemOpen
  \bibfield  {author} {\bibinfo {author} {\bibfnamefont {M.}~\bibnamefont
  {Gao}}, \bibinfo {author} {\bibfnamefont {D.}~\bibnamefont {Nakajima~An}},
  \bibinfo {author} {\bibfnamefont {J.~M.}\ \bibnamefont {Parks}},\ and\
  \bibinfo {author} {\bibfnamefont {J.}~\bibnamefont {Skolnick}},\ }\bibfield
  {title} {\bibinfo {title} {Af2complex predicts direct physical interactions
  in multimeric proteins with deep learning},\ }\href@noop {} {\bibfield
  {journal} {\bibinfo  {journal} {Nat. Commun.}\ }\textbf {\bibinfo {volume}
  {13}},\ \bibinfo {pages} {1744} (\bibinfo {year} {2022})}\BibitemShut
  {NoStop}%
\bibitem [{\citenamefont {Abramson}\ \emph {et~al.}(2024)\citenamefont
  {Abramson}, \citenamefont {Adler}, \citenamefont {Dunger}, \citenamefont
  {Evans}, \citenamefont {Green}, \citenamefont {Pritzel}, \citenamefont
  {Ronneberger}, \citenamefont {Willmore}, \citenamefont {Ballard},
  \citenamefont {Bambrick} \emph {et~al.}}]{abramson2024accurate}%
  \BibitemOpen
  \bibfield  {author} {\bibinfo {author} {\bibfnamefont {J.}~\bibnamefont
  {Abramson}}, \bibinfo {author} {\bibfnamefont {J.}~\bibnamefont {Adler}},
  \bibinfo {author} {\bibfnamefont {J.}~\bibnamefont {Dunger}}, \bibinfo
  {author} {\bibfnamefont {R.}~\bibnamefont {Evans}}, \bibinfo {author}
  {\bibfnamefont {T.}~\bibnamefont {Green}}, \bibinfo {author} {\bibfnamefont
  {A.}~\bibnamefont {Pritzel}}, \bibinfo {author} {\bibfnamefont
  {O.}~\bibnamefont {Ronneberger}}, \bibinfo {author} {\bibfnamefont
  {L.}~\bibnamefont {Willmore}}, \bibinfo {author} {\bibfnamefont {A.~J.}\
  \bibnamefont {Ballard}}, \bibinfo {author} {\bibfnamefont {J.}~\bibnamefont
  {Bambrick}}, \emph {et~al.},\ }\bibfield  {title} {\bibinfo {title} {Accurate
  structure prediction of biomolecular interactions with alphafold 3},\
  }\href@noop {} {\bibfield  {journal} {\bibinfo  {journal} {Nature}\ }\textbf
  {\bibinfo {volume} {630}},\ \bibinfo {pages} {493} (\bibinfo {year}
  {2024})}\BibitemShut {NoStop}%
\bibitem [{\citenamefont {Lin}\ \emph {et~al.}(2023)\citenamefont {Lin},
  \citenamefont {Akin}, \citenamefont {Rao}, \citenamefont {Hie}, \citenamefont
  {Zhu}, \citenamefont {Lu}, \citenamefont {Smetanin}, \citenamefont {Verkuil},
  \citenamefont {Kabeli}, \citenamefont {Shmueli}, \citenamefont {Dos
  Santos~Costa}, \citenamefont {Fazel-Zarandi}, \citenamefont {Sercu},
  \citenamefont {Candido},\ and\ \citenamefont {Rives}}]{lin2023evolutionary}%
  \BibitemOpen
  \bibfield  {author} {\bibinfo {author} {\bibfnamefont {Z.}~\bibnamefont
  {Lin}}, \bibinfo {author} {\bibfnamefont {H.}~\bibnamefont {Akin}}, \bibinfo
  {author} {\bibfnamefont {R.}~\bibnamefont {Rao}}, \bibinfo {author}
  {\bibfnamefont {B.}~\bibnamefont {Hie}}, \bibinfo {author} {\bibfnamefont
  {Z.}~\bibnamefont {Zhu}}, \bibinfo {author} {\bibfnamefont {W.}~\bibnamefont
  {Lu}}, \bibinfo {author} {\bibfnamefont {N.}~\bibnamefont {Smetanin}},
  \bibinfo {author} {\bibfnamefont {R.}~\bibnamefont {Verkuil}}, \bibinfo
  {author} {\bibfnamefont {O.}~\bibnamefont {Kabeli}}, \bibinfo {author}
  {\bibfnamefont {Y.}~\bibnamefont {Shmueli}}, \bibinfo {author} {\bibfnamefont
  {A.}~\bibnamefont {Dos Santos~Costa}}, \bibinfo {author} {\bibfnamefont
  {M.}~\bibnamefont {Fazel-Zarandi}}, \bibinfo {author} {\bibfnamefont
  {T.}~\bibnamefont {Sercu}}, \bibinfo {author} {\bibfnamefont
  {S.}~\bibnamefont {Candido}},\ and\ \bibinfo {author} {\bibfnamefont
  {A.}~\bibnamefont {Rives}},\ }\bibfield  {title} {\bibinfo {title}
  {Evolutionary-scale prediction of atomic-level protein structure with a
  language model},\ }\href@noop {} {\bibfield  {journal} {\bibinfo  {journal}
  {Science}\ }\textbf {\bibinfo {volume} {379}},\ \bibinfo {pages} {1123}
  (\bibinfo {year} {2023})}\BibitemShut {NoStop}%
\bibitem [{\citenamefont {Baek}\ \emph {et~al.}(2021)\citenamefont {Baek},
  \citenamefont {DiMaio}, \citenamefont {Anishchenko}, \citenamefont
  {Dauparas}, \citenamefont {Ovchinnikov}, \citenamefont {Lee}, \citenamefont
  {Wang}, \citenamefont {Cong}, \citenamefont {Kinch}, \citenamefont
  {Schaeffer}, \citenamefont {Mil{\'a}n}, \citenamefont {Park}, \citenamefont
  {Adams}, \citenamefont {Glassman}, \citenamefont {DeGiovanni}, \citenamefont
  {Pereira}, \citenamefont {Rodrigues}, \citenamefont {van Dijk}, \citenamefont
  {Ebrecht}, \citenamefont {Opperman}, \citenamefont {Sagmeister},
  \citenamefont {Buhlheller}, \citenamefont {Pavkov-Keller}, \citenamefont
  {Rathinaswamy}, \citenamefont {Dalwadi}, \citenamefont {Yip}, \citenamefont
  {Burke}, \citenamefont {Garcia}, \citenamefont {Grishin}, \citenamefont
  {Adams}, \citenamefont {Read},\ and\ \citenamefont
  {Baker}}]{baek2021accurate}%
  \BibitemOpen
  \bibfield  {author} {\bibinfo {author} {\bibfnamefont {M.}~\bibnamefont
  {Baek}}, \bibinfo {author} {\bibfnamefont {F.}~\bibnamefont {DiMaio}},
  \bibinfo {author} {\bibfnamefont {I.}~\bibnamefont {Anishchenko}}, \bibinfo
  {author} {\bibfnamefont {J.}~\bibnamefont {Dauparas}}, \bibinfo {author}
  {\bibfnamefont {S.}~\bibnamefont {Ovchinnikov}}, \bibinfo {author}
  {\bibfnamefont {G.~R.}\ \bibnamefont {Lee}}, \bibinfo {author} {\bibfnamefont
  {J.}~\bibnamefont {Wang}}, \bibinfo {author} {\bibfnamefont {Q.}~\bibnamefont
  {Cong}}, \bibinfo {author} {\bibfnamefont {L.~N.}\ \bibnamefont {Kinch}},
  \bibinfo {author} {\bibfnamefont {R.~D.}\ \bibnamefont {Schaeffer}}, \bibinfo
  {author} {\bibfnamefont {C.}~\bibnamefont {Mil{\'a}n}}, \bibinfo {author}
  {\bibfnamefont {H.}~\bibnamefont {Park}}, \bibinfo {author} {\bibfnamefont
  {C.}~\bibnamefont {Adams}}, \bibinfo {author} {\bibfnamefont {C.~R.}\
  \bibnamefont {Glassman}}, \bibinfo {author} {\bibfnamefont {A.}~\bibnamefont
  {DeGiovanni}}, \bibinfo {author} {\bibfnamefont {J.~H.}\ \bibnamefont
  {Pereira}}, \bibinfo {author} {\bibfnamefont {A.~V.}\ \bibnamefont
  {Rodrigues}}, \bibinfo {author} {\bibfnamefont {A.~A.}\ \bibnamefont {van
  Dijk}}, \bibinfo {author} {\bibfnamefont {A.~C.}\ \bibnamefont {Ebrecht}},
  \bibinfo {author} {\bibfnamefont {D.~J.}\ \bibnamefont {Opperman}}, \bibinfo
  {author} {\bibfnamefont {T.}~\bibnamefont {Sagmeister}}, \bibinfo {author}
  {\bibfnamefont {C.}~\bibnamefont {Buhlheller}}, \bibinfo {author}
  {\bibfnamefont {T.}~\bibnamefont {Pavkov-Keller}}, \bibinfo {author}
  {\bibfnamefont {M.~K.}\ \bibnamefont {Rathinaswamy}}, \bibinfo {author}
  {\bibfnamefont {U.}~\bibnamefont {Dalwadi}}, \bibinfo {author} {\bibfnamefont
  {C.~K.}\ \bibnamefont {Yip}}, \bibinfo {author} {\bibfnamefont {J.~E.}\
  \bibnamefont {Burke}}, \bibinfo {author} {\bibfnamefont {K.~C.}\ \bibnamefont
  {Garcia}}, \bibinfo {author} {\bibfnamefont {N.~V.}\ \bibnamefont {Grishin}},
  \bibinfo {author} {\bibfnamefont {P.~D.}\ \bibnamefont {Adams}}, \bibinfo
  {author} {\bibfnamefont {R.~J.}\ \bibnamefont {Read}},\ and\ \bibinfo
  {author} {\bibfnamefont {D.}~\bibnamefont {Baker}},\ }\bibfield  {title}
  {\bibinfo {title} {Accurate prediction of protein structures and interactions
  using a three-track neural network},\ }\href@noop {} {\bibfield  {journal}
  {\bibinfo  {journal} {Science}\ }\textbf {\bibinfo {volume} {373}},\ \bibinfo
  {pages} {871} (\bibinfo {year} {2021})}\BibitemShut {NoStop}%
\bibitem [{\citenamefont {Krishna}\ \emph {et~al.}(2024)\citenamefont
  {Krishna}, \citenamefont {Wang}, \citenamefont {Ahern}, \citenamefont
  {Sturmfels}, \citenamefont {Venkatesh}, \citenamefont {Kalvet}, \citenamefont
  {Lee}, \citenamefont {Morey-Burrows}, \citenamefont {Anishchenko},
  \citenamefont {Humphreys}, \citenamefont {McHugh}, \citenamefont {Vafeados},
  \citenamefont {Li}, \citenamefont {Sutherland}, \citenamefont {Hitchcock},
  \citenamefont {Hunter}, \citenamefont {Kang}, \citenamefont {Brackenbrough},
  \citenamefont {Bera}, \citenamefont {Baek}, \citenamefont {DiMaio},\ and\
  \citenamefont {Baker}}]{krishna2024generalized}%
  \BibitemOpen
  \bibfield  {author} {\bibinfo {author} {\bibfnamefont {R.}~\bibnamefont
  {Krishna}}, \bibinfo {author} {\bibfnamefont {J.}~\bibnamefont {Wang}},
  \bibinfo {author} {\bibfnamefont {W.}~\bibnamefont {Ahern}}, \bibinfo
  {author} {\bibfnamefont {P.}~\bibnamefont {Sturmfels}}, \bibinfo {author}
  {\bibfnamefont {P.}~\bibnamefont {Venkatesh}}, \bibinfo {author}
  {\bibfnamefont {I.}~\bibnamefont {Kalvet}}, \bibinfo {author} {\bibfnamefont
  {G.~R.}\ \bibnamefont {Lee}}, \bibinfo {author} {\bibfnamefont {F.~S.}\
  \bibnamefont {Morey-Burrows}}, \bibinfo {author} {\bibfnamefont
  {I.}~\bibnamefont {Anishchenko}}, \bibinfo {author} {\bibfnamefont {I.~R.}\
  \bibnamefont {Humphreys}}, \bibinfo {author} {\bibfnamefont {R.}~\bibnamefont
  {McHugh}}, \bibinfo {author} {\bibfnamefont {D.}~\bibnamefont {Vafeados}},
  \bibinfo {author} {\bibfnamefont {X.}~\bibnamefont {Li}}, \bibinfo {author}
  {\bibfnamefont {G.~A.}\ \bibnamefont {Sutherland}}, \bibinfo {author}
  {\bibfnamefont {A.}~\bibnamefont {Hitchcock}}, \bibinfo {author}
  {\bibfnamefont {C.~N.}\ \bibnamefont {Hunter}}, \bibinfo {author}
  {\bibfnamefont {A.}~\bibnamefont {Kang}}, \bibinfo {author} {\bibfnamefont
  {E.}~\bibnamefont {Brackenbrough}}, \bibinfo {author} {\bibfnamefont {A.~K.}\
  \bibnamefont {Bera}}, \bibinfo {author} {\bibfnamefont {M.}~\bibnamefont
  {Baek}}, \bibinfo {author} {\bibfnamefont {F.}~\bibnamefont {DiMaio}},\ and\
  \bibinfo {author} {\bibfnamefont {D.}~\bibnamefont {Baker}},\ }\bibfield
  {title} {\bibinfo {title} {Generalized biomolecular modeling and design with
  rosettafold all-atom},\ }\href@noop {} {\bibfield  {journal} {\bibinfo
  {journal} {Science}\ }\textbf {\bibinfo {volume} {384}},\ \bibinfo {pages}
  {eadl2528} (\bibinfo {year} {2024})}\BibitemShut {NoStop}%
\bibitem [{\citenamefont {Watson}\ \emph {et~al.}(2023)\citenamefont {Watson},
  \citenamefont {Juergens}, \citenamefont {Bennett}, \citenamefont {Trippe},
  \citenamefont {Yim}, \citenamefont {Eisenach}, \citenamefont {Ahern},
  \citenamefont {Borst}, \citenamefont {Ragotte}, \citenamefont {Milles},
  \citenamefont {Wicky}, \citenamefont {Hanikel}, \citenamefont {Pellock},
  \citenamefont {Courbet}, \citenamefont {Sheffler}, \citenamefont {Wang},
  \citenamefont {Venkatesh}, \citenamefont {Sappington}, \citenamefont
  {V{\'a}zquez~Torres}, \citenamefont {Lauko}, \citenamefont {De~Bortoli},
  \citenamefont {Mathieu}, \citenamefont {Ovchinnikov}, \citenamefont
  {Barzilay}, \citenamefont {Jaakola}, \citenamefont {DiMaio}, \citenamefont
  {Baek},\ and\ \citenamefont {Baker}}]{watson2023novo}%
  \BibitemOpen
  \bibfield  {author} {\bibinfo {author} {\bibfnamefont {J.~L.}\ \bibnamefont
  {Watson}}, \bibinfo {author} {\bibfnamefont {D.}~\bibnamefont {Juergens}},
  \bibinfo {author} {\bibfnamefont {N.~R.}\ \bibnamefont {Bennett}}, \bibinfo
  {author} {\bibfnamefont {B.~L.}\ \bibnamefont {Trippe}}, \bibinfo {author}
  {\bibfnamefont {J.}~\bibnamefont {Yim}}, \bibinfo {author} {\bibfnamefont
  {H.~E.}\ \bibnamefont {Eisenach}}, \bibinfo {author} {\bibfnamefont
  {W.}~\bibnamefont {Ahern}}, \bibinfo {author} {\bibfnamefont {A.~J.}\
  \bibnamefont {Borst}}, \bibinfo {author} {\bibfnamefont {R.~J.}\ \bibnamefont
  {Ragotte}}, \bibinfo {author} {\bibfnamefont {L.~F.}\ \bibnamefont {Milles}},
  \bibinfo {author} {\bibfnamefont {B.~I.~M.}\ \bibnamefont {Wicky}}, \bibinfo
  {author} {\bibfnamefont {N.}~\bibnamefont {Hanikel}}, \bibinfo {author}
  {\bibfnamefont {S.~J.}\ \bibnamefont {Pellock}}, \bibinfo {author}
  {\bibfnamefont {A.}~\bibnamefont {Courbet}}, \bibinfo {author} {\bibfnamefont
  {W.}~\bibnamefont {Sheffler}}, \bibinfo {author} {\bibfnamefont
  {J.}~\bibnamefont {Wang}}, \bibinfo {author} {\bibfnamefont {P.}~\bibnamefont
  {Venkatesh}}, \bibinfo {author} {\bibfnamefont {I.}~\bibnamefont
  {Sappington}}, \bibinfo {author} {\bibfnamefont {S.}~\bibnamefont
  {V{\'a}zquez~Torres}}, \bibinfo {author} {\bibfnamefont {A.}~\bibnamefont
  {Lauko}}, \bibinfo {author} {\bibfnamefont {V.}~\bibnamefont {De~Bortoli}},
  \bibinfo {author} {\bibfnamefont {E.}~\bibnamefont {Mathieu}}, \bibinfo
  {author} {\bibfnamefont {S.}~\bibnamefont {Ovchinnikov}}, \bibinfo {author}
  {\bibfnamefont {R.}~\bibnamefont {Barzilay}}, \bibinfo {author}
  {\bibfnamefont {T.~S.}\ \bibnamefont {Jaakola}}, \bibinfo {author}
  {\bibfnamefont {F.}~\bibnamefont {DiMaio}}, \bibinfo {author} {\bibfnamefont
  {M.}~\bibnamefont {Baek}},\ and\ \bibinfo {author} {\bibfnamefont
  {D.}~\bibnamefont {Baker}},\ }\bibfield  {title} {\bibinfo {title} {De novo
  design of protein structure and function with rfdiffusion},\ }\href@noop {}
  {\bibfield  {journal} {\bibinfo  {journal} {Nature}\ }\textbf {\bibinfo
  {volume} {620}},\ \bibinfo {pages} {1089} (\bibinfo {year}
  {2023})}\BibitemShut {NoStop}%
\bibitem [{\citenamefont {Petrovi{\'c}}\ \emph {et~al.}(2018)\citenamefont
  {Petrovi{\'c}}, \citenamefont {Risso}, \citenamefont {Kamerlin},\ and\
  \citenamefont {Sanchez-Ruiz}}]{petrovic2018conformational}%
  \BibitemOpen
  \bibfield  {author} {\bibinfo {author} {\bibfnamefont {D.}~\bibnamefont
  {Petrovi{\'c}}}, \bibinfo {author} {\bibfnamefont {V.~A.}\ \bibnamefont
  {Risso}}, \bibinfo {author} {\bibfnamefont {S.~C.~L.}\ \bibnamefont
  {Kamerlin}},\ and\ \bibinfo {author} {\bibfnamefont {J.~M.}\ \bibnamefont
  {Sanchez-Ruiz}},\ }\bibfield  {title} {\bibinfo {title} {Conformational
  dynamics and enzyme evolution},\ }\href@noop {} {\bibfield  {journal}
  {\bibinfo  {journal} {J. Roy. Soc. Interface}\ }\textbf {\bibinfo {volume}
  {15}},\ \bibinfo {pages} {20180330} (\bibinfo {year} {2018})}\BibitemShut
  {NoStop}%
\bibitem [{\citenamefont {Campitelli}\ \emph {et~al.}(2020)\citenamefont
  {Campitelli}, \citenamefont {Modi}, \citenamefont {Kumar},\ and\
  \citenamefont {Ozkan}}]{campitelli2020role}%
  \BibitemOpen
  \bibfield  {author} {\bibinfo {author} {\bibfnamefont {P.}~\bibnamefont
  {Campitelli}}, \bibinfo {author} {\bibfnamefont {T.}~\bibnamefont {Modi}},
  \bibinfo {author} {\bibfnamefont {S.}~\bibnamefont {Kumar}},\ and\ \bibinfo
  {author} {\bibfnamefont {S.~B.}\ \bibnamefont {Ozkan}},\ }\bibfield  {title}
  {\bibinfo {title} {The role of conformational dynamics and allostery in
  modulating protein evolution},\ }\href@noop {} {\bibfield  {journal}
  {\bibinfo  {journal} {Ann. Rev. Biophys.}\ }\textbf {\bibinfo {volume}
  {49}},\ \bibinfo {pages} {267} (\bibinfo {year} {2020})}\BibitemShut
  {NoStop}%
\bibitem [{\citenamefont {Ayaz}\ \emph {et~al.}(2023)\citenamefont {Ayaz},
  \citenamefont {Lyczek}, \citenamefont {Paung}, \citenamefont {Mingione},
  \citenamefont {Iacob}, \citenamefont {de~Waal}, \citenamefont {Engen},
  \citenamefont {Seeliger}, \citenamefont {Shan},\ and\ \citenamefont
  {Shaw}}]{ayaz2023}%
  \BibitemOpen
  \bibfield  {author} {\bibinfo {author} {\bibfnamefont {P.}~\bibnamefont
  {Ayaz}}, \bibinfo {author} {\bibfnamefont {A.}~\bibnamefont {Lyczek}},
  \bibinfo {author} {\bibfnamefont {Y.}~\bibnamefont {Paung}}, \bibinfo
  {author} {\bibfnamefont {V.~R.}\ \bibnamefont {Mingione}}, \bibinfo {author}
  {\bibfnamefont {R.~E.}\ \bibnamefont {Iacob}}, \bibinfo {author}
  {\bibfnamefont {P.~W.}\ \bibnamefont {de~Waal}}, \bibinfo {author}
  {\bibfnamefont {J.~R.}\ \bibnamefont {Engen}}, \bibinfo {author}
  {\bibfnamefont {M.~A.}\ \bibnamefont {Seeliger}}, \bibinfo {author}
  {\bibfnamefont {Y.}~\bibnamefont {Shan}},\ and\ \bibinfo {author}
  {\bibfnamefont {D.~E.}\ \bibnamefont {Shaw}},\ }\bibfield  {title} {\bibinfo
  {title} {Structural mechanism of a drug-binding process involving a large
  conformational change of the protein target},\ }\href@noop {} {\bibfield
  {journal} {\bibinfo  {journal} {Nat. Commun.}\ }\textbf {\bibinfo {volume}
  {14}},\ \bibinfo {pages} {1885} (\bibinfo {year} {2023})}\BibitemShut
  {NoStop}%
\bibitem [{\citenamefont {Berman}\ \emph {et~al.}(2000)\citenamefont {Berman},
  \citenamefont {Westbrook}, \citenamefont {Feng}, \citenamefont {Gilliland},
  \citenamefont {Bhat}, \citenamefont {Weissig}, \citenamefont {Shindyalov},\
  and\ \citenamefont {Bourne}}]{PDB}%
  \BibitemOpen
  \bibfield  {author} {\bibinfo {author} {\bibfnamefont {H.~M.}\ \bibnamefont
  {Berman}}, \bibinfo {author} {\bibfnamefont {J.}~\bibnamefont {Westbrook}},
  \bibinfo {author} {\bibfnamefont {Z.}~\bibnamefont {Feng}}, \bibinfo {author}
  {\bibfnamefont {G.}~\bibnamefont {Gilliland}}, \bibinfo {author}
  {\bibfnamefont {T.~N.}\ \bibnamefont {Bhat}}, \bibinfo {author}
  {\bibfnamefont {H.}~\bibnamefont {Weissig}}, \bibinfo {author} {\bibfnamefont
  {I.~N.}\ \bibnamefont {Shindyalov}},\ and\ \bibinfo {author} {\bibfnamefont
  {P.~E.}\ \bibnamefont {Bourne}},\ }\bibfield  {title} {\bibinfo {title} {{The
  Protein Data Bank}},\ }\href {https://doi.org/10.1093/nar/28.1.235}
  {\bibfield  {journal} {\bibinfo  {journal} {Nucleic Acids Res.}\ }\textbf
  {\bibinfo {volume} {28}},\ \bibinfo {pages} {235} (\bibinfo {year} {2000})},\
  \Eprint
  {https://arxiv.org/abs/https://academic.oup.com/nar/article-pdf/28/1/235/9895144/280235.pdf}
  {https://academic.oup.com/nar/article-pdf/28/1/235/9895144/280235.pdf}
  \BibitemShut {NoStop}%
\bibitem [{\citenamefont {Br{\"a}nd{\'e}n}\ and\ \citenamefont
  {Neutze}(2021)}]{branden2021advances}%
  \BibitemOpen
  \bibfield  {author} {\bibinfo {author} {\bibfnamefont {G.}~\bibnamefont
  {Br{\"a}nd{\'e}n}}\ and\ \bibinfo {author} {\bibfnamefont {R.}~\bibnamefont
  {Neutze}},\ }\bibfield  {title} {\bibinfo {title} {Advances and challenges in
  time-resolved macromolecular crystallography},\ }\href@noop {} {\bibfield
  {journal} {\bibinfo  {journal} {Science}\ }\textbf {\bibinfo {volume}
  {373}},\ \bibinfo {pages} {eaba0954} (\bibinfo {year} {2021})}\BibitemShut
  {NoStop}%
\bibitem [{\citenamefont {Hekstra}(2023)}]{hekstra2023emerging}%
  \BibitemOpen
  \bibfield  {author} {\bibinfo {author} {\bibfnamefont {D.~R.}\ \bibnamefont
  {Hekstra}},\ }\bibfield  {title} {\bibinfo {title} {Emerging time-resolved
  x-ray diffraction approaches for protein dynamics},\ }\href@noop {}
  {\bibfield  {journal} {\bibinfo  {journal} {Ann. Rev. Biophys.}\ }\textbf
  {\bibinfo {volume} {52}},\ \bibinfo {pages} {255} (\bibinfo {year}
  {2023})}\BibitemShut {NoStop}%
\bibitem [{\citenamefont {Nettels}\ \emph {et~al.}(2024)\citenamefont
  {Nettels}, \citenamefont {Galvanetto}, \citenamefont {Ivanovi{\'c}},
  \citenamefont {N{\"u}esch}, \citenamefont {Yang},\ and\ \citenamefont
  {Schuler}}]{nettels2024single}%
  \BibitemOpen
  \bibfield  {author} {\bibinfo {author} {\bibfnamefont {D.}~\bibnamefont
  {Nettels}}, \bibinfo {author} {\bibfnamefont {N.}~\bibnamefont {Galvanetto}},
  \bibinfo {author} {\bibfnamefont {M.~T.}\ \bibnamefont {Ivanovi{\'c}}},
  \bibinfo {author} {\bibfnamefont {M.}~\bibnamefont {N{\"u}esch}}, \bibinfo
  {author} {\bibfnamefont {T.}~\bibnamefont {Yang}},\ and\ \bibinfo {author}
  {\bibfnamefont {B.}~\bibnamefont {Schuler}},\ }\bibfield  {title} {\bibinfo
  {title} {Single-molecule fret for probing nanoscale biomolecular dynamics},\
  }\href@noop {} {\bibfield  {journal} {\bibinfo  {journal} {Nat. Rev. Phys.}\
  }\textbf {\bibinfo {volume} {6}},\ \bibinfo {pages} {587} (\bibinfo {year}
  {2024})}\BibitemShut {NoStop}%
\bibitem [{\citenamefont {Hohng}\ \emph {et~al.}(2014)\citenamefont {Hohng},
  \citenamefont {Lee}, \citenamefont {Lee},\ and\ \citenamefont
  {Jo}}]{hohng2014maximizing}%
  \BibitemOpen
  \bibfield  {author} {\bibinfo {author} {\bibfnamefont {S.}~\bibnamefont
  {Hohng}}, \bibinfo {author} {\bibfnamefont {S.}~\bibnamefont {Lee}}, \bibinfo
  {author} {\bibfnamefont {J.}~\bibnamefont {Lee}},\ and\ \bibinfo {author}
  {\bibfnamefont {M.~H.}\ \bibnamefont {Jo}},\ }\bibfield  {title} {\bibinfo
  {title} {Maximizing information content of single-molecule fret experiments:
  multi-color fret and fret combined with force or torque},\ }\href@noop {}
  {\bibfield  {journal} {\bibinfo  {journal} {Chem. Soc. Rev.}\ }\textbf
  {\bibinfo {volume} {43}},\ \bibinfo {pages} {1007} (\bibinfo {year}
  {2014})}\BibitemShut {NoStop}%
\bibitem [{\citenamefont {G{\"o}tz}\ \emph {et~al.}(2016)\citenamefont
  {G{\"o}tz}, \citenamefont {Wortmann}, \citenamefont {Schmid},\ and\
  \citenamefont {Hugel}}]{gotz2016multicolor}%
  \BibitemOpen
  \bibfield  {author} {\bibinfo {author} {\bibfnamefont {M.}~\bibnamefont
  {G{\"o}tz}}, \bibinfo {author} {\bibfnamefont {P.}~\bibnamefont {Wortmann}},
  \bibinfo {author} {\bibfnamefont {S.}~\bibnamefont {Schmid}},\ and\ \bibinfo
  {author} {\bibfnamefont {T.}~\bibnamefont {Hugel}},\ }\bibfield  {title}
  {\bibinfo {title} {A multicolor single-molecule fret approach to study
  protein dynamics and interactions simultaneously},\ }in\ \href@noop {} {\emph
  {\bibinfo {booktitle} {Methods in enzymology}}},\ Vol.\ \bibinfo {volume}
  {581}\ (\bibinfo  {publisher} {Elsevier},\ \bibinfo {year} {2016})\ pp.\
  \bibinfo {pages} {487--516}\BibitemShut {NoStop}%
\bibitem [{\citenamefont {Chung}\ and\ \citenamefont
  {Eaton}(2018)}]{chung2018protein}%
  \BibitemOpen
  \bibfield  {author} {\bibinfo {author} {\bibfnamefont {H.~S.}\ \bibnamefont
  {Chung}}\ and\ \bibinfo {author} {\bibfnamefont {W.~A.}\ \bibnamefont
  {Eaton}},\ }\bibfield  {title} {\bibinfo {title} {Protein folding transition
  path times from single molecule fret},\ }\href@noop {} {\bibfield  {journal}
  {\bibinfo  {journal} {Current opinion in structural biology}\ }\textbf
  {\bibinfo {volume} {48}},\ \bibinfo {pages} {30} (\bibinfo {year}
  {2018})}\BibitemShut {NoStop}%
\bibitem [{\citenamefont {Yoo}\ \emph {et~al.}(2020)\citenamefont {Yoo},
  \citenamefont {Kim}, \citenamefont {Louis}, \citenamefont {Gopich},\ and\
  \citenamefont {Chung}}]{yoo2020fast}%
  \BibitemOpen
  \bibfield  {author} {\bibinfo {author} {\bibfnamefont {J.}~\bibnamefont
  {Yoo}}, \bibinfo {author} {\bibfnamefont {J.-Y.}\ \bibnamefont {Kim}},
  \bibinfo {author} {\bibfnamefont {J.~M.}\ \bibnamefont {Louis}}, \bibinfo
  {author} {\bibfnamefont {I.~V.}\ \bibnamefont {Gopich}},\ and\ \bibinfo
  {author} {\bibfnamefont {H.~S.}\ \bibnamefont {Chung}},\ }\bibfield  {title}
  {\bibinfo {title} {Fast three-color single-molecule fret using statistical
  inference},\ }\href@noop {} {\bibfield  {journal} {\bibinfo  {journal} {Nat.
  Commun.}\ }\textbf {\bibinfo {volume} {11}},\ \bibinfo {pages} {3336}
  (\bibinfo {year} {2020})}\BibitemShut {NoStop}%
\bibitem [{\citenamefont {Barducci}\ \emph {et~al.}(2008)\citenamefont
  {Barducci}, \citenamefont {Bussi},\ and\ \citenamefont
  {Parrinello}}]{barducci2008well}%
  \BibitemOpen
  \bibfield  {author} {\bibinfo {author} {\bibfnamefont {A.}~\bibnamefont
  {Barducci}}, \bibinfo {author} {\bibfnamefont {G.}~\bibnamefont {Bussi}},\
  and\ \bibinfo {author} {\bibfnamefont {M.}~\bibnamefont {Parrinello}},\
  }\bibfield  {title} {\bibinfo {title} {Well-tempered metadynamics: a smoothly
  converging and tunable free-energy method},\ }\href@noop {} {\bibfield
  {journal} {\bibinfo  {journal} {Phys. Rev. Lett.}\ }\textbf {\bibinfo
  {volume} {100}},\ \bibinfo {pages} {020603} (\bibinfo {year}
  {2008})}\BibitemShut {NoStop}%
\bibitem [{\citenamefont {Zwier}\ \emph {et~al.}(2015)\citenamefont {Zwier},
  \citenamefont {Adelman}, \citenamefont {Kaus}, \citenamefont {Pratt},
  \citenamefont {Wong}, \citenamefont {Rego}, \citenamefont {Su{\'a}rez},
  \citenamefont {Lettieri}, \citenamefont {Wang}, \citenamefont {Grabe},
  \citenamefont {Zuckerman},\ and\ \citenamefont {Chong}}]{zwier2015westpa}%
  \BibitemOpen
  \bibfield  {author} {\bibinfo {author} {\bibfnamefont {M.~C.}\ \bibnamefont
  {Zwier}}, \bibinfo {author} {\bibfnamefont {J.~L.}\ \bibnamefont {Adelman}},
  \bibinfo {author} {\bibfnamefont {J.~W.}\ \bibnamefont {Kaus}}, \bibinfo
  {author} {\bibfnamefont {A.~J.}\ \bibnamefont {Pratt}}, \bibinfo {author}
  {\bibfnamefont {K.~F.}\ \bibnamefont {Wong}}, \bibinfo {author}
  {\bibfnamefont {N.~B.}\ \bibnamefont {Rego}}, \bibinfo {author}
  {\bibfnamefont {E.}~\bibnamefont {Su{\'a}rez}}, \bibinfo {author}
  {\bibfnamefont {S.}~\bibnamefont {Lettieri}}, \bibinfo {author}
  {\bibfnamefont {D.~W.}\ \bibnamefont {Wang}}, \bibinfo {author}
  {\bibfnamefont {M.}~\bibnamefont {Grabe}}, \bibinfo {author} {\bibfnamefont
  {D.~M.}\ \bibnamefont {Zuckerman}},\ and\ \bibinfo {author} {\bibfnamefont
  {L.~T.}\ \bibnamefont {Chong}},\ }\bibfield  {title} {\bibinfo {title}
  {Westpa: An interoperable, highly scalable software package for weighted
  ensemble simulation and analysis},\ }\href@noop {} {\bibfield  {journal}
  {\bibinfo  {journal} {J. Chem. Theory Comput.}\ }\textbf {\bibinfo {volume}
  {11}},\ \bibinfo {pages} {800} (\bibinfo {year} {2015})}\BibitemShut
  {NoStop}%
\bibitem [{\citenamefont {Lindorff-Larsen}\ \emph {et~al.}(2011)\citenamefont
  {Lindorff-Larsen}, \citenamefont {Piana}, \citenamefont {Dror},\ and\
  \citenamefont {Shaw}}]{lindorff2011fast}%
  \BibitemOpen
  \bibfield  {author} {\bibinfo {author} {\bibfnamefont {K.}~\bibnamefont
  {Lindorff-Larsen}}, \bibinfo {author} {\bibfnamefont {S.}~\bibnamefont
  {Piana}}, \bibinfo {author} {\bibfnamefont {R.~O.}\ \bibnamefont {Dror}},\
  and\ \bibinfo {author} {\bibfnamefont {D.~E.}\ \bibnamefont {Shaw}},\
  }\bibfield  {title} {\bibinfo {title} {How fast-folding proteins fold},\
  }\href@noop {} {\bibfield  {journal} {\bibinfo  {journal} {Science}\ }\textbf
  {\bibinfo {volume} {334}},\ \bibinfo {pages} {517} (\bibinfo {year}
  {2011})}\BibitemShut {NoStop}%
\bibitem [{\citenamefont {Lindorff-Larsen}\ \emph {et~al.}(2016)\citenamefont
  {Lindorff-Larsen}, \citenamefont {Maragakis}, \citenamefont {Piana},\ and\
  \citenamefont {Shaw}}]{lindorff2016picosecond}%
  \BibitemOpen
  \bibfield  {author} {\bibinfo {author} {\bibfnamefont {K.}~\bibnamefont
  {Lindorff-Larsen}}, \bibinfo {author} {\bibfnamefont {P.}~\bibnamefont
  {Maragakis}}, \bibinfo {author} {\bibfnamefont {S.}~\bibnamefont {Piana}},\
  and\ \bibinfo {author} {\bibfnamefont {D.~E.}\ \bibnamefont {Shaw}},\
  }\bibfield  {title} {\bibinfo {title} {Picosecond to millisecond structural
  dynamics in human ubiquitin},\ }\href@noop {} {\bibfield  {journal} {\bibinfo
   {journal} {The Journal of Physical Chemistry B}\ }\textbf {\bibinfo {volume}
  {120}},\ \bibinfo {pages} {8313} (\bibinfo {year} {2016})}\BibitemShut
  {NoStop}%
\bibitem [{\citenamefont {Shaw}\ \emph {et~al.}(2021)\citenamefont {Shaw},
  \citenamefont {Adams}, \citenamefont {Azaria}, \citenamefont {Bank},
  \citenamefont {Batson}, \citenamefont {Bell}, \citenamefont {Bergdorf},
  \citenamefont {Bhatt}, \citenamefont {Butts}, \citenamefont {Correia} \emph
  {et~al.}}]{shaw2021}%
  \BibitemOpen
  \bibfield  {author} {\bibinfo {author} {\bibfnamefont {D.~E.}\ \bibnamefont
  {Shaw}}, \bibinfo {author} {\bibfnamefont {P.~J.}\ \bibnamefont {Adams}},
  \bibinfo {author} {\bibfnamefont {A.}~\bibnamefont {Azaria}}, \bibinfo
  {author} {\bibfnamefont {J.~A.}\ \bibnamefont {Bank}}, \bibinfo {author}
  {\bibfnamefont {B.}~\bibnamefont {Batson}}, \bibinfo {author} {\bibfnamefont
  {A.}~\bibnamefont {Bell}}, \bibinfo {author} {\bibfnamefont {M.}~\bibnamefont
  {Bergdorf}}, \bibinfo {author} {\bibfnamefont {J.}~\bibnamefont {Bhatt}},
  \bibinfo {author} {\bibfnamefont {J.~A.}\ \bibnamefont {Butts}}, \bibinfo
  {author} {\bibfnamefont {T.}~\bibnamefont {Correia}}, \emph {et~al.},\
  }\bibfield  {title} {\bibinfo {title} {Anton 3: Twenty microseconds of
  molecular dynamics simulation before lunch},\ }in\ \href@noop {} {\emph
  {\bibinfo {booktitle} {Proceedings of the International Conference for High
  Performance Computing, Networking, Storage and Analysis}}}\ (\bibinfo {year}
  {2021})\ pp.\ \bibinfo {pages} {1--11}\BibitemShut {NoStop}%
\bibitem [{\citenamefont {Zimmerman}\ \emph {et~al.}(2021)\citenamefont
  {Zimmerman}, \citenamefont {Porter}, \citenamefont {Ward}, \citenamefont
  {Singh}, \citenamefont {Vithani}, \citenamefont {Meller}, \citenamefont
  {Mallimadugula}, \citenamefont {Kuhn}, \citenamefont {Borowsky},
  \citenamefont {Wiewiora} \emph {et~al.}}]{zimmerman2021sars}%
  \BibitemOpen
  \bibfield  {author} {\bibinfo {author} {\bibfnamefont {M.}~\bibnamefont
  {Zimmerman}}, \bibinfo {author} {\bibfnamefont {J.}~\bibnamefont {Porter}},
  \bibinfo {author} {\bibfnamefont {M.}~\bibnamefont {Ward}}, \bibinfo {author}
  {\bibfnamefont {S.}~\bibnamefont {Singh}}, \bibinfo {author} {\bibfnamefont
  {N.}~\bibnamefont {Vithani}}, \bibinfo {author} {\bibfnamefont
  {A.}~\bibnamefont {Meller}}, \bibinfo {author} {\bibfnamefont
  {U.}~\bibnamefont {Mallimadugula}}, \bibinfo {author} {\bibfnamefont
  {C.}~\bibnamefont {Kuhn}}, \bibinfo {author} {\bibfnamefont {J.}~\bibnamefont
  {Borowsky}}, \bibinfo {author} {\bibfnamefont {R.}~\bibnamefont {Wiewiora}},
  \emph {et~al.},\ }\bibfield  {title} {\bibinfo {title} {Sars-cov-2
  simulations go exascale to predict dramatic spike opening and cryptic pockets
  across the proteome},\ }\href@noop {} {\bibfield  {journal} {\bibinfo
  {journal} {Nat. Chem.}\ }\textbf {\bibinfo {volume} {13}},\ \bibinfo {pages}
  {651} (\bibinfo {year} {2021})}\BibitemShut {NoStop}%
\bibitem [{\citenamefont {Brotzakis}\ and\ \citenamefont
  {Parrinello}(2018)}]{brotzakis2018enhanced}%
  \BibitemOpen
  \bibfield  {author} {\bibinfo {author} {\bibfnamefont {Z.~F.}\ \bibnamefont
  {Brotzakis}}\ and\ \bibinfo {author} {\bibfnamefont {M.}~\bibnamefont
  {Parrinello}},\ }\bibfield  {title} {\bibinfo {title} {Enhanced sampling of
  protein conformational transitions via dynamically optimized collective
  variables},\ }\href@noop {} {\bibfield  {journal} {\bibinfo  {journal} {J.
  Chem. Theory Comput.}\ }\textbf {\bibinfo {volume} {15}},\ \bibinfo {pages}
  {1393} (\bibinfo {year} {2018})}\BibitemShut {NoStop}%
\bibitem [{\citenamefont {Bonati}\ \emph {et~al.}(2023)\citenamefont {Bonati},
  \citenamefont {Trizio}, \citenamefont {Rizzi},\ and\ \citenamefont
  {Parrinello}}]{bonati2023unified}%
  \BibitemOpen
  \bibfield  {author} {\bibinfo {author} {\bibfnamefont {L.}~\bibnamefont
  {Bonati}}, \bibinfo {author} {\bibfnamefont {E.}~\bibnamefont {Trizio}},
  \bibinfo {author} {\bibfnamefont {A.}~\bibnamefont {Rizzi}},\ and\ \bibinfo
  {author} {\bibfnamefont {M.}~\bibnamefont {Parrinello}},\ }\bibfield  {title}
  {\bibinfo {title} {A unified framework for machine learning collective
  variables for enhanced sampling simulations: mlcolvar},\ }\href@noop {}
  {\bibfield  {journal} {\bibinfo  {journal} {J. Chem. Phys.}\ }\textbf
  {\bibinfo {volume} {159}} (\bibinfo {year} {2023})}\BibitemShut {NoStop}%
\bibitem [{\citenamefont {Shmilovich}\ and\ \citenamefont
  {Ferguson}(2023)}]{shmilovich2023girsanov}%
  \BibitemOpen
  \bibfield  {author} {\bibinfo {author} {\bibfnamefont {K.}~\bibnamefont
  {Shmilovich}}\ and\ \bibinfo {author} {\bibfnamefont {A.~L.}\ \bibnamefont
  {Ferguson}},\ }\bibfield  {title} {\bibinfo {title} {Girsanov reweighting
  enhanced sampling technique (grest): On-the-fly data-driven discovery of and
  enhanced sampling in slow collective variables},\ }\href@noop {} {\bibfield
  {journal} {\bibinfo  {journal} {J. Phys. Chem. A}\ }\textbf {\bibinfo
  {volume} {127}},\ \bibinfo {pages} {3497} (\bibinfo {year}
  {2023})}\BibitemShut {NoStop}%
\bibitem [{\citenamefont {Rydzewski}\ \emph {et~al.}(2022)\citenamefont
  {Rydzewski}, \citenamefont {Chen}, \citenamefont {Ghosh},\ and\ \citenamefont
  {Valsson}}]{rydzewski2022reweighted}%
  \BibitemOpen
  \bibfield  {author} {\bibinfo {author} {\bibfnamefont {J.}~\bibnamefont
  {Rydzewski}}, \bibinfo {author} {\bibfnamefont {M.}~\bibnamefont {Chen}},
  \bibinfo {author} {\bibfnamefont {T.~K.}\ \bibnamefont {Ghosh}},\ and\
  \bibinfo {author} {\bibfnamefont {O.}~\bibnamefont {Valsson}},\ }\bibfield
  {title} {\bibinfo {title} {Reweighted manifold learning of collective
  variables from enhanced sampling simulations},\ }\href@noop {} {\bibfield
  {journal} {\bibinfo  {journal} {J. Chem. Theory Comput.}\ }\textbf {\bibinfo
  {volume} {18}},\ \bibinfo {pages} {7179} (\bibinfo {year}
  {2022})}\BibitemShut {NoStop}%
\bibitem [{\citenamefont {Mehdi}\ \emph {et~al.}(2024)\citenamefont {Mehdi},
  \citenamefont {Smith}, \citenamefont {Herron}, \citenamefont {Zou},\ and\
  \citenamefont {Tiwary}}]{mehdi2024enhanced}%
  \BibitemOpen
  \bibfield  {author} {\bibinfo {author} {\bibfnamefont {S.}~\bibnamefont
  {Mehdi}}, \bibinfo {author} {\bibfnamefont {Z.}~\bibnamefont {Smith}},
  \bibinfo {author} {\bibfnamefont {L.}~\bibnamefont {Herron}}, \bibinfo
  {author} {\bibfnamefont {Z.}~\bibnamefont {Zou}},\ and\ \bibinfo {author}
  {\bibfnamefont {P.}~\bibnamefont {Tiwary}},\ }\bibfield  {title} {\bibinfo
  {title} {Enhanced sampling with machine learning},\ }\href@noop {} {\bibfield
   {journal} {\bibinfo  {journal} {Ann. Rev. Phys. Chem.}\ }\textbf {\bibinfo
  {volume} {75}} (\bibinfo {year} {2024})}\BibitemShut {NoStop}%
\bibitem [{\citenamefont {Shen}\ \emph {et~al.}(2024)\citenamefont {Shen},
  \citenamefont {Wan}, \citenamefont {Li}, \citenamefont {Gao},\ and\
  \citenamefont {Shi}}]{shen2024adaptive}%
  \BibitemOpen
  \bibfield  {author} {\bibinfo {author} {\bibfnamefont {W.}~\bibnamefont
  {Shen}}, \bibinfo {author} {\bibfnamefont {K.}~\bibnamefont {Wan}}, \bibinfo
  {author} {\bibfnamefont {D.}~\bibnamefont {Li}}, \bibinfo {author}
  {\bibfnamefont {H.}~\bibnamefont {Gao}},\ and\ \bibinfo {author}
  {\bibfnamefont {X.}~\bibnamefont {Shi}},\ }\bibfield  {title} {\bibinfo
  {title} {Adaptive cvgen: Leveraging reinforcement learning for advanced
  sampling in protein folding and chemical reactions},\ }\href@noop {}
  {\bibfield  {journal} {\bibinfo  {journal} {Proc. Natl. Acad. Sci. U.S.A}\
  }\textbf {\bibinfo {volume} {121}},\ \bibinfo {pages} {e2414205121} (\bibinfo
  {year} {2024})}\BibitemShut {NoStop}%
\bibitem [{\citenamefont {Sauer}\ and\ \citenamefont
  {Heyden}(2023)}]{sauer2023frequency}%
  \BibitemOpen
  \bibfield  {author} {\bibinfo {author} {\bibfnamefont {M.~A.}\ \bibnamefont
  {Sauer}}\ and\ \bibinfo {author} {\bibfnamefont {M.}~\bibnamefont {Heyden}},\
  }\bibfield  {title} {\bibinfo {title} {Frequency-selective anharmonic mode
  analysis of thermally excited vibrations in proteins},\ }\href@noop {}
  {\bibfield  {journal} {\bibinfo  {journal} {J. Chem. Theory Comput.}\
  }\textbf {\bibinfo {volume} {19}},\ \bibinfo {pages} {5481} (\bibinfo {year}
  {2023})}\BibitemShut {NoStop}%
\bibitem [{\citenamefont {Brooks}\ and\ \citenamefont
  {Karplus}(1983)}]{brooks1983harmonic}%
  \BibitemOpen
  \bibfield  {author} {\bibinfo {author} {\bibfnamefont {B.}~\bibnamefont
  {Brooks}}\ and\ \bibinfo {author} {\bibfnamefont {M.}~\bibnamefont
  {Karplus}},\ }\bibfield  {title} {\bibinfo {title} {Harmonic dynamics of
  proteins: normal modes and fluctuations in bovine pancreatic trypsin
  inhibitor.},\ }\href@noop {} {\bibfield  {journal} {\bibinfo  {journal}
  {Proc. Natl. Acad. Sci. U.S.A}\ }\textbf {\bibinfo {volume} {80}},\ \bibinfo
  {pages} {6571} (\bibinfo {year} {1983})}\BibitemShut {NoStop}%
\bibitem [{\citenamefont {Yang}\ \emph {et~al.}(2007)\citenamefont {Yang},
  \citenamefont {Song},\ and\ \citenamefont {Jernigan}}]{yang2007well}%
  \BibitemOpen
  \bibfield  {author} {\bibinfo {author} {\bibfnamefont {L.}~\bibnamefont
  {Yang}}, \bibinfo {author} {\bibfnamefont {G.}~\bibnamefont {Song}},\ and\
  \bibinfo {author} {\bibfnamefont {R.~L.}\ \bibnamefont {Jernigan}},\
  }\bibfield  {title} {\bibinfo {title} {How well can we understand large-scale
  protein motions using normal modes of elastic network models?},\ }\href@noop
  {} {\bibfield  {journal} {\bibinfo  {journal} {Biophys. J.}\ }\textbf
  {\bibinfo {volume} {93}},\ \bibinfo {pages} {920} (\bibinfo {year}
  {2007})}\BibitemShut {NoStop}%
\bibitem [{\citenamefont {Mahajan}\ and\ \citenamefont
  {Sanejouand}(2017)}]{mahajan2017jumping}%
  \BibitemOpen
  \bibfield  {author} {\bibinfo {author} {\bibfnamefont {S.}~\bibnamefont
  {Mahajan}}\ and\ \bibinfo {author} {\bibfnamefont {Y.-H.}\ \bibnamefont
  {Sanejouand}},\ }\bibfield  {title} {\bibinfo {title} {Jumping between
  protein conformers using normal modes},\ }\href@noop {} {\bibfield  {journal}
  {\bibinfo  {journal} {J. Comput. Chem.}\ }\textbf {\bibinfo {volume} {38}},\
  \bibinfo {pages} {1622} (\bibinfo {year} {2017})}\BibitemShut {NoStop}%
\bibitem [{\citenamefont {Costa}\ \emph {et~al.}(2023)\citenamefont {Costa},
  \citenamefont {Batista}, \citenamefont {Gomes}, \citenamefont {Bastos},
  \citenamefont {Louet}, \citenamefont {Floquet}, \citenamefont {Bisch},\ and\
  \citenamefont {Perahia}}]{costa2023}%
  \BibitemOpen
  \bibfield  {author} {\bibinfo {author} {\bibfnamefont {M.~G.}\ \bibnamefont
  {Costa}}, \bibinfo {author} {\bibfnamefont {P.~R.}\ \bibnamefont {Batista}},
  \bibinfo {author} {\bibfnamefont {A.}~\bibnamefont {Gomes}}, \bibinfo
  {author} {\bibfnamefont {L.~S.}\ \bibnamefont {Bastos}}, \bibinfo {author}
  {\bibfnamefont {M.}~\bibnamefont {Louet}}, \bibinfo {author} {\bibfnamefont
  {N.}~\bibnamefont {Floquet}}, \bibinfo {author} {\bibfnamefont {P.~M.}\
  \bibnamefont {Bisch}},\ and\ \bibinfo {author} {\bibfnamefont
  {D.}~\bibnamefont {Perahia}},\ }\bibfield  {title} {\bibinfo {title}
  {Mdexciter: Enhanced sampling molecular dynamics by excited normal modes or
  principal components obtained from experiments},\ }\href@noop {} {\bibfield
  {journal} {\bibinfo  {journal} {J. Chem. Theory Comput.}\ }\textbf {\bibinfo
  {volume} {19}},\ \bibinfo {pages} {412} (\bibinfo {year} {2023})}\BibitemShut
  {NoStop}%
\bibitem [{\citenamefont {Mondal}\ \emph {et~al.}(2024)\citenamefont {Mondal},
  \citenamefont {Sauer},\ and\ \citenamefont {Heyden}}]{mondal2024exploring}%
  \BibitemOpen
  \bibfield  {author} {\bibinfo {author} {\bibfnamefont {S.}~\bibnamefont
  {Mondal}}, \bibinfo {author} {\bibfnamefont {M.~A.}\ \bibnamefont {Sauer}},\
  and\ \bibinfo {author} {\bibfnamefont {M.}~\bibnamefont {Heyden}},\
  }\bibfield  {title} {\bibinfo {title} {Exploring conformational landscapes
  along anharmonic low-frequency vibrations},\ }\href@noop {} {\bibfield
  {journal} {\bibinfo  {journal} {J. Phys. Chem. B}\ }\textbf {\bibinfo
  {volume} {128}},\ \bibinfo {pages} {7112} (\bibinfo {year}
  {2024})}\BibitemShut {NoStop}%
\bibitem [{\citenamefont {Huang}\ \emph {et~al.}(2018)\citenamefont {Huang},
  \citenamefont {McCammon},\ and\ \citenamefont {Miao}}]{huang2018replica}%
  \BibitemOpen
  \bibfield  {author} {\bibinfo {author} {\bibfnamefont {Y.~M.}\ \bibnamefont
  {Huang}}, \bibinfo {author} {\bibfnamefont {J.~A.}\ \bibnamefont
  {McCammon}},\ and\ \bibinfo {author} {\bibfnamefont {Y.}~\bibnamefont
  {Miao}},\ }\bibfield  {title} {\bibinfo {title} {Replica exchange gaussian
  accelerated molecular dynamics: Improved enhanced sampling and free energy
  calculation},\ }\href@noop {} {\bibfield  {journal} {\bibinfo  {journal} {J.
  Chem. Theory Comput.}\ }\textbf {\bibinfo {volume} {14}},\ \bibinfo {pages}
  {1853} (\bibinfo {year} {2018})}\BibitemShut {NoStop}%
\bibitem [{\citenamefont {Benabderrahmane}\ \emph {et~al.}(2020)\citenamefont
  {Benabderrahmane}, \citenamefont {Bureau}, \citenamefont {Voisin-Chiret},\
  and\ \citenamefont {Sopkova-de
  Oliveira~Santos}}]{benabderrahmane2020insights}%
  \BibitemOpen
  \bibfield  {author} {\bibinfo {author} {\bibfnamefont {M.}~\bibnamefont
  {Benabderrahmane}}, \bibinfo {author} {\bibfnamefont {R.}~\bibnamefont
  {Bureau}}, \bibinfo {author} {\bibfnamefont {A.~S.}\ \bibnamefont
  {Voisin-Chiret}},\ and\ \bibinfo {author} {\bibfnamefont {J.}~\bibnamefont
  {Sopkova-de Oliveira~Santos}},\ }\bibfield  {title} {\bibinfo {title}
  {Insights into mcl-1 conformational states and allosteric inhibition
  mechanism from molecular dynamics simulations, enhanced sampling, and pocket
  crosstalk analysis},\ }\href@noop {} {\bibfield  {journal} {\bibinfo
  {journal} {J. Chem. Inf. Model.}\ }\textbf {\bibinfo {volume} {60}},\
  \bibinfo {pages} {3172} (\bibinfo {year} {2020})}\BibitemShut {NoStop}%
\bibitem [{\citenamefont {Ren}\ \emph {et~al.}(2021)\citenamefont {Ren},
  \citenamefont {Dokainish}, \citenamefont {Shinobu}, \citenamefont {Oshima},\
  and\ \citenamefont {Sugita}}]{ren2021unraveling}%
  \BibitemOpen
  \bibfield  {author} {\bibinfo {author} {\bibfnamefont {W.}~\bibnamefont
  {Ren}}, \bibinfo {author} {\bibfnamefont {H.~M.}\ \bibnamefont {Dokainish}},
  \bibinfo {author} {\bibfnamefont {A.}~\bibnamefont {Shinobu}}, \bibinfo
  {author} {\bibfnamefont {H.}~\bibnamefont {Oshima}},\ and\ \bibinfo {author}
  {\bibfnamefont {Y.}~\bibnamefont {Sugita}},\ }\bibfield  {title} {\bibinfo
  {title} {Unraveling the coupling between conformational changes and ligand
  binding in ribose binding protein using multiscale molecular dynamics and
  free-energy calculations},\ }\href@noop {} {\bibfield  {journal} {\bibinfo
  {journal} {J. Phys. Chem. B}\ }\textbf {\bibinfo {volume} {125}},\ \bibinfo
  {pages} {2898} (\bibinfo {year} {2021})}\BibitemShut {NoStop}%
\bibitem [{\citenamefont {Chen}\ \emph {et~al.}(2021)\citenamefont {Chen},
  \citenamefont {Zhang}, \citenamefont {Wang}, \citenamefont {Pang},
  \citenamefont {Zhang},\ and\ \citenamefont {Liu}}]{chen2021mutation}%
  \BibitemOpen
  \bibfield  {author} {\bibinfo {author} {\bibfnamefont {J.}~\bibnamefont
  {Chen}}, \bibinfo {author} {\bibfnamefont {S.}~\bibnamefont {Zhang}},
  \bibinfo {author} {\bibfnamefont {W.}~\bibnamefont {Wang}}, \bibinfo {author}
  {\bibfnamefont {L.}~\bibnamefont {Pang}}, \bibinfo {author} {\bibfnamefont
  {Q.}~\bibnamefont {Zhang}},\ and\ \bibinfo {author} {\bibfnamefont
  {X.}~\bibnamefont {Liu}},\ }\bibfield  {title} {\bibinfo {title}
  {Mutation-induced impacts on the switch transformations of the gdp-and
  gtp-bound k-ras: insights from multiple replica gaussian accelerated
  molecular dynamics and free energy analysis},\ }\href@noop {} {\bibfield
  {journal} {\bibinfo  {journal} {J. Chem. Inf. Model.}\ }\textbf {\bibinfo
  {volume} {61}},\ \bibinfo {pages} {1954} (\bibinfo {year}
  {2021})}\BibitemShut {NoStop}%
\bibitem [{\citenamefont {De~Simone}\ \emph {et~al.}(2013)\citenamefont
  {De~Simone}, \citenamefont {Montalvao}, \citenamefont {Dobson},\ and\
  \citenamefont {Vendruscolo}}]{desimone2013characterization}%
  \BibitemOpen
  \bibfield  {author} {\bibinfo {author} {\bibfnamefont {A.}~\bibnamefont
  {De~Simone}}, \bibinfo {author} {\bibfnamefont {R.~W.}\ \bibnamefont
  {Montalvao}}, \bibinfo {author} {\bibfnamefont {C.~M.}\ \bibnamefont
  {Dobson}},\ and\ \bibinfo {author} {\bibfnamefont {M.}~\bibnamefont
  {Vendruscolo}},\ }\bibfield  {title} {\bibinfo {title} {Characterization of
  the interdomain motions in hen lysozyme using residual dipolar couplings as
  replica-averaged structural restraints in molecular dynamics simulations},\
  }\href@noop {} {\bibfield  {journal} {\bibinfo  {journal} {Biochemistry}\
  }\textbf {\bibinfo {volume} {52}},\ \bibinfo {pages} {6480} (\bibinfo {year}
  {2013})}\BibitemShut {NoStop}%
\bibitem [{\citenamefont {Branduardi}\ \emph {et~al.}(2012)\citenamefont
  {Branduardi}, \citenamefont {Bussi},\ and\ \citenamefont
  {Parrinello}}]{branduardi2012}%
  \BibitemOpen
  \bibfield  {author} {\bibinfo {author} {\bibfnamefont {D.}~\bibnamefont
  {Branduardi}}, \bibinfo {author} {\bibfnamefont {G.}~\bibnamefont {Bussi}},\
  and\ \bibinfo {author} {\bibfnamefont {M.}~\bibnamefont {Parrinello}},\
  }\bibfield  {title} {\bibinfo {title} {Metadynamics with adaptive
  gaussians},\ }\href@noop {} {\bibfield  {journal} {\bibinfo  {journal} {J.
  Chem. Theory Comput.}\ }\textbf {\bibinfo {volume} {8}},\ \bibinfo {pages}
  {2247} (\bibinfo {year} {2012})}\BibitemShut {NoStop}%
\bibitem [{\citenamefont {Levitt}\ \emph {et~al.}(1985)\citenamefont {Levitt},
  \citenamefont {Sander},\ and\ \citenamefont {Stern}}]{levitt1985protein}%
  \BibitemOpen
  \bibfield  {author} {\bibinfo {author} {\bibfnamefont {M.}~\bibnamefont
  {Levitt}}, \bibinfo {author} {\bibfnamefont {C.}~\bibnamefont {Sander}},\
  and\ \bibinfo {author} {\bibfnamefont {P.~S.}\ \bibnamefont {Stern}},\
  }\bibfield  {title} {\bibinfo {title} {Protein normal-mode dynamics: trypsin
  inhibitor, crambin, ribonuclease and lysozyme},\ }\href@noop {} {\bibfield
  {journal} {\bibinfo  {journal} {J. Mol. Biol.}\ }\textbf {\bibinfo {volume}
  {181}},\ \bibinfo {pages} {423} (\bibinfo {year} {1985})}\BibitemShut
  {NoStop}%
\bibitem [{\citenamefont {Chennubhotla}\ \emph {et~al.}(2005)\citenamefont
  {Chennubhotla}, \citenamefont {Rader}, \citenamefont {Yang},\ and\
  \citenamefont {Bahar}}]{chennubhotla2005elastic}%
  \BibitemOpen
  \bibfield  {author} {\bibinfo {author} {\bibfnamefont {C.}~\bibnamefont
  {Chennubhotla}}, \bibinfo {author} {\bibfnamefont {A.}~\bibnamefont {Rader}},
  \bibinfo {author} {\bibfnamefont {L.-W.}\ \bibnamefont {Yang}},\ and\
  \bibinfo {author} {\bibfnamefont {I.}~\bibnamefont {Bahar}},\ }\bibfield
  {title} {\bibinfo {title} {Elastic network models for understanding
  biomolecular machinery: from enzymes to supramolecular assemblies},\
  }\href@noop {} {\bibfield  {journal} {\bibinfo  {journal} {Phys. Biol.}\
  }\textbf {\bibinfo {volume} {2}},\ \bibinfo {pages} {S173} (\bibinfo {year}
  {2005})}\BibitemShut {NoStop}%
\bibitem [{\citenamefont {Consortium}(2019)}]{uniprot2019uniprot}%
  \BibitemOpen
  \bibfield  {author} {\bibinfo {author} {\bibfnamefont {U.}~\bibnamefont
  {Consortium}},\ }\bibfield  {title} {\bibinfo {title} {Uniprot: a worldwide
  hub of protein knowledge},\ }\href@noop {} {\bibfield  {journal} {\bibinfo
  {journal} {Nucleic Acids Res.}\ }\textbf {\bibinfo {volume} {47}},\ \bibinfo
  {pages} {D506} (\bibinfo {year} {2019})}\BibitemShut {NoStop}%
\bibitem [{\citenamefont {Janson}\ \emph {et~al.}(2023)\citenamefont {Janson},
  \citenamefont {Valdes-Garcia}, \citenamefont {Heo},\ and\ \citenamefont
  {Feig}}]{janson2023direct}%
  \BibitemOpen
  \bibfield  {author} {\bibinfo {author} {\bibfnamefont {G.}~\bibnamefont
  {Janson}}, \bibinfo {author} {\bibfnamefont {G.}~\bibnamefont
  {Valdes-Garcia}}, \bibinfo {author} {\bibfnamefont {L.}~\bibnamefont {Heo}},\
  and\ \bibinfo {author} {\bibfnamefont {M.}~\bibnamefont {Feig}},\ }\bibfield
  {title} {\bibinfo {title} {Direct generation of protein conformational
  ensembles via machine learning},\ }\href@noop {} {\bibfield  {journal}
  {\bibinfo  {journal} {Nat. Commun.}\ }\textbf {\bibinfo {volume} {14}},\
  \bibinfo {pages} {774} (\bibinfo {year} {2023})}\BibitemShut {NoStop}%
\bibitem [{\citenamefont {Tesei}\ \emph {et~al.}(2024)\citenamefont {Tesei},
  \citenamefont {Trolle}, \citenamefont {Jonsson}, \citenamefont {Betz},
  \citenamefont {Knudsen}, \citenamefont {Pesce}, \citenamefont {Johansson},\
  and\ \citenamefont {Lindorff-Larsen}}]{tesei2024conformational}%
  \BibitemOpen
  \bibfield  {author} {\bibinfo {author} {\bibfnamefont {G.}~\bibnamefont
  {Tesei}}, \bibinfo {author} {\bibfnamefont {A.~I.}\ \bibnamefont {Trolle}},
  \bibinfo {author} {\bibfnamefont {N.}~\bibnamefont {Jonsson}}, \bibinfo
  {author} {\bibfnamefont {J.}~\bibnamefont {Betz}}, \bibinfo {author}
  {\bibfnamefont {F.~E.}\ \bibnamefont {Knudsen}}, \bibinfo {author}
  {\bibfnamefont {F.}~\bibnamefont {Pesce}}, \bibinfo {author} {\bibfnamefont
  {K.~E.}\ \bibnamefont {Johansson}},\ and\ \bibinfo {author} {\bibfnamefont
  {K.}~\bibnamefont {Lindorff-Larsen}},\ }\bibfield  {title} {\bibinfo {title}
  {Conformational ensembles of the human intrinsically disordered proteome},\
  }\href@noop {} {\bibfield  {journal} {\bibinfo  {journal} {Nature}\ }\textbf
  {\bibinfo {volume} {626}},\ \bibinfo {pages} {897} (\bibinfo {year}
  {2024})}\BibitemShut {NoStop}%
\bibitem [{\citenamefont {Lotthammer}\ \emph {et~al.}(2024)\citenamefont
  {Lotthammer}, \citenamefont {Ginell}, \citenamefont {Griffith}, \citenamefont
  {Emenecker},\ and\ \citenamefont {Holehouse}}]{lotthammer2024direct}%
  \BibitemOpen
  \bibfield  {author} {\bibinfo {author} {\bibfnamefont {J.~M.}\ \bibnamefont
  {Lotthammer}}, \bibinfo {author} {\bibfnamefont {G.~M.}\ \bibnamefont
  {Ginell}}, \bibinfo {author} {\bibfnamefont {D.}~\bibnamefont {Griffith}},
  \bibinfo {author} {\bibfnamefont {R.}~\bibnamefont {Emenecker}},\ and\
  \bibinfo {author} {\bibfnamefont {A.~S.}\ \bibnamefont {Holehouse}},\
  }\bibfield  {title} {\bibinfo {title} {Direct prediction of intrinsically
  disordered protein conformational properties from sequence},\ }\href@noop {}
  {\bibfield  {journal} {\bibinfo  {journal} {Nat. Methods}\ }\textbf {\bibinfo
  {volume} {21}},\ \bibinfo {pages} {465} (\bibinfo {year} {2024})}\BibitemShut
  {NoStop}%
\bibitem [{\citenamefont {Jennewein}\ \emph {et~al.}(2023)\citenamefont
  {Jennewein}, \citenamefont {Lee}, \citenamefont {Kurtz}, \citenamefont
  {Dizon}, \citenamefont {Shaeffer}, \citenamefont {Chapman}, \citenamefont
  {Chiquete}, \citenamefont {Burks}, \citenamefont {Carlson}, \citenamefont
  {Mason}, \citenamefont {Kobwala}, \citenamefont {Jagadeesan}, \citenamefont
  {Barghav}, \citenamefont {Battelle}, \citenamefont {Belshe}, \citenamefont
  {McCaffrey}, \citenamefont {Brazil}, \citenamefont {Inumella}, \citenamefont
  {Kuznia}, \citenamefont {Buzinski}, \citenamefont {Dudley}, \citenamefont
  {Shah}, \citenamefont {Speyer},\ and\ \citenamefont {Yalim}}]{sol}%
  \BibitemOpen
  \bibfield  {author} {\bibinfo {author} {\bibfnamefont {D.~M.}\ \bibnamefont
  {Jennewein}}, \bibinfo {author} {\bibfnamefont {J.}~\bibnamefont {Lee}},
  \bibinfo {author} {\bibfnamefont {C.}~\bibnamefont {Kurtz}}, \bibinfo
  {author} {\bibfnamefont {W.}~\bibnamefont {Dizon}}, \bibinfo {author}
  {\bibfnamefont {I.}~\bibnamefont {Shaeffer}}, \bibinfo {author}
  {\bibfnamefont {A.}~\bibnamefont {Chapman}}, \bibinfo {author} {\bibfnamefont
  {A.}~\bibnamefont {Chiquete}}, \bibinfo {author} {\bibfnamefont
  {J.}~\bibnamefont {Burks}}, \bibinfo {author} {\bibfnamefont
  {A.}~\bibnamefont {Carlson}}, \bibinfo {author} {\bibfnamefont
  {N.}~\bibnamefont {Mason}}, \bibinfo {author} {\bibfnamefont
  {A.}~\bibnamefont {Kobwala}}, \bibinfo {author} {\bibfnamefont
  {T.}~\bibnamefont {Jagadeesan}}, \bibinfo {author} {\bibfnamefont
  {P.}~\bibnamefont {Barghav}}, \bibinfo {author} {\bibfnamefont
  {T.}~\bibnamefont {Battelle}}, \bibinfo {author} {\bibfnamefont
  {R.}~\bibnamefont {Belshe}}, \bibinfo {author} {\bibfnamefont
  {D.}~\bibnamefont {McCaffrey}}, \bibinfo {author} {\bibfnamefont
  {M.}~\bibnamefont {Brazil}}, \bibinfo {author} {\bibfnamefont
  {C.}~\bibnamefont {Inumella}}, \bibinfo {author} {\bibfnamefont
  {K.}~\bibnamefont {Kuznia}}, \bibinfo {author} {\bibfnamefont
  {J.}~\bibnamefont {Buzinski}}, \bibinfo {author} {\bibfnamefont
  {S.}~\bibnamefont {Dudley}}, \bibinfo {author} {\bibfnamefont
  {D.}~\bibnamefont {Shah}}, \bibinfo {author} {\bibfnamefont {G.}~\bibnamefont
  {Speyer}},\ and\ \bibinfo {author} {\bibfnamefont {J.}~\bibnamefont
  {Yalim}},\ }\bibfield  {title} {\bibinfo {title} {{The Sol Supercomputer at
  Arizona State University}},\ }in\ \href
  {https://doi.org/10.1145/3569951.3597573} {\emph {\bibinfo {booktitle}
  {Practice and Experience in Advanced Research Computing}}},\ \bibinfo {series
  and number} {PEARC '23}\ (\bibinfo  {publisher} {Association for Computing
  Machinery},\ \bibinfo {address} {New York, NY, USA},\ \bibinfo {year}
  {2023})\ pp.\ \bibinfo {pages} {296--301}\BibitemShut {NoStop}%
\bibitem [{\citenamefont {Abraham}\ \emph {et~al.}(2015)\citenamefont
  {Abraham}, \citenamefont {Murtola}, \citenamefont {Schulz}, \citenamefont
  {P{\'a}ll}, \citenamefont {Smith}, \citenamefont {Hess},\ and\ \citenamefont
  {Lindahl}}]{abraham15}%
  \BibitemOpen
  \bibfield  {author} {\bibinfo {author} {\bibfnamefont {M.~J.}\ \bibnamefont
  {Abraham}}, \bibinfo {author} {\bibfnamefont {T.}~\bibnamefont {Murtola}},
  \bibinfo {author} {\bibfnamefont {R.}~\bibnamefont {Schulz}}, \bibinfo
  {author} {\bibfnamefont {S.}~\bibnamefont {P{\'a}ll}}, \bibinfo {author}
  {\bibfnamefont {J.~C.}\ \bibnamefont {Smith}}, \bibinfo {author}
  {\bibfnamefont {B.}~\bibnamefont {Hess}},\ and\ \bibinfo {author}
  {\bibfnamefont {E.}~\bibnamefont {Lindahl}},\ }\bibfield  {title} {\bibinfo
  {title} {Gromacs: High performance molecular simulations through multi-level
  parallelism from laptops to supercomputers},\ }\href@noop {} {\bibfield
  {journal} {\bibinfo  {journal} {SoftwareX}\ }\textbf {\bibinfo {volume}
  {1}},\ \bibinfo {pages} {19} (\bibinfo {year} {2015})}\BibitemShut {NoStop}%
\bibitem [{\citenamefont {Jorgensen}\ \emph {et~al.}(1983)\citenamefont
  {Jorgensen}, \citenamefont {Chandrasekhar}, \citenamefont {Madura},
  \citenamefont {Impey},\ and\ \citenamefont {Klein}}]{jorgensen83}%
  \BibitemOpen
  \bibfield  {author} {\bibinfo {author} {\bibfnamefont {W.-L.}\ \bibnamefont
  {Jorgensen}}, \bibinfo {author} {\bibfnamefont {J.}~\bibnamefont
  {Chandrasekhar}}, \bibinfo {author} {\bibfnamefont {J.-D.}\ \bibnamefont
  {Madura}}, \bibinfo {author} {\bibfnamefont {R.-W.}\ \bibnamefont {Impey}},\
  and\ \bibinfo {author} {\bibfnamefont {M.-L.}\ \bibnamefont {Klein}},\
  }\bibfield  {title} {\bibinfo {title} {Comparison of simple potential
  functions for simulating liquid water},\ }\href@noop {} {\bibfield  {journal}
  {\bibinfo  {journal} {J. Chem. Phys.}\ }\textbf {\bibinfo {volume} {79}},\
  \bibinfo {pages} {926} (\bibinfo {year} {1983})}\BibitemShut {NoStop}%
\bibitem [{\citenamefont {MacKerell~Jr}\ \emph {et~al.}(1998)\citenamefont
  {MacKerell~Jr}, \citenamefont {Bashford}, \citenamefont {Bellott},
  \citenamefont {Dunbrack~Jr}, \citenamefont {Evanseck}, \citenamefont {Field},
  \citenamefont {Fischer}, \citenamefont {Gao}, \citenamefont {Guo},
  \citenamefont {Ha}, \citenamefont {Joseph-McCarthy}, \citenamefont {Kuchnir},
  \citenamefont {Kuczera}, \citenamefont {Lau}, \citenamefont {Mattos},
  \citenamefont {Michnick}, \citenamefont {Ngo}, \citenamefont {Nguyen},
  \citenamefont {Prodhom}, \citenamefont {Reiher}, \citenamefont {Roux},
  \citenamefont {Schlenkrich}, \citenamefont {Smith}, \citenamefont {Stote},
  \citenamefont {Straub}, \citenamefont {Watanabe}, \citenamefont
  {Wi{\'o}rkiewicz-Kuczera}, \citenamefont {Yin},\ and\ \citenamefont
  {Karplus}}]{mackerell98}%
  \BibitemOpen
  \bibfield  {author} {\bibinfo {author} {\bibfnamefont {A.~D.}\ \bibnamefont
  {MacKerell~Jr}}, \bibinfo {author} {\bibfnamefont {D.}~\bibnamefont
  {Bashford}}, \bibinfo {author} {\bibfnamefont {M.~L. D.~R.}\ \bibnamefont
  {Bellott}}, \bibinfo {author} {\bibfnamefont {R.~L.}\ \bibnamefont
  {Dunbrack~Jr}}, \bibinfo {author} {\bibfnamefont {J.~D.}\ \bibnamefont
  {Evanseck}}, \bibinfo {author} {\bibfnamefont {M.~J.}\ \bibnamefont {Field}},
  \bibinfo {author} {\bibfnamefont {S.}~\bibnamefont {Fischer}}, \bibinfo
  {author} {\bibfnamefont {J.}~\bibnamefont {Gao}}, \bibinfo {author}
  {\bibfnamefont {H.}~\bibnamefont {Guo}}, \bibinfo {author} {\bibfnamefont
  {S.}~\bibnamefont {Ha}}, \bibinfo {author} {\bibfnamefont {D.}~\bibnamefont
  {Joseph-McCarthy}}, \bibinfo {author} {\bibfnamefont {L.}~\bibnamefont
  {Kuchnir}}, \bibinfo {author} {\bibfnamefont {K.}~\bibnamefont {Kuczera}},
  \bibinfo {author} {\bibfnamefont {F.~T.~K.}\ \bibnamefont {Lau}}, \bibinfo
  {author} {\bibfnamefont {C.}~\bibnamefont {Mattos}}, \bibinfo {author}
  {\bibfnamefont {S.}~\bibnamefont {Michnick}}, \bibinfo {author}
  {\bibfnamefont {T.}~\bibnamefont {Ngo}}, \bibinfo {author} {\bibfnamefont
  {D.~T.}\ \bibnamefont {Nguyen}}, \bibinfo {author} {\bibfnamefont
  {B.}~\bibnamefont {Prodhom}}, \bibinfo {author} {\bibfnamefont
  {I.}~\bibnamefont {Reiher}, \bibfnamefont {W.~E.}}, \bibinfo {author}
  {\bibfnamefont {B.}~\bibnamefont {Roux}}, \bibinfo {author} {\bibfnamefont
  {M.}~\bibnamefont {Schlenkrich}}, \bibinfo {author} {\bibfnamefont {J.~C.}\
  \bibnamefont {Smith}}, \bibinfo {author} {\bibfnamefont {R.}~\bibnamefont
  {Stote}}, \bibinfo {author} {\bibfnamefont {J.}~\bibnamefont {Straub}},
  \bibinfo {author} {\bibfnamefont {M.}~\bibnamefont {Watanabe}}, \bibinfo
  {author} {\bibfnamefont {J.}~\bibnamefont {Wi{\'o}rkiewicz-Kuczera}},
  \bibinfo {author} {\bibfnamefont {D.}~\bibnamefont {Yin}},\ and\ \bibinfo
  {author} {\bibfnamefont {M.}~\bibnamefont {Karplus}},\ }\bibfield  {title}
  {\bibinfo {title} {All-atom empirical potential for molecular modeling and
  dynamics studies of proteins},\ }\href@noop {} {\bibfield  {journal}
  {\bibinfo  {journal} {J. Phys. Chem. B}\ }\textbf {\bibinfo {volume} {102}},\
  \bibinfo {pages} {3586} (\bibinfo {year} {1998})}\BibitemShut {NoStop}%
\bibitem [{\citenamefont {Hornak}\ \emph {et~al.}(2006)\citenamefont {Hornak},
  \citenamefont {Abel}, \citenamefont {Okur}, \citenamefont {Strockbine},
  \citenamefont {Roitberg},\ and\ \citenamefont
  {Simmerling}}]{hornak2006amber99sb}%
  \BibitemOpen
  \bibfield  {author} {\bibinfo {author} {\bibfnamefont {V.}~\bibnamefont
  {Hornak}}, \bibinfo {author} {\bibfnamefont {R.}~\bibnamefont {Abel}},
  \bibinfo {author} {\bibfnamefont {A.}~\bibnamefont {Okur}}, \bibinfo {author}
  {\bibfnamefont {B.}~\bibnamefont {Strockbine}}, \bibinfo {author}
  {\bibfnamefont {A.}~\bibnamefont {Roitberg}},\ and\ \bibinfo {author}
  {\bibfnamefont {C.}~\bibnamefont {Simmerling}},\ }\bibfield  {title}
  {\bibinfo {title} {Comparison of multiple amber force fields and development
  of improved protein backbone parameters},\ }\href@noop {} {\bibfield
  {journal} {\bibinfo  {journal} {Proteins Struct. Funct. Bioinf.}\ }\textbf
  {\bibinfo {volume} {65}},\ \bibinfo {pages} {712} (\bibinfo {year}
  {2006})}\BibitemShut {NoStop}%
\bibitem [{\citenamefont {Maier}\ \emph {et~al.}(2015)\citenamefont {Maier},
  \citenamefont {Martinez}, \citenamefont {Kasavajhala}, \citenamefont
  {Wickstrom}, \citenamefont {Hauser},\ and\ \citenamefont
  {Simmerling}}]{maier15amber14sb}%
  \BibitemOpen
  \bibfield  {author} {\bibinfo {author} {\bibfnamefont {J.~A.}\ \bibnamefont
  {Maier}}, \bibinfo {author} {\bibfnamefont {C.}~\bibnamefont {Martinez}},
  \bibinfo {author} {\bibfnamefont {K.}~\bibnamefont {Kasavajhala}}, \bibinfo
  {author} {\bibfnamefont {L.}~\bibnamefont {Wickstrom}}, \bibinfo {author}
  {\bibfnamefont {K.~E.}\ \bibnamefont {Hauser}},\ and\ \bibinfo {author}
  {\bibfnamefont {C.}~\bibnamefont {Simmerling}},\ }\bibfield  {title}
  {\bibinfo {title} {ff14sb: Improving the accuracy of protein side chain and
  backbone parameters from ff99sb},\ }\href
  {https://doi.org/10.1021/acs.jctc.5b00255} {\bibfield  {journal} {\bibinfo
  {journal} {J. Chem. Theory Comput.}\ }\textbf {\bibinfo {volume} {11}},\
  \bibinfo {pages} {3696} (\bibinfo {year} {2015})},\ \bibinfo {note} {pMID:
  26574453}\BibitemShut {NoStop}%
\bibitem [{\citenamefont {Huang}\ \emph {et~al.}(2017)\citenamefont {Huang},
  \citenamefont {Rauscher}, \citenamefont {Nawrocki}, \citenamefont {Ting~Ran},
  \citenamefont {de~Groot}, \citenamefont {Grubm{\"u}ller},\ and\ \citenamefont
  {Jr}}]{charmm36m}%
  \BibitemOpen
  \bibfield  {author} {\bibinfo {author} {\bibfnamefont {J.}~\bibnamefont
  {Huang}}, \bibinfo {author} {\bibfnamefont {S.}~\bibnamefont {Rauscher}},
  \bibinfo {author} {\bibfnamefont {G.}~\bibnamefont {Nawrocki}}, \bibinfo
  {author} {\bibfnamefont {M.~F.}\ \bibnamefont {Ting~Ran}}, \bibinfo {author}
  {\bibfnamefont {B.~L.}\ \bibnamefont {de~Groot}}, \bibinfo {author}
  {\bibfnamefont {H.}~\bibnamefont {Grubm{\"u}ller}},\ and\ \bibinfo {author}
  {\bibfnamefont {A.~D.~M.}\ \bibnamefont {Jr}},\ }\bibfield  {title} {\bibinfo
  {title} {Charmm36m: an improved force field for folded and intrinsically
  disordered proteins},\ }\href@noop {} {\bibfield  {journal} {\bibinfo
  {journal} {Nat. Methods}\ }\textbf {\bibinfo {volume} {14}},\ \bibinfo
  {pages} {71} (\bibinfo {year} {2017})}\BibitemShut {NoStop}%
\bibitem [{\citenamefont {Bussi}\ \emph {et~al.}(2007)\citenamefont {Bussi},
  \citenamefont {Donadio},\ and\ \citenamefont {Parrinello}}]{bussi07}%
  \BibitemOpen
  \bibfield  {author} {\bibinfo {author} {\bibfnamefont {G.}~\bibnamefont
  {Bussi}}, \bibinfo {author} {\bibfnamefont {D.}~\bibnamefont {Donadio}},\
  and\ \bibinfo {author} {\bibfnamefont {M.}~\bibnamefont {Parrinello}},\
  }\bibfield  {title} {\bibinfo {title} {Canonical sampling through velocity
  rescaling},\ }\href@noop {} {\bibfield  {journal} {\bibinfo  {journal} {J.
  Chem. Phys.}\ }\textbf {\bibinfo {volume} {126}},\ \bibinfo {pages} {014101}
  (\bibinfo {year} {2007})}\BibitemShut {NoStop}%
\bibitem [{\citenamefont {Bernetti}\ and\ \citenamefont
  {Bussi}(2020)}]{bernetti2020}%
  \BibitemOpen
  \bibfield  {author} {\bibinfo {author} {\bibfnamefont {M.}~\bibnamefont
  {Bernetti}}\ and\ \bibinfo {author} {\bibfnamefont {G.}~\bibnamefont
  {Bussi}},\ }\bibfield  {title} {\bibinfo {title} {Pressure control using
  stochastic cell rescaling},\ }\bibfield  {journal} {\bibinfo  {journal} {J.
  Chem. Phys.}\ }\textbf {\bibinfo {volume} {153}},\ \href
  {https://doi.org/10.1063/5.0020514} {10.1063/5.0020514} (\bibinfo {year}
  {2020})\BibitemShut {NoStop}%
\bibitem [{\citenamefont {Nos{\'e}}(1984)}]{nose1984}%
  \BibitemOpen
  \bibfield  {author} {\bibinfo {author} {\bibfnamefont {S.}~\bibnamefont
  {Nos{\'e}}},\ }\bibfield  {title} {\bibinfo {title} {A molecular dynamics
  method for simulations in the canonical ensemble},\ }\href@noop {} {\bibfield
   {journal} {\bibinfo  {journal} {Mol. Phys.}\ }\textbf {\bibinfo {volume}
  {52}},\ \bibinfo {pages} {255} (\bibinfo {year} {1984})}\BibitemShut
  {NoStop}%
\bibitem [{\citenamefont {Hoover}(1985)}]{hoover1985}%
  \BibitemOpen
  \bibfield  {author} {\bibinfo {author} {\bibfnamefont {W.~G.}\ \bibnamefont
  {Hoover}},\ }\bibfield  {title} {\bibinfo {title} {Canonical dynamics:
  Equilibrium phase-space distributions},\ }\href@noop {} {\bibfield  {journal}
  {\bibinfo  {journal} {Phys. Rev. A}\ }\textbf {\bibinfo {volume} {31}},\
  \bibinfo {pages} {1695} (\bibinfo {year} {1985})}\BibitemShut {NoStop}%
\bibitem [{\citenamefont {Parrinello}\ and\ \citenamefont
  {Rahman}(1981)}]{parrinello1981}%
  \BibitemOpen
  \bibfield  {author} {\bibinfo {author} {\bibfnamefont {M.}~\bibnamefont
  {Parrinello}}\ and\ \bibinfo {author} {\bibfnamefont {A.}~\bibnamefont
  {Rahman}},\ }\bibfield  {title} {\bibinfo {title} {Polymorphic transitions in
  single crystals: A new molecular dynamics method},\ }\href@noop {} {\bibfield
   {journal} {\bibinfo  {journal} {J. Appl. Phys.}\ }\textbf {\bibinfo {volume}
  {52}},\ \bibinfo {pages} {7182} (\bibinfo {year} {1981})}\BibitemShut
  {NoStop}%
\bibitem [{\citenamefont {Hess}\ \emph {et~al.}(1997)\citenamefont {Hess},
  \citenamefont {Bekker}, \citenamefont {Berendsen},\ and\ \citenamefont
  {Fraaije}}]{hess1997}%
  \BibitemOpen
  \bibfield  {author} {\bibinfo {author} {\bibfnamefont {B.}~\bibnamefont
  {Hess}}, \bibinfo {author} {\bibfnamefont {H.}~\bibnamefont {Bekker}},
  \bibinfo {author} {\bibfnamefont {H.~J.}\ \bibnamefont {Berendsen}},\ and\
  \bibinfo {author} {\bibfnamefont {J.~G.}\ \bibnamefont {Fraaije}},\
  }\bibfield  {title} {\bibinfo {title} {Lincs: A linear constraint solver for
  molecular simulations},\ }\href@noop {} {\bibfield  {journal} {\bibinfo
  {journal} {J. Comput. Chem.}\ }\textbf {\bibinfo {volume} {18}},\ \bibinfo
  {pages} {1463} (\bibinfo {year} {1997})}\BibitemShut {NoStop}%
\bibitem [{\citenamefont {Darden}\ \emph {et~al.}(1993)\citenamefont {Darden},
  \citenamefont {York},\ and\ \citenamefont {Pedersen}}]{darden1993}%
  \BibitemOpen
  \bibfield  {author} {\bibinfo {author} {\bibfnamefont {T.}~\bibnamefont
  {Darden}}, \bibinfo {author} {\bibfnamefont {D.}~\bibnamefont {York}},\ and\
  \bibinfo {author} {\bibfnamefont {L.}~\bibnamefont {Pedersen}},\ }\bibfield
  {title} {\bibinfo {title} {Particle mesh ewald: An n log (n) method for ewald
  sums in large systems},\ }\href@noop {} {\bibfield  {journal} {\bibinfo
  {journal} {J. Chem. Phys.}\ }\textbf {\bibinfo {volume} {98}},\ \bibinfo
  {pages} {10089} (\bibinfo {year} {1993})}\BibitemShut {NoStop}%
\bibitem [{\citenamefont {Bonomi}\ \emph {et~al.}(2009)\citenamefont {Bonomi},
  \citenamefont {Branduardi}, \citenamefont {Bussi}, \citenamefont {Camilloni},
  \citenamefont {Provasi}, \citenamefont {Raiteri}, \citenamefont {Donadio},
  \citenamefont {Marinelli}, \citenamefont {Pietrucci}, \citenamefont
  {Broglia},\ and\ \citenamefont {Parrinello}}]{plumed}%
  \BibitemOpen
  \bibfield  {author} {\bibinfo {author} {\bibfnamefont {M.}~\bibnamefont
  {Bonomi}}, \bibinfo {author} {\bibfnamefont {D.}~\bibnamefont {Branduardi}},
  \bibinfo {author} {\bibfnamefont {G.}~\bibnamefont {Bussi}}, \bibinfo
  {author} {\bibfnamefont {C.}~\bibnamefont {Camilloni}}, \bibinfo {author}
  {\bibfnamefont {D.}~\bibnamefont {Provasi}}, \bibinfo {author} {\bibfnamefont
  {P.}~\bibnamefont {Raiteri}}, \bibinfo {author} {\bibfnamefont
  {D.}~\bibnamefont {Donadio}}, \bibinfo {author} {\bibfnamefont
  {F.}~\bibnamefont {Marinelli}}, \bibinfo {author} {\bibfnamefont
  {F.}~\bibnamefont {Pietrucci}}, \bibinfo {author} {\bibfnamefont {R.~A.}\
  \bibnamefont {Broglia}},\ and\ \bibinfo {author} {\bibfnamefont
  {M.}~\bibnamefont {Parrinello}},\ }\bibfield  {title} {\bibinfo {title}
  {Plumed: A portable plugin for free-energy calculations with molecular
  dynamics},\ }\href
  {https://doi.org/https://doi.org/10.1016/j.cpc.2009.05.011} {\bibfield
  {journal} {\bibinfo  {journal} {Comput. Phys. Commun.}\ }\textbf {\bibinfo
  {volume} {180}},\ \bibinfo {pages} {1961} (\bibinfo {year}
  {2009})}\BibitemShut {NoStop}%
\bibitem [{\citenamefont {Tribello}\ \emph {et~al.}(2014)\citenamefont
  {Tribello}, \citenamefont {Bonomi}, \citenamefont {Branduardi}, \citenamefont
  {Camilloni},\ and\ \citenamefont {Bussi}}]{tribello2014}%
  \BibitemOpen
  \bibfield  {author} {\bibinfo {author} {\bibfnamefont {G.~A.}\ \bibnamefont
  {Tribello}}, \bibinfo {author} {\bibfnamefont {M.}~\bibnamefont {Bonomi}},
  \bibinfo {author} {\bibfnamefont {D.}~\bibnamefont {Branduardi}}, \bibinfo
  {author} {\bibfnamefont {C.}~\bibnamefont {Camilloni}},\ and\ \bibinfo
  {author} {\bibfnamefont {G.}~\bibnamefont {Bussi}},\ }\bibfield  {title}
  {\bibinfo {title} {Plumed 2: New feathers for an old bird},\ }\href@noop {}
  {\bibfield  {journal} {\bibinfo  {journal} {Comput. Phys. Commun.}\ }\textbf
  {\bibinfo {volume} {185}},\ \bibinfo {pages} {604} (\bibinfo {year}
  {2014})}\BibitemShut {NoStop}%
\end{thebibliography}
%

\end{document}